\documentclass[12pt]{article}
\usepackage{graphics}
\usepackage{graphicx}
\usepackage{color}

\usepackage{float}

\usepackage{amsfonts}

\usepackage{amsmath}
\usepackage{amstext}
\usepackage{amsopn}
\usepackage{amsbsy}
\usepackage{amscd}
\usepackage{amsxtra}
\usepackage{amsthm}
\usepackage{enumerate}
\usepackage{ulem}
\usepackage{psfrag}
\allowdisplaybreaks

\numberwithin{equation}{section}

\renewcommand{\abstractname}

\title{Multi-scale turbulence modeling and maximum information principle. Part 2}

\author{L. Tao\thanks{ Department of Aerospace Engineering, Indian Institute of Technology Madras, Chennai 600 036, India.
 Email: luoyitao@iitm.ac.in}\ \ and 
M. Ramakrishna\thanks{ Department of Aerospace Engineering, Indian Institute of Technology Madras, Chennai 600 036, India.
Email: krishna@ae.iitm.ac.in} }
\date{}

\newcommand{\bunderline}[2][4]{\underline{#2\mkern-#1mu}\mkern#1mu }
\begin{document}
\maketitle
\def\s{\!}
\def\ss{\!\!}
\def\sss{\!\!\!}
\def\l{\left}
\def\r{\right}
\def\bx{{\bf x}}
\def\by{{\bf y}}
\def\bz{{\bf z}}
\def\ba{{\bf a}}
\def\bw{{\bf w}}
\def\tbw{\tilde{\bf w}}
\def\bV{{\bf V}}
\def\Vi{V_i}
\def\Vj{V_j}
\def\Vk{V_k}
\def\Vij{\Vi,_j}
\def\Vji{\Vj,_i}
\def\Vkl{\Vk,_l}
\def\bv{{\bf v}}
\def\vi{v_i}
\def\vj{v_j}
\def\tw{\tilde{w}}
\def\wi{w_i}
\def\twi{\tilde{w}_i}
\def\wj{w_j}
\def\twj{\tilde{w}_j}
\def\wk{w_k}
\def\twk{\tilde{w}_k}
\def\wl{w_l}
\def\twl{\tilde{w}_l}
\def\bm{{\bf m}}
\def\bmp{\bm^{\prime}}
\def\bmpp{\bm^{\prime\prime}}
\def\mi{m_i}
\def\bbm{(\bm)}
\def\bbmp{\s\l(\bmp\r)}
\def\bM{{\bf M}}
\def\bbM{\s\l(\bM;\bx\r)}
\def\bn{{\bf n}}
\def\bnp{\bn^{\prime}}
\def\bnpp{\bn^{\prime\prime}}
\def\bbn{(\bn)}
\def\bbnp{\s\l(\bnp\r)}
\def\bk{{\bf k}}
\def\bkp{\bk^{\prime}}
\def\bbk{(\bk)}
\def\bbkp{\s\l(\bkp\r)}
\def\bK{{\bf K}}
\def\bbK{\s\l(\bK;\bx\r)}
\def\bl{{\bf l}}
\def\blp{\bl^{\prime}}
\def\bbl{(\bl)}
\def\bblp{\s\l(\blp\r)}
\def\bi{{\bf i}}
\def\bip{\bi^{\prime}}
\def\bbi{\s\l(\bi\r)}
\def\bbip{\s\l(\bip\r)}
\def\bj{{\bf j}}
\def\bjp{\bj^{\prime}}
\def\bbj{(\bj)}
\def\bbjp{\s\l(\bjp\r)}
\def\bL{{\bf L}}
\def\bp{{\bf p}}
\def\bpp{\bp^{\prime}}
\def\bbp{(\bp)}
\def\bbpp{(\bpp)}
\def\bq{{\bf q}}
\def\bqp{\bq^{\prime}}
\def\p{p}
\def\bbq{(\bq)}
\def\bbqp{\s\l(\bqp\r)}
\def\P{P}
\def\q{q}
\def\tq{\tilde{q}}
\def\barwiwj{\overline{\wi \wj}}
\def\barwiwjwk{\overline{\wi \wj \wk}}
\def\barwixwix{\overline{\wi(\bx)\, \wi(\bx)}}
\def\barwixwjx{\overline{\wi(\bx)\, \wj(\bx)}}
\def\barwixwjy{\overline{\wi(\bx)\, \wj(\by)}}
\def\barwjywkz{\overline{\wj(\by)\, \wk(\bz)}}
\def\barwjywjy{\overline{\wj(\by)\, \wj(\by)}}
\def\barwkzwkz{\overline{\wk(\bz)\, \wk(\bz)}}
\def\barwixwjywkz{\overline{\wi(\bx)\,\wj(\by)\,\wk(\bz)}}
\def\barwixwjxwkx{\overline{\wi(\bx)\,\wj(\bx)\,\wk(\bx)}}
\def\barwipxwjpywkpz{\overline{\wi,_{i'}\ss(\bx)\,\wj,_{j'}\ss(\by)\,\wk,_{k'}\ss(\bz)}}
\def\barwixwjywkzwla{\overline{\wi(\bx,t)\,\wj(\by,t)\,\wk(\bz,t)\,\wl(\ba,t)}}
\def\barbwbw{\overline{\bw\bw}}
\def\bartwimtwjn{\overline{\twi\bbm\twj\bbn}}
\def\bartwimtwln{\overline{\twi\bbm\twl\bbn}}
\def\bartwiktwjl{\overline{\twi\bbk\,\twj\bbl}}
\def\bartwititwjtj{\overline{\twi(t;\bi)\twj(t;\bj)}}
\def\bartwitktwjtl{\overline{\twi(t;\bk)\twj(t;\bl)}}
\def\bartwititwjtjtwktk{\overline{\twi(t;\bi)\,\twj(t;\bj)\,\twk(t;\bk)}}
\def\B{{\cal B}}
\def\DD{{\cal D}}
\def\DDi{\DD\s\l(i\r)}
\def\DDj{\DD\s\l(j\r)}
\def\DDk{\DD\s\l(k\r)}
\def\n{{\rm N}}
\def\nO{{\cal O}\s\l(\n\r)}
\def\b{b}
\def\Aj{A_j}
\def\Al{A_l}
\def\Ak{A_k}
\def\Bi{B_i}
\def\Bj{B_j}
\def\Bk{B_k}
\def\Bl{B_l}
\def\Cji{C_{ji}}
\def\Cli{C_{li}}
\def\Clk{C_{lk}}
\def\Cjk{C_{jk}}
\def\Clm{C_{lm}}
\def\Cnm{C_{nm}}
\def\D{D}
\def\E{E}
\def\EE{{\cal E}}
\def\H{H}
\def\Hij{\H_{ij}}
\def\Wi{W_i}
\def\Wk{W_k}
\def\Wl{W_l}
\def\W{W}
\def\Wij{W_{ij}}
\def\Wil{W_{il}}
\def\Wjl{W_{jl}}
\def\Wkl{W_{kl}}
\def\WiHOMS{\overline{W}_i}
\def\WkHOMS{\overline{W}_k}
\def\WlHOMS{\overline{W}_l}
\def\W{W}
\def\WHOMS{\overline{\W}}
\def\Wij{W_{ij}}
\def\Wil{W_{il}}
\def\Wjl{W_{jl}}
\def\Wkl{W_{kl}}
\def\WijHOMS{\overline{W}_{ij}}
\def\WilHOMS{\overline{W}_{il}}
\def\WjlHOMS{\overline{W}_{jl}}
\def\WklHOMS{\overline{W}_{kl}}
\def\soc{\beta}
\def\socR{\soc^{(R)}}
\def\socI{\soc^{(I)}}
\def\socDim{\soc^{(R)}}
\def\socDimO{\soc^{(a)}}
\def\socS{\Pi}
\def\toc{\gamma}
\def\tocR{\toc^{(R)}}
\def\tocI{\toc^{(I)}}
\def\tocDim{\toc}
\def\tocDimR{\toc^{(R)}}
\def\tocDimI{\toc^{(I)}}
\def\tocDimIO{\toc^{(a)}}
\def\tocDimIOpm{\toc^{(Ia)\pm}}
\def\tocDimIOmp{\toc^{(Ia)\mp}}
\def\tocDimIOp{\toc^{(Ia)+}}
\def\tocDimIOm{\toc^{(Ia)-}}
\def\foc{\delta}
\def\focDim{\hat{\foc}}
\def\foocG{\delta^G}
\def\siocG{\zeta^G}
\def\eiocG{\theta^G}
\def\bN{{\bf N}}
\def\bH{{\bf H}}
\def\bR{\mathbb{R}}
\def\bX{{\bf X}}
\def\barbwbwbw{\overline{\bw\bw\bw}}
\def\G{{\cal G}}
\def\xi{x_i}
\def\xj{x_j}
\def\xk{x_k}
\def\xl{x_l}
\def\balpha{{{\bf \alpha}}}
\def\mbi{b_i}
\def\mbj{b_j}
\def\mbk{b_k}
\def\inform{I}
\def\pdf{f}
\def\pdfG{f_G}
\def\pdfD{f_D}
\def\pdfL{f^{(L)}}
\def\pdfH{f^{(H)}}
\def\varvec{\hat{\hat{\bw}}}
\def\Reprevarvec{\hat{\hat{\bw}}}
\def\ReprevarvecL{\hat{\hat{\bw}}^{(L)}}
\def\ReprevarvecH{\hat{\hat{\bw}}^{(H)}}
\def\bkH{{\bf k}^{(H)}}
\def\LagMultiplier{\lambda}
\def\bK{{\bf K}}
\def\hhw{\hat{\hat{w}}}
\def\Det{\inform_D}
\def\K{\inform_T}
\def\vort{\omega}
\def\BETA{\pmb{\beta}}
\def\ip{i^{\prime}}
\def\jp{j^{\prime}}
\def\kp{k^{\prime}}
\def\lp{l^{\prime}}
\def\twip{\tilde{w}_{\ip}}
\def\twjp{\tilde{w}_{\jp}}
\def\twkp{\tilde{w}_{\kp}}
\def\twlp{\tilde{w}_{\lp}}
\def\imaginary{\imath}
\def\Real{\text{RE}}
\def\Imag{\text{IM}}
\def\VGrad{V}
\def\twl{\tilde{w}_l}
\def\twm{\tilde{w}_m}
\def\twn{\tilde{w}_n}
\def\twmp{\tilde{w}_{m'}}
\def\twnp{\tilde{w}_{n'}}
\def\twmpp{\tilde{w}_{m''}}
\def\twnpp{\tilde{w}_{n''}}
\def\bm{{\bf m}}
\def\bmp{\bm^{\prime}}
\def\bmpp{\bm^{\prime\prime}}
\def\mi{m_i}
\def\bbm{(\bm)}
\def\bbmp{\s\l(\bmp\r)}
\def\bM{{\bf M}}
\def\bbM{\s\l(\bM;\bx\r)}
\def\bn{{\bf n}}
\def\bnp{\bn^{\prime}}
\def\bnpp{\bn^{\prime\prime}}
\def\bbn{(\bn)}
\def\bbnp{\s\l(\bnp\r)}
\def\bk{{\bf k}}
\def\bkp{\bk^{\prime}}
\def\bbk{(\bk)}
\def\bbkp{\s\l(\bkp\r)}
\def\bK{{\bf K}}
\def\bbK{\s\l(\bK;\bx\r)}
\def\bl{{\bf l}}
\def\blp{\bl^{\prime}}
\def\bbl{(\bl)}
\def\bblp{\s\l(\blp\r)}
\def\bi{{\bf i}}
\def\bip{\bi^{\prime}}
\def\bbi{\s\l(\bi\r)}
\def\bbip{\s\l(\bip\r)}
\def\bj{{\bf j}}
\def\bjp{\bj^{\prime}}
\def\bbj{(\bj)}
\def\bbjp{\s\l(\bjp\r)}
\def\bL{{\bf L}}
\def\bp{{\bf p}}
\def\bpp{\bp^{\prime}}
\def\bbp{(\bp)}
\def\bbpp{(\bpp)}
\def\bq{{\bf q}}
\def\bqp{\bq^{\prime}}
\def\p{p}
\def\bbq{(\bq)}
\def\bbqp{\s\l(\bqp\r)}
\def\P{P}
\def\q{q}
\def\tq{\tilde{q}}
\def\barqq{\overline{\q \q}}
\def\barwiwj{\overline{\wi \wj}}
\def\barwiwjwk{\overline{\wi \wj \wk}}
\def\barwiwjwkwl{\overline{\wi \wj \wk \wl}}
\def\barwixwjy{\overline{\wi(\bx)\, \wj(\by)}}
\def\barwkxwkx{\overline{\wk(\bx)\, \wk(\bx)}}
\def\barwixwjywkz{\overline{\wi(\bx)\,\wj(\by)\,\wk(\bz)}}
\def\barwixwjywkzwla{\overline{\wi(\bx)\,\wj(\by)\,\wk(\bz)\,\wl(\ba)}}
\def\barbwbw{\overline{\bw\bw}}
\def\bartwimtwjn{\overline{\twi\bbm\twj\bbn}}
\def\bartwimtwln{\overline{\twi\bbm\twl\bbn}}
\def\bartwiktwjl{\overline{\twi\bbk\,\twj\bbl}}
\def\bartwititwjtj{\overline{\twi(t;\bi)\twj(t;\bj)}}
\def\bartwitktwjtl{\overline{\twi(t;\bk)\twj(t;\bl)}}
\def\bartwititwjtjtwktk{\overline{\twi(t;\bi)\,\twj(t;\bj)\,\twk(t;\bk)}}
\def\bartwiitwjj{\overline{\twi(\bi)\twj(\bj)}}
\def\bartwiktwjl{\overline{\twi(\bk)\twj(\bl)}}
\def\bartwiitwjjtwkk{\overline{\twi(\bi)\,\twj(\bj)\,\twk(\bk)}}
\def\ReynoldsNo{\text{Re}}
\def\tDim{\hat{t}}
\def\WNS{{\cal W}}
\def\transformedsoc{\hat{\soc}}
\def\transformedsocDim{\hat{\soc}}
\def\transformedsocDimO{\hat{\soc}^{(a)}}
\def\transformedtocDim{\hat{\toc}}
\def\transformedtocDimR{\hat{\toc}^{(R)}}
\def\transformedtocDimI{\hat{\toc}^{(I)}}
\def\transformedtocDimRO{\hat{\toc}^{(Ra)}}
\def\transformedtocDimIO{\hat{\toc}^{(Ia)}}
\def\transformedbk{\hat{\mathbf{k}}}
\def\transformedbm{\hat{\mathbf{m}}}
\def\transformedbn{\hat{\mathbf{n}}}
\def\transformedbl{\hat{\mathbf{l}}}
\def\transformedbj{\hat{\mathbf{j}}}
\def\transformedk{\hat{k}}
\def\transformedm{\hat{m}}
\def\transformedn{\hat{n}}
\def\transformedl{\hat{l}}
\def\betaT{\sqrt{\s\s\beta}}
\def\g{\gamma}
\def\gO{\g^{(a)}}
\def\bO{\betaT^{(a)}}
\def\B{B}
\def\G{G}

\def\br{\mathbf{r}}
\def\bs{\mathbf{s}}
\def\wm{w_m}
\def\wn{w_n}
\def\W{U}
\def\Wij{\W_{ij}}
\def\Wkj{\W_{kj}}
\def\Wji{\W_{ji}}
\def\Wikj{\W_{ikj}}
\def\Wkij{\W_{kij}}
\def\Wjki{\W_{jki}}
\def\Wkji{\W_{kji}}
\def\Wjlk{\W_{jlk}}
\def\Wlkj{\W_{lkj}}
\def\Q{Q}
\def\Qi{\Q_i}
\def\Qj{\Q_j}
\def\Qk{\Q_k}
\def\tW{\tilde{U}}
\def\tWij{\tW_{ij}}
\def\tWkj{\tW_{kj}}
\def\tWji{\tW_{ji}}
\def\tWikj{\tW_{ikj}}
\def\tWkij{\tW_{kij}}
\def\tWjki{\tW_{jki}}
\def\tWkji{\tW_{kji}}
\def\tWjlk{\tW_{jlk}}
\def\tWjli{\tW_{jli}}
\def\tWlkj{\tW_{lkj}}
\def\tWlki{\tW_{lki}}
\def\tQ{\tilde{Q}}
\def\tQi{\tQ_i}
\def\tQj{\tQ_j}
\def\tQk{\tQ_k}
\def\tWImag{\tilde{U}^{(I)}}

\def\tWOneOneDim{\beta^{(R)}}
\def\tWOneOneOneDim{\gamma^{(I)}_{1}}
\def\tWOneTwoOneDim{\gamma^{(I)}_{2}}
\def\ok{\overline{k}}
\def\obk{\overline{\bk}}
\def\obr{\overline{\br}}
\def\ot{\overline{t}}

\def\socAsy{\beta^{(a)}}
\def\tocAsy{\gamma^{(a)}}
\def\focAsy{\delta^{(a)}}
\def\foc{\delta}
\def\fociAsy{\foc^{(a)}_i}
\def\focIAsy{\foc^{(a)}_1}
\def\focIIAsy{\foc^{(a)}_2}
\def\focIIIAsy{\foc^{(a)}_3}
\def\foci{\foc_i}
\def\focI{\foc_1}
\def\focII{\foc_2}
\def\focIII{\foc_3}
\def\QAsy{Q^{(a)}}
\def\tQAsy{\tQ^{(a)}}
\def\WAsy{\W^{(a)}}
\def\tWAsy{\tW^{(a)}}
\def\kp{k^{\prime}}
\def\lp{l^{\prime}}
\def\bkp{\bk^{\prime}}
\def\blp{\bl^{\prime}}
\def\Kinfty{K^{\infty}}
\def\psiAsy{\psi^{(a)}}
\def\tWtocAsy{\tW^{(Ia)}}

\def\gW{\dot{\W}}
\def\gQ{\dot{\Q}}
\def\gsoc{\dot{\soc}}
\def\gtoc{\dot{\toc}}
\def\gfoc{\dot{\foc}}
\def\gfocI{\dot{\focI}}
\def\gfocII{\dot{\focII}}
\def\gsocAsy{\gsoc^{(a)}}
\def\gtocAsy{\gtoc^{(a)}}
\def\gfocAsy{\dot{\foc}^{(a)}}
\def\gfocIAsy{\dot{\foc}^{(a)}_1}
\def\gfocIIAsy{\dot{\foc}^{(a)}_2}
\def\ggfoc{\ddot\foc}

\def\dsoc{\dot{\soc}}
\def\dtoc{\dot{\toc}}
\def\dfoc{\dot{\foc}}
\def\dfocI{\dot{\foc}_1}
\def\dfocII{\dot{\foc}_2}
\def\dsocAsy{\dot{\soc}^{(a)}}
\def\dtocAsy{\dot{\toc}^{(a)}}
\def\dfocAsy{\dot{\foc}^{(a)}}
\def\dfocIAsy{\dot{\foc}^{(a)}_1}
\def\dfocIIAsy{\dot{\foc}^{(a)}_2}

\def\w{w}
\def\tw{\tilde{\w}}
\def\underlinei{\bunderline[2]{i}}
\def\underlinej{\bunderline{j}}
\def\underlinek{\bunderline{k}}
\def\underlinel{\bunderline[2]{l}}
\def\underlinem{\bunderline{m}}
\def\underlinen{\bunderline{n}}

\def\tk{\tilde{k}}
\def\tbk{\tilde{\bk}}

\def\SS{s}
\def\CC{c}
\def\Kinfty{K^{\infty}}

\def\MPeak{k_{2P}}
\def\MValley{k_{2V}}
\def\MMiddle{k_{20}(k_1;\sigma)}
\def\SupportNbkp{{\cal S}_L(\sigma)}
\def\Supportfbkp{{\cal S}_f(\sigma)}
\def\unitkone{\bar{k}_1}
\def\unitktwo{\bar{k}'_2}
\def\unitlone{\bar{l}_1}
\def\unitltwo{\bar{l}_2}

\def\konemin{k_{1min}(\sigma)}
\def\koneminzero{k_{1min}(0)}
\def\MExtremeDomain{{\cal S}_M(\sigma)}
\def\MExtremeDomainVoid{{\cal S}_M}
\def\MaxSupportGGbkpbl{{\cal S}_G(\sigma)}
\def\MaxSupportNbkp{{\cal S}_L(\sigma)}
\def\MaxSupportNbkpVoid{{\cal S}_L}
\def\MaxSupportNbkpNeg{{\cal S}_L^{-}(\sigma)}
\def\MaxSupportNbkpNegVoid{{\cal S}_L^{-}}
\def\MaxSupportNbkpPos{{\cal S}_L^{+}(\sigma)}
\def\MaxSupportNbkpPosVoid{{\cal S}_L^{+}}
\def\MaxSupportGbkpblNodes{{\cal N}_G(\sigma)}
\def\MaxSupportNbkpNodes{{\cal N}_{L}(\sigma)}
\def\MaxSupportNbkpNegNodes{{\cal N}_{L}^{-}(\sigma)}
\def\MaxSupportNbkpPosNodes{{\cal N}_{L}^{+}(\sigma)}
\def\MaxSupportGbkpblFvalues{{\cal G}(\sigma)}
\def\MaxSupportGbkpblFvaluesNeg{{\cal G}^{-}(\sigma)}
\def\MaxSupportGbkpblFvaluesPos{{\cal G}^{+}(\sigma)}

\def\I{I}
\def\J{J}
\def\M{M}
\def\Incrementione{\hat{\I}_1}
\def\Incrementitwo{\hat{\I}_2}
\def\Incrementjone{\hat{\J}_1}
\def\Incrementjtwo{\hat{\J}_2}
\def\MaxSupportFNbkp{{\cal S}_{L_G}(\sigma)}
\def\MaxSupportFNbkpNeg{{\cal S}_{L_G}^{-}(\sigma)}
\def\MaxSupportFNbkpPos{{\cal S}_{L_G}^{+}(\sigma)}
\def\MaxSupportGbkpbl{{\cal S}_{\gamma}(\sigma)}
\def\MaxSupportGNbkp{{\cal S}_{L_{\gamma}}(\sigma)}
\def\CharacteristicFunction{\chi_{\MaxSupportNbkp}}

\def\MaxSupportLbk{{\cal S}_{L}(\sigma)}
\def\MaxSupportLbkLvalues{{\cal L}(\sigma)}
\def\MaxSupportLbkLvaluesNeg{{\cal L}^{-}(\sigma)}
\def\MaxSupportLbkLvaluesPos{{\cal L}^{+}(\sigma)}
\def\MaxSupportLbkNodes{{\cal L}(\sigma)}

\def\konemin{k_{1min}(\sigma)}
\def\koneminzero{k_{1min}(0)}
\def\MExtremeDomain{{\cal S}_M(\sigma)}
\def\MExtremeDomainVoid{{\cal S}_M}
\def\MaxSupportGbkpbl{{\cal S}_G(\sigma)}
\def\MaxSupportGmax{{\cal S}_G}
\def\MaxSupportNbkp{{\cal S}_L(\sigma)}
\def\MaxSupportNbkpVoid{{\cal S}_L}
\def\MaxSupportNbkpNeg{{\cal S}_L^{-}(\sigma)}
\def\MaxSupportNbkpNegVoid{{\cal S}_L^{-}}
\def\MaxSupportNbkpPos{{\cal S}_L^{+}(\sigma)}
\def\MaxSupportNbkpPosVoid{{\cal S}_L^{+}}
\def\MaxSupportGbkpblNodes{{\cal N}_G(\sigma)}
\def\MaxSupportNbkpNodes{{\cal N}_{L}(\sigma)}
\def\MaxSupportNbkpNegNodes{{\cal N}_{L}^{-}(\sigma)}
\def\MaxSupportNbkpPosNodes{{\cal N}_{L}^{+}(\sigma)}
\def\MaxSupportGbkpblFvalues{{\cal G}(\sigma)}
\def\MaxSupportGbkpblFvaluesNeg{{\cal G}^{-}(\sigma)}
\def\MaxSupportGbkpblFvaluesPos{{\cal G}^{+}(\sigma)}

\def\I{I}
\def\J{J}
\def\M{M}
\def\Incrementione{\hat{\I}_1}
\def\Incrementitwo{\hat{\I}_2}
\def\Incrementjone{\hat{\J}_1}
\def\Incrementjtwo{\hat{\J}_2}
\def\MaxSupportFNbkp{{\cal S}_{L_G}(\sigma)}
\def\MaxSupportFNbkpNeg{{\cal S}_{L_G}^{-}(\sigma)}
\def\MaxSupportFNbkpPos{{\cal S}_{L_G}^{+}(\sigma)}
\def\MaxSupportGbkpbl{{\cal S}_{\gamma}(\sigma)}
\def\MaxSupportgammamax{{\cal S}_{\gamma}(0)}
\def\MaxSupportGNbkp{{\cal S}_{L_{\gamma}}(\sigma)}
\def\CharacteristicFunction{\chi_{\MaxSupportNbkp}}
\def\CharacteristicFunctionNeg{\chi_{\MaxSupportNbkpNeg}}
\def\MaxSupportDbkpblbm{{\cal S}_{\delta}(\sigma)}

\def\LPObjectiveFunctionCoefficient{c}
\def\LPConstraintCoefficient{a}
\def\MaxSupportbeta{{\cal S}^{-}_{\beta}(\sigma)}
\def\MaxSupportbetaVoid{{\cal S}^{-}_{\beta}}
\def\MaxSupportbetaUB{k_{2U\s B}}

\def\NodeSNNeg{N^{-}}
\def\TriangleSNNeg{T^{-}}
\def\PointMatrixSNNeg{{\cal P}^{-}_L(\sigma)}
\def\PointMatrixSNNegVoid{{\cal P}^{-}_L}
\def\ConnectivityMatrixSNNeg{{\cal C}^{-}_L(\sigma)}
\def\ConnectivityMatrixSNNegVoid{{\cal C}^{-}_L}
\def\NodeNumberTotalSNNeg{N_{}{\raisebox{-1pt}{\text{\tiny $\PointMatrixSNNegVoid$}}}}  \def\TriangleNumberTotalSNNeg{N_{}{\raisebox{-1pt}{\text{\tiny $\ConnectivityMatrixSNNegVoid$}}}}  \def\MaxSupportNbkpNegTriangles{{\cal T}_{L}^{-}(\sigma)}
\def\MaxSupportNbkpNegTrianglesVoid{{\cal T}_{L}^{-}}
\def\NodeSNPos{N^{+}}
\def\TriangleSNPos{T^{+}}
\def\PointMatrixSNPos{{\cal P}^{+}_L(\sigma)}
\def\PointMatrixSNPosVoid{{\cal P}^{+}_L}
\def\ConnectivityMatrixSNPos{{\cal C}^{+}_L(\sigma)}
\def\ConnectivityMatrixSNPosVoid{{\cal C}^{+}_L}
\def\NodeNumberTotalSNPos{N_{\text{\tiny $\PointMatrixSNPosVoid$}}}
\def\TriangleNumberTotalSNPos{N_{\text{\tiny $\ConnectivityMatrixSNPosVoid$}}}
\def\MaxSupportNbkpPosTriangles{{\cal T}_{L}^{+}(\sigma)}
\def\MaxSupportNbkpPosTrianglesVoid{{\cal T}_{L}^{+}}
\def\MaxSupportNbkpTriangles{{\cal T}_{L}(\sigma)}

\def\PointMatrixSN{{\cal P}_L(\sigma)}
\def\PointMatrixSNVoid{{\cal P}_L}
\def\ConnectivityMatrixSN{{\cal C}_L(\sigma)}
\def\ConnectivityMatrixSNVoid{{\cal C}_L}
\def\MaxSupportGbkpblTrangles{{\cal T}_G(\sigma)}
\def\MaxSupportGbkpblTranglesVoid{{\cal T}_G}

\def\ShapeFunction{\varphi}
\def\NodeSNPosNeg{N^{\pm}}
\def\TriangleSNPosNeg{T^{\pm}}

\def\CharacteristicFunctionTriangleSNPosi{\chi_{\TriangleSNPos_i}}
\def\CharacteristicFunctionTriangleSNPosj{\chi_{\TriangleSNPos_j}}
\def\CharacteristicFunctionTriangleSNPosk{\chi_{\TriangleSNPos_k}}
\def\CharacteristicFunctionTriangleSNPosl{\chi_{\TriangleSNPos_l}}
\def\CharacteristicFunctionTriangleSNPosm{\chi_{\TriangleSNPos_m}}

\def\CharacteristicFunctionTriangleSNNegi{\chi_{\TriangleSNNeg_i}}
\def\CharacteristicFunctionTriangleSNNegj{\chi_{\TriangleSNNeg_j}}
\def\CharacteristicFunctionTriangleSNNegk{\chi_{\TriangleSNNeg_k}}
\def\CharacteristicFunctionTriangleSNNegl{\chi_{\TriangleSNNeg_l}}
\def\CharacteristicFunctionTriangleSNNegm{\chi_{\TriangleSNNeg_m}}

\def\LAsy{L^{(a)}}
\def\GAsy{G^{(a)}}
\def\bo{\mathbf{0}}

\def\dAtocAsy{\dtocAsy}
\def\gAtocAsy{\dtocAsy}

\begin{abstract}
We consider two-dimensional homogeneous shear turbulence within the context of optimal control, a multi-scale turbulence model containing the fluctuation velocity and pressure correlations up to the fourth order; The model is formulated on the basis of the Navier-Stokes equations, Reynolds average, the constraints of inequality from both physical and mathematical considerations, the turbulent energy density as the objective to be maximized, and the fourth order correlations as the control variables.  Without imposing the maximization and the constraints, the resultant equations of motion in the Fourier wave number space are formally solved  to obtain the transient state solutions, the asymptotic state solutions and the evolution of a transient toward an asymptotic under certain conditions. The asymptotic state solutions are characterized by the dimensionless exponential time rate of growth $2\sigma$ which has an upper bound of $2\sigma_{\max}$ $=$ $0$; The asymptotic solutions can be obtained from a linear objective convex programming. For the asymptotic state solutions of the reduced model containing the correlations up to the third order, the optimal control problem reduces to linear programming with the primary component of the third order correlations or a related integral quantity as the control variable; the supports of the second and third order correlations are estimated for the sake of numerical simulation; the existence of feasible solutions is demonstrated when the related quantity is the control variable. The relevance of the formulation to flow stability analysis is suggested.
\end{abstract}

\section{Introduction}
 \ \ \ \ 
In \cite{TaoRamakrishna2010Part1} (to be referred to as PART I hereafter), we have presented a framework of multi-scale turbulence modeling with the correlations up to the fourth order, based on the Navier-Stokes equations, Reynolds average, the constraints of inequality from the physical considerations and the Cauchy-Schwarz inequality and so on, the maximum information principle and the alternative objective function such as turbulent energy contained in the flow. The model is an optimal control problem with the fourth order correlations as the control variables.

We have adopted the notion of the information $\inform$ and the maximum information principle, unlike that of
 Edwards and McComb \cite{EdwardsMcComb1969} who resorted to the entropy method to fix certain response functions of an isotropic homogeneous model through the maximization of entropy.
 The interpretation of the information $\inform$ as a thermodynamic entropy raises an interesting issue; If we view the Navier-Stokes equations as a consequence of the second law of thermodynamics in that $\mu\geq 0$ or $\nu\geq 0$, the question arises on how to justify $\inform$ as another entropy of thermodynamic nature, in addition to the one leading to $\nu\geq 0$. As an alternative,
 one may view the information as the mixing entropy as done in \cite{RobertSommeria1992} (of macro-scales).
 The next important question is how to make the evaluation of $\inform$ computationally feasible under certain constraints such as the equations of evolution for the correlations and the positive semi-definiteness of the Reynolds stress listed in PART I. 
 From the point of view of modeling, the maximization of the information  $\inform$ under the constraints reflects the uncertainty in our inference based on the data and information available and specified, a ground for our adoption of the notion. 

 To understand the mathematical challenges faced by the formulation of PART I, we apply it to two-dimensional homogeneous shear turbulence in this work.
 We need to modify the formulation slightly, especially the alternative objective function, in order to cope with the infinite domain of motion; the turbulent energy density is used as the objective to be maximized.
 On the basis of the supposed homogeneity, Fourier transforms are applied to the correlations, and two primary integro-differential equations are obtained in the Fourier wave number space, one for the second order correlations and the other for the third order correlations.
 Without imposing the objective maximization and the constraints of inequality, these two equations can be solved formally by the method of characteristics and by the separation of variables, respectively: (i) The solutions of the former hold for rather general initial conditions and describe the corresponding evolution of the motion (the transient state solutions). (ii) The solutions from the latter hold for some special initial conditions and have an exponential dependence on time with spatial supports (the asymptotic state solutions). (iii) Under certain conditions yet to be studied, a transient solution evolves, at great time, into a corresponding asymptotic state solution, and this evolution process involves the turbulent energy transfer among different wave numbers or different spatial scales.
 The asymptotic state solutions are characterized by the dimensionless exponential time rate of growth, $2\sigma$, compatible with the studies of three-dimensional homogeneous shear turbulence (\cite{BernardPseziale1992}, \cite{DeSouzaetal1995}, \cite{IsazaCollins2009}, \cite{Isazaetal2009}, \cite{Leeetal1990}, \cite{Piquet1999}, \cite{Rohretal1988}, \cite{SagautCambon2008}, \cite{Tavoularis1985}, \cite{TavoularisKarnik1989}),
 and the rate of growth is bounded from above by $\sigma_{\max}$ $=$ $0$,
 as argued mathematically with the help of certain constraints of inequality; 
 The existence of such an upper bound in the associated three-dimensional shear turbulence will be explored in a report forthcoming.
  The asymptotic solutions of the fourth order model are to be obtained from convex programming, with mathematical proofs  to argue for the convexity of the quadratic constraints on the basis of linearization.
 
 For the asymptotic state solutions of the reduced model with the correlations up to the third order,
 the objective and all the constraints are linear, and the optimization reduces to a linear programming problem,
with the possibility of either the primary component $\dtocAsy(\bk,\bl)$ of the third order correlations or an associated  integral quantity $\LAsy(\bk)$ as the optimal control variable.
  For the sake of exploring the multi-scale structure of the turbulent motion,
 we relax mathematically the restriction of $\sigma_{\max}$ $=$ $0$ to $\sigma_{\max}$ $\leq$ $1/2$,
 which is justified and allowed by the two additional arguments for the existence of $\sigma_{\max}>0$ in the reduced model. 
At a specific $\sigma \in [0, \sigma_{\max})$, the asymptotic solutions of the correlations are effectively nontrivial only inside certain bounded domains of the wave number spaces; and the sizes of the domains shrink as $\sigma$ increases from $0$ to $\sigma_{\max}$. In the case that $\LAsy(\bk)$ is the optimal control variable, there exist feasible solutions for any $\sigma\in[0,0.5)$, implying that the reduced model may be inadequate to simulate the transient states which do not decay.

The homogeneous turbulence modeling problem concerned may also be viewed as a stability problem due to the averaged flow field is held constant. It raises the possibility that the framework of optimization developed here may have relevance to flow stability analysis.
 Further works need to be done to assess the adequacy and the feasibility of the idea.
 
This paper is organized as follows.
 In Section~\ref{sec:HomogeneousTurbulence}, we develop the differential equations, the constraints of inequality and the objective function in physical and Fourier wave number spaces. 
 In Section~\ref{Sec:FormalSolutionsWithoutEnforcingConstraints}, without enforcing the maximization of the objective function and the constraints of inequality, we present the formal solutions, both transient and asymptotic, to the primary integro-differential equations.
 Also, we discuss the effects of  bounded solutions at finite time on the distributions of the correlations in the wave number spaces, the intrinsic equalities of zero sum balance for some integral quantities, and the evolution of a transient state solution to an asymptotic state solution under certain conditions.
 We also address the relevance of the formulation of turbulence modeling as optimal control to flow stability analysis.
 In Section~\ref{sec:AsymptoticStateSolution}, we analyze in detail the asymptotic state solutions, especially for the case of the reduced model. The convexity of the quadratic constraints is demonstrated;
 Various restrictions on the exponential growth rate are discussed;
 The possible structures of the correlations in the wave number space are explored
 and two possible implementations of numerical approximations are presented.

\section{\label{sec:HomogeneousTurbulence}Basic Formulation}
\ \ \ \ 
To examine how challenging the formulation proposed in PART I is mathematically and whether it can produce adequate results, we consider the homogeneous shear turbulence in $\DD=$ $\mathbb{R}^2$ with an average velocity field of
\begin{align}
V_1=S x_2,\quad
V_2=0
\label{HST_AverageVelocityField}
\end{align}
where $S$ is a nontrivial constant. Since the average flow field of $\Vi$ and $\P$ is not affected by the correlations, we need to consider only the fluctuation fields of $\wi(\bx, t)$ and $\q(\bx,t)$ governed by  
\begin{align}
\frac{\partial\wk}{\partial x_k}=0
\label{HST_DivergenceFreeInPhysicalSpace_Dim}
\end{align}
\begin{align}
\frac{\partial \wi}{\partial t}
+S x_2 \frac{\partial \wi}{\partial x_1}
+S \delta_{i1} w_2
+\frac{\partial (\wi \wk)}{\partial x_k}
=
-\frac{\partial \q}{\partial x_i}
+\nu \frac{\partial^2 \wi}{\partial x_k \partial x_k}
\label{HST_CLMInPhysicalSpace_Dim}
\end{align}
and
\begin{align}
\frac{\partial^2 \q}{\partial x_k \partial x_k}
=-2 S\frac{\partial w_2}{\partial x_1}
-\frac{\partial^2 (\wl \wk)}{\partial x_k \partial x_l}
\label{HST_PressureInPhysicalSpace_Dim}
\end{align}

Due to the symmetry of the flows associated with $S<0$ and $S>0$, we will restrict to $S>0$ in this work. Under this restriction we can introduce the dimensionless quantities through
\begin{align}
t=\frac{t^{\,\prime}}{S},\quad
x_i=\sqrt{\frac{\nu}{S}}\,x^{\,\prime}_i,\quad
w_i=\sqrt{\nu S}\,w^{\,\prime}_i,\quad
\q=\nu S q^{\,\prime}
\end{align}
and non-dimensionalize the above equations of motion to obtain the forms of
\begin{align}
\frac{\partial\wk}{\partial x_k}=0,
\label{HST_DivergenceFreeInPhysicalSpace}
\end{align}
\begin{align}
\frac{\partial \wi}{\partial t}
+ x_2 \frac{\partial \wi}{\partial x_1}
+ \delta_{i1} w_2
+\frac{\partial (\wi \wk)}{\partial x_k}
=
-\frac{\partial \q}{\partial x_i}
+ \frac{\partial^2 \wi}{\partial x_k \partial x_k}
\label{HST_CLMInPhysicalSpace}
\end{align}
and
\begin{align}
\frac{\partial^2 \q}{\partial x_k \partial x_k}
=-2 \frac{\partial w_2}{\partial x_1}
-\frac{\partial^2 (\wl \wk)}{\partial x_k \partial x_l}
\label{HST_PressureInPhysicalSpace}
\end{align}
Here, we have removed the accent $'$ for the sake of brevity.

\subsection{Evolution Equations and Homogeneity}
\ \ \ \
Considering that the probability density function $\pdf$ will not be present explicitly in the optimization problem, we can incorporate the supposed homogeneity in the first place in order to simplify the mathematical treatment. To this end, we construct, on the basis of \eqref{HST_DivergenceFreeInPhysicalSpace} through \eqref{HST_PressureInPhysicalSpace}, the following equations for the evolution of the multi-point correlations up to the fourth order,
\begin{align}
&
\frac{\partial}{\partial x_k}\overline{\wk(\bx) \wj(\by)}=0,\quad
\frac{\partial}{\partial x_k}\overline{\wk(\bx) \wj(\by) \wl(\bz)}=0,\quad
\frac{\partial}{\partial x_i}\overline{\wi(\bx) \wj(\by) \wk(\bz) \wl(\bz')}=0,
\notag\\[4pt]
&
\frac{\partial}{\partial x_k}\overline{\wk(\bx) \q(\by)}=0,\quad
\frac{\partial}{\partial x_k}\overline{\wk(\bx) \wl(\by) \q(\bz)}=0
\label{HST_DivergenceFreeInPhysicalSpace_2p_3p}
\end{align}
\begin{align}
&
\frac{\partial}{\partial t}\overline{\wi(\bx) \wj(\by)}
+ \bigg(x_2\frac{\partial }{\partial x_1}+y_2\frac{\partial }{\partial y_1}\bigg) \overline{\wi(\bx) \wj(\by)}
+ \delta_{i1} \overline{w_2(\bx) \wj(\by)}
\notag\\[4pt]
&
+ \delta_{j1} \overline{\wi(\bx) w_2(\by)}
+\frac{\partial}{\partial x_k}\overline{\wi(\bx) \wk(\bx) \wj(\by)}
+\frac{\partial}{\partial y_k}\overline{\wi(\bx) \wk(\by) \wj(\by)}
\notag\\[4pt]
=&
-\frac{\partial }{\partial x_i}\overline{\q(\bx) \wj(\by)}
-\frac{\partial }{\partial y_j}\overline{\wi(\bx) \q(\by)}
+ \bigg(\frac{\partial^2}{\partial x_k \partial x_k}+\frac{\partial^2}{\partial y_k\partial y_k}\bigg)\overline{\wi(\bx) \wj(\by)}
\label{HST_CLMInPhysicalSpace_2p}
\end{align}
\begin{align}
&
 \frac{\partial }{\partial t}\overline{\wi(\bx)\wj(\by)\wk(\bz)}
+ x_2 \frac{\partial }{\partial x_1}\overline{\wi(\bx)\wj(\by)\wk(\bz)}
+ y_2 \frac{\partial }{\partial y_1}\overline{\wj(\by)\wi(\bx)\wk(\bz)}
\notag\\[4pt]
&
+ z_2 \frac{\partial }{\partial z_1}\overline{\wi(\bx)\wj(\by)\wk(\bz)}
+ \delta_{i1} \overline{w_2(\bx)\wj(\by)\wk(\bz)}
+ \delta_{j1} \overline{w_2(\by)\wi(\bx)\wk(\bz)}
\notag\\[4pt]
&
+ \delta_{k1} \overline{\wi(\bx)\wj(\by)w_2(\bz)}
+\frac{\partial }{\partial x_l}\overline{\wi(\bx) \wl(\bx)\wj(\by)\wk(\bz)}
+\frac{\partial }{\partial y_l}\overline{\wj(\by) \wl(\by)\wi(\bx)\wk(\bz)}
\notag\\[4pt]
&
+\frac{\partial }{\partial z_l}\overline{\wk(\bz) \wl(\bz)\wi(\bx)\wj(\by)}
=
-\frac{\partial }{\partial x_i}\overline{\q(\bx) \wj(\by)\wk(\bz)}
-\frac{\partial }{\partial y_j}\overline{\q(\by) \wi(\bx)\wk(\bz)}
\notag\\[4pt]
&
-\frac{\partial }{\partial z_k}\overline{\q(\bz) \wi(\bx)\wj(\by)}
+ \bigg(
 \frac{\partial^2 }{\partial x_l\partial x_l}
+\frac{\partial^2 }{\partial y_l\partial y_l}
+\frac{\partial^2 }{\partial z_l\partial z_l}\bigg)\overline{\wi(\bx)\wj(\by)\wk(\bz)}
\label{HST_CLMInPhysicalSpace_3p}
\end{align}
\begin{align}
\frac{\partial^2 }{\partial x_k x_k}\overline{\q(\bx) \wj(\by)}
=-2 \frac{\partial }{\partial x_1}\overline{w_2(\bx) \wj(\by)}
-\frac{\partial^2}{\partial x_k \partial x_l}\overline{\wl(\bx) \wk(\bx) \wj(\by)}
\label{HST_PressureInPhysicalSpace_2p}
\end{align}
\begin{align}
\frac{\partial^2 }{\partial x_l x_l}\overline{\q(\bx)\wj(\by)\wk(\bz)}
=-2 \frac{\partial }{\partial x_1}\overline{w_2(\bx)\wj(\by)\wk(\bz)}
-\frac{\partial^2 }{\partial x_m \partial x_l}\overline{\wl(\bx) \wm(\bx)\wj(\by)\wk(\bz)}
\label{HST_PressureInPhysicalSpace_3p}
\end{align}
and
\begin{align}
\frac{\partial^2 }{\partial y_k\partial y_k}\overline{\q(\bx)\,\q(\by)}
=-2 \frac{\partial}{\partial y_1}\overline{\q(\bx) w_2(\by)}
-\frac{\partial^2}{\partial y_k \partial y_l}\overline{\q(\bx) \wk(\by) \wl(\by)}
\label{HST_PressureInPhysicalSpace_qq_2p}
\end{align}
 Here and below the dependence of the fluctuations and correlations on $t$ is suppressed for the sake of brevity.

We now apply the homogeneity to the multi-point correlations involved in \eqref{HST_DivergenceFreeInPhysicalSpace_2p_3p} through \eqref{HST_PressureInPhysicalSpace_qq_2p},
\begin{align}
&
\overline{\wi(\bx) \wj(\by)}=\overline{\wi(\mathbf{0}) \wj(\by-\bx)}=:\Wij(\br),
\quad 
\overline{\wi(\bx) \wj(\by) \wk(\bz)}=\overline{\wi(\mathbf{0}) \wj(\br) \wk(\bs)}=:\W_{ijk}(\br,\bs),
\notag\\[4pt]
&
\overline{\wi(\bx) \wj(\by) \wk(\bz) \wl(\bz')}=\overline{\wi(\mathbf{0}) \wj(\mathbf{\br}) \wk(\bs) \wl(\bs')}=:\W_{ijkl}(\br,\bs,\bs'),
\notag\\[4pt]
&
\overline{\q(\bx) \,\q(\by)}=\overline{\q(\mathbf{0})\, \q(\by-\bx)}=:\Q(\br),
\quad
\overline{\q(\bx) \wj(\by)}=\overline{\q(\mathbf{0}) \wj(\by-\bx)}=:\Qj(\br),
\notag\\[4pt]
&
\overline{\q(\bx)\wj(\by)\wk(\bz)}=\overline{\q(\mathbf{0})\wj(\by-\bx)\wk(\bz-\bx)}=:\Q_{jk}(\br,\bs)
\label{Homogeneity}
\end{align}
where $\br:=\by-\bx$, $\bs:=\bz-\bx$ and $\bs':=\bz'-\bx$.
  Obviously, there are symmetric relations from the definitions above such as
\begin{align}
&
\Wij(\br)=\Wji(-\br),\quad
\W_{ijk}(\br,\bs)=\W_{ikj}(\bs,\br)=\W_{jik}(-\br,\bs-\br)=\W_{kij}(-\bs,\br-\bs),
\notag\\[4pt]
&
\W_{ijkl}(\br,\bs,\bs')=\W_{ijlk}(\br,\bs',\bs)=\W_{ilkj}(\bs',\bs,\br)=\W_{ikjl}(\bs,\br,\bs')=\W_{jikl}(-\br,\bs-\br,\bs'-\br)
\notag\\[4pt]
&
=\W_{kijl}(-\bs,\br-\bs,\bs'-\bs)=\W_{lijk}(-\bs',\br-\bs',\bs-\bs'),
\quad
\Q(\br)=\Q(-\br),
\quad
\Q_{jk}(\br,\bs)=\Q_{kj}(\bs,\br)
\label{Homogeneity_Symmetry}
\end{align}

The domain of motion and the averaged flow field \eqref{HST_AverageVelocityField} are symmetric under the coordinate transformation of $\bx\rightarrow-\bx$.
 Further, it can be verified directly that, if $\{w_i(\bx)$, $\q(\bx)\}$ is a solution of \eqref{HST_DivergenceFreeInPhysicalSpace} through \eqref{HST_PressureInPhysicalSpace}, $\{-w_i(-\bx)$, $\q(-\bx)\}$ is also a solution, that is,  the solution satisfies the symmetry of inversion,
\begin{align}
w_i(\bx)=-w_i(-\bx),\quad
\q(\bx)=\q(-\bx)
\label{SymmetryOfInversion}
\end{align}
 provided that the initial condition is adequate, such as \eqref{SymmetryOfInversion} holding at $t=0$.
 It is interesting to notice that the adoption of \eqref{SymmetryOfInversion} implies that $w_i(\mathbf{0})=0$. That is, $\bx=\mathbf{0}$ is a peculiar point at which the velocity fluctuation remains zero under the symmetry of the exact solutions for the corresponding initial conditions; This result has the non-physical consequence of
$\Wij(\mathbf{0})=0$. It follows that the above symmetry does not hold for all the realizable individual solutions since the initial conditions do not possess such a symmetry. We will still adopt, however, the symmetry in a statistical sense as formulated in \eqref{StatisticalSymmetryOfInversion}, which may be justified from the aspect of the coordinate transformation for the flow due to its geometric and kinematic symmetries. For instance, if we rotate the Cartesian coordinate system under $\bx\rightarrow-\bx$, we have 
\begin{align*}
\{\Vi, P, \wi, \q\} \rightarrow \{-\Vi,  P, -\wi, \q\}
\end{align*}
and we expect that the statistical correlations transform accordingly as specified in \eqref{StatisticalSymmetryOfInversion} below.  
  We now impose the statistical symmetry of inversion,
\begin{align}
&
\overline{w_i(\bx) w_j(\by)}=\overline{(-w_i(-\bx))\, (-w_j(-\by))},
\quad
\overline{w_i(\bx) w_j(\by) w_k(\bz)}=\overline{(-w_i(-\bx))\, (-w_j(-\by))\, (-w_k(-\bz))},
\notag\\[4pt]
&
\overline{w_i(\bx) w_j(\by) w_k(\bz) w_l(\bz')}=\overline{(-w_i(-\bx))\, (-w_j(-\by))\, (-w_k(-\bz))\, (-w_l(-\bz'))},
\notag\\[4pt]
&
\overline{\q(\bx\,) \q(\by)}=\overline{\q(-\bx) \q(-\by)},\quad
\overline{\q(\bx) \wj(\by)}=\overline{\q(-\bx) (-\wj(-\by))},
\notag\\[4pt]
&
\overline{\q(\bx) \wj(\by) \wk(\bz)}=\overline{\q(-\bx) (-\wj(-\by)) (-\wk(-\bz))}
\label{StatisticalSymmetryOfInversion}
\end{align}
or
\begin{align}
&
\W_{ij}(\br)=\W_{ij}(-\br),\quad
\W_{ijk}(\br,\bs)=-\W_{ijk}(-\br,-\bs),\quad
\W_{ijkl}(\br,\bs,\bs')=\W_{ijkl}(-\br,-\bs,-\bs'),
\notag\\[4pt]
&
\Q(\br)=\Q(-\br),\quad
\Q_j(\br)=-\Q_j(-\br),\quad
\Q_{jk}(\br,\bs)=\Q_{jk}(-\br,-\bs)
\label{Homogeneity_Inversion}
\end{align}

 We can substitute \eqref{Homogeneity} into \eqref{HST_DivergenceFreeInPhysicalSpace_2p_3p} through \eqref{HST_PressureInPhysicalSpace_qq_2p} to get
\begin{align}
&
\frac{\partial}{\partial r_k}\W_{kj}(\br)=0,\quad
\frac{\partial}{\partial r_j}\W_{kj}(\br)=0,
\notag\\[4pt] &
\bigg(\frac{\partial}{\partial r_k}+\frac{\partial}{\partial s_k}\bigg)\W_{kjl}(\br,\bs)=0,\quad
\frac{\partial}{\partial r_j}\W_{kjl}(\br,\bs)=0,\quad
\frac{\partial}{\partial s_l}\W_{kjl}(\br,\bs)=0,
\notag\\[4pt]
&
\bigg(\frac{\partial}{\partial r_i}+\frac{\partial}{\partial s_i}+\frac{\partial}{\partial s'_i}\bigg)\W_{ijkl}(\br,\bs,\bs')=0,\quad
\frac{\partial}{\partial r_j}\W_{ijkl}(\br,\bs,\bs')=0,\quad
\frac{\partial}{\partial s_k}\W_{ijkl}(\br,\bs,\bs')=0,
\notag\\[4pt] &
\frac{\partial}{\partial s'_l}\W_{ijkl}(\br,\bs,\bs')=0,\quad
\frac{\partial}{\partial r_k}\Q_{k}(\br)=0,\quad
\frac{\partial}{\partial r_k}\Q_{kl}(\br,\bs)=0,\quad
\frac{\partial}{\partial s_l}\Q_{kl}(\br,\bs)=0
\label{HST_DivergenceFreeInPhysicalSpace_2p_3p_rs}
\end{align}
\begin{align}
&
\frac{\partial}{\partial t}\W_{ij}(\br)
+ r_2 \frac{\partial }{\partial r_1} \W_{ij}(\br)
+ \delta_{i1} \W_{2j}(\br)
+ \delta_{j1} \W_{i2}(\br)
-\frac{\partial}{\partial r_k}\W_{ikj}(\mathbf{0},\br)
+\frac{\partial}{\partial r_k}\W_{jki}(\mathbf{0},-\br)
\notag\\[4pt]
= &\,
 \frac{\partial }{\partial r_i}\Q_j(\br)
-\frac{\partial }{\partial r_j}\Q_i(-\br)
+2  \frac{\partial^2}{\partial r_k\partial  r_k}\W_{ij}(\br)
\label{HST_CLMInPhysicalSpace_2p_r}
\end{align}
\begin{align}
&
 \frac{\partial }{\partial t} \W_{ijk}(\br,\bs)            
+ r_2 \frac{\partial }{\partial r_1} \W_{ijk}(\br,\bs)      
+ s_2 \frac{\partial }{\partial s_1} \W_{ijk}(\br,\bs)      
+ \delta_{i1}   \W_{2jk}(\br,\bs)    
+ \delta_{j1}   \W_{i2k}(\br,\bs)    
\notag\\[4pt]
&
+ \delta_{k1}   \W_{ij2}(\br,\bs)     
-\bigg(\frac{\partial }{\partial r_l}+\frac{\partial }{\partial s_l}\bigg) \W_{iljk}(\mathbf{0},\br,\bs) 
+\frac{\partial }{\partial r_l} \W_{jlik}(\mathbf{0},-\br,\bs-\br)    
\notag\\[4pt]
&
+\frac{\partial }{\partial s_l} \W_{klij}(\mathbf{0},-\bs,\br-\bs)       
=
 \frac{\partial }{\partial r_i} \Q_{jk}(\br,\bs)  
+\frac{\partial }{\partial s_i} \Q_{jk}(\br,\bs)  
-\frac{\partial }{\partial r_j} \Q_{ik}(-\br,\bs-\br)   
\notag\\[4pt]
&
-\frac{\partial }{\partial s_k} \Q_{ij}(-\bs,\br-\bs)         
+2 \bigg(
 \frac{\partial^2 }{\partial r_l\partial  r_l}
+\frac{\partial^2 }{\partial s_l\partial  s_l}
+\frac{\partial^2}{\partial r_l \partial s_l}
\bigg) \W_{ijk}(\br,\bs)  
\label{HST_CLMInPhysicalSpace_3p_rs}
\end{align}
\begin{align}
\frac{\partial^2 }{\partial r_k\partial  r_k} \Qj(\br) 
= 2 \frac{\partial }{\partial r_1} \W_{2j}(\br)  
-\frac{\partial^2}{\partial r_k \partial r_l} \W_{lkj}(\mathbf{0},\br)
\label{HST_PressureInPhysicalSpace_2p_r}
\end{align}
\begin{align}
&
\bigg(\frac{\partial}{\partial r_l}+\frac{\partial}{\partial s_l}\bigg)\bigg(\frac{\partial}{\partial r_l}+\frac{\partial}{\partial s_l}\bigg)\Q_{jk}(\br,\bs) 
=2 \bigg(\frac{\partial }{\partial r_1}+\frac{\partial }{\partial s_1}\bigg)  \W_{2jk}(\br,\bs)
\notag\\[4pt]
&
-\bigg(\frac{\partial}{\partial r_m}+\frac{\partial}{\partial s_m}\bigg)\bigg(\frac{\partial}{\partial r_l}+\frac{\partial}{\partial s_l}\bigg) \W_{lmjk}(\mathbf{0},\br,\bs)
\label{HST_PressureInPhysicalSpace_3p_rs}
\end{align}
and
\begin{align}
\frac{\partial^2 }{\partial r_k r_k}\Q(\br)
=-2 \frac{\partial}{\partial r_1}\Q_2(\br)
-\frac{\partial^2}{\partial r_k \partial r_l}\Q_{kl}(\br,\br)
\label{HST_PressureInPhysicalSpace_qq_2p_r}
\end{align}

\subsection{Fourier Transforms}
\ \ \ \
It is convenient to formulate the mathematical problem with the help of Fourier transforms in $\mathbb{R}^n$, $n=2$, $4$, $6$. With this adoption of an infinite domain of flow, we need to modify our treatment presented in PART I accordingly, as to be mentioned in the appropriate places below. We  adopt the Fourier transforms of
\begin{align}
&
\Wij(\br)=\int_{\mathbb{R}^2} \tWij(\bk)\, \exp(\imaginary\, \bk\s\cdot\s\br)\, d\bk,
\quad
\W_{ijk}(\br,\bs)=\int_{\mathbb{R}^2\times\mathbb{R}^2} \tW_{ijk}(\bk,\bl)\, \exp\s\l[\imaginary\, (\bk\s\cdot\s\br+\bl\s\cdot\s\bs)\r]\, d\bk\, d\bl,
\notag\\[4pt]
&
\W_{ijkl}(\br,\bs,\bs')=\int_{\mathbb{R}^2\times\mathbb{R}^2\times\mathbb{R}^2} 
 \tW_{ijkl}(\bk,\bl,\bm)\, \exp\s\l[\imaginary\, (\bk\s\cdot\s\br+\bl\s\cdot\s\bs+\bm\s\cdot\s\bs')\r]\, d\bk\, d\bl\, d\bm,
\notag\\[4pt]
&
\Q(\br)=\int_{\mathbb{R}^2} \tQ(\bk)\, \exp(\imaginary\, \bk\s\cdot\s\br)\, d\bk,
\quad 
\Qj(\br)=\int_{\mathbb{R}^2} \tQj(\bk)\, \exp(\imaginary\, \bk\s\cdot\s\br)\, d\bk,
\notag\\[4pt]
&
\Q_{jk}(\br,\bs)=\int_{\mathbb{R}^2\times\mathbb{R}^2} \tQ_{jk}(\bk,\bl)\, \exp\s\l[\imaginary\, (\bk\s\cdot\s\br+\bl\s\cdot\s\bs)\r]\, d\bk \,d\bl
\label{FourierTransform}
\end{align}
That the one-point and multi-point correlations in the physical space are real requires that
\begin{align}
&
\tW^*_{ij}(\bk) =\tW_{ij}(-\bk),
\quad
\tW^*_{ijk}(\bk,\bl) =\tW_{ijk}(-\bk,-\bl) ,
\quad
\tW^*_{ijkl}(\bk,\bl,\bm) =\tW_{ijkl}(-\bk,-\bl,-\bm),
\notag\\[4pt]
&
\tQ^*(\bk)=\tQ(-\bk) ,
\quad
\tQ^*_j(\bk)=\tQ_j(-\bk) ,
\quad
\tQ^*_{jk}(\bk,\bl)=\tQ_{jk}(-\bk,-\bl)
\label{RealCorrelations_Inps}
\end{align}
where the superscript $*$ denotes the complex conjugate operation.

Combining \eqref{Homogeneity_Symmetry}, \eqref{Homogeneity_Inversion}, \eqref{FourierTransform} and \eqref{RealCorrelations_Inps}, we get
\begin{align}
&
\tW_{ij}(\bk)=\tW_{ji}(\bk)=\tW_{ij}(-\bk)=\tW^*_{ij}(\bk),\quad
\notag\\[6pt]
&
\tW_{ijk}(\bk,\bl)=\tW_{ikj}(\bl,\bk)=\tW_{jik}(-\bk-\bl,\bl)=\tW_{kij}(-\bk-\bl,\bk)=-\tW_{ijk}(-\bk,-\bl)=-\tW^*_{ijk}(\bk,\bl),
\notag\\[6pt]
&
\tW_{ijkl}(\bk,\bl,\bm)=\tW_{ijlk}(\bk,\bm,\bl)=\tW_{ilkj}(\bm,\bl,\bk)=\tW_{ikjl}(\bl,\bk,\bm)
=\tW_{jikl}(-\bk-\bl-\bm,\bl,\bm)
\notag\\[3pt]
&
=\tW_{kijl}(-\bk-\bl-\bm,\bk,\bm)
=\tW_{lijk}(-\bk-\bl-\bm,\bk,\bl)
=\tW_{ijkl}(-\bk,-\bl,-\bm)=\tW^*_{ijkl}(\bk,\bl,\bm),
\notag\\[6pt]
&
\tQ(\bk)=\tQ(-\bk)=\tQ^*(\bk),
\quad
\tQ_j(\bk)=-\tQ_j(-\bk)=-\tQ^*_j(\bk),
\notag\\[6pt]
&
\tQ_{jk}(\bk,\bl)=\tQ_{jk}(-\bk,-\bl)=\tQ^*_{jk}(\bk,\bl)=\tQ_{kj}(\bl,\bk)
\label{Homogeneity_Symmetry_Inversion_fs}
\end{align}
It then follows that $\tW_{ij}(\bk)$, $\tW_{ijkl}(\bk,\bl)$, $\tQ(\bk)$ and $\tQ_{ij}(\bk,\bl)$ are real and $\tW_{ijk}(\bk,\bl)$ and $\tQ_j(\bk)$ are purely imaginary, i.e.,
\begin{align}
&
\tW_{ijk}(\bk,\bl)=\imaginary\,\,\tW^{(I)}_{ijk}(\bk,\bl),\quad
\tW^{(I)}_{ijk}(-\bk,-\bl)=-\tW^{(I)}_{ijk}(\bk,\bl),
\notag\\[6pt]
&
\tQ_j(\bk)=\imaginary\,\,\tQ^{(I)}_j(\bk),\quad
\tQ^{(I)}_j(-\bk)=-\tQ^{(I)}_j(\bk)
\label{wiwjwk_PurelyImaginary_fs}
\end{align}

We now transform \eqref{HST_DivergenceFreeInPhysicalSpace_2p_3p_rs} through \eqref{HST_PressureInPhysicalSpace_qq_2p_r} in the physical space to their corresponding relations in the wave number space of $\bk\in\mathbb{R}^2$ and so on, 
\begin{align}
&
k_k\,\tW_{kj}(\bk) =0,\quad
k_j\,\tW_{kj}(\bk) =0,\quad
 \big(k_k+l_k\big)\,\tW^{(I)}_{kjl}(\bk,\bl)=0,\quad
k_j\,\tW^{(I)}_{kjl}(\bk,\bl)=0,\quad
l_l\,\tW^{(I)}_{kjl}(\bk,\bl)=0,
\notag\\[6pt] &
(k_i+l_i+m_i)\,\tW_{ijkl}(\bk,\bl,\bm)=0,\quad
k_j\,\tW_{ijkl}(\bk,\bl,\bm)=0,\quad
l_k\,\tW_{ijkl}(\bk,\bl,\bm)=0,
\notag\\[6pt]&
m_l\,\tW_{ijkl}(\bk,\bl,\bm)=0,\quad
k_k\,\tQ^{(I)}_{k}(\bk)=0,\quad
k_k\,\tQ_{kl}(\bk,\bl)=0,\quad
l_l\,\tQ_{kl}(\bk,\bl)=0
\label{HST_DivergenceFreeInPhysicalSpace_qandw_fs}
\end{align}
\begin{align}
 \tQ(\bk) 
=
-\frac{2\, k_1}{|\bk|^2}\,\tQ^{(I)}_2(\bk)   
-\frac{k_k\,k_l}{|\bk|^2}\,\int_{\mathbb{R}^2}\tQ_{kl}(\bk-\bl,\bl)\,d\bl
\label{HST_PressureInPhysicalSpace_qq_fs}
\end{align}
\begin{align}
\tQ^{(I)}_j(\bk) = 
 -\frac{2\,k_1}{|\bk|^2}\,\tW_{2j}(\bk) 
-\frac{k_k\,k_l}{|\bk|^2}\,\int_{\mathbb{R}^2} \tW^{(I)}_{lkj}(\bl,\bk)\, d\bl  
\label{HST_PressureInPhysicalSpace_qw_fs}
\end{align}
\begin{align}
\tQ_{jk}(\bk,\bl)  =\frac{2\, (k_1+l_1)}{|\bk+\bl|^2}\,\tW^{(I)}_{2jk}(\bk,\bl)
 -\frac{(k_m+l_m)(k_l+l_l)}{|\bk+\bl|^2}\,\int_{\mathbb{R}^2}\tW_{lmjk}(\bm,\bk,\bl)\,d\bm 
\label{HST_PressureInPhysicalSpace_qww_fs}
\end{align}
\begin{align}
&
 \frac{\partial}{\partial t}\tW_{ij}(\bk)    
+2\,|\bk|^2\,\tW_{ij}(\bk)   
- k_1\,\frac{\partial}{\partial k_2}\tW_{ij}(\bk)   
+ \delta_{i1} \tW_{2j}(\bk)+\delta_{j1} \tW_{i2}(\bk) 
\notag\\[4pt]
= &\,
-k_i\,\tQj^{(I)}(\bk)
+k_j\,\tQi^{(I)}(-\bk) 
-k_k \int_{\mathbb{R}^2} \Big(\tWImag_{ijk}(\bk,\bl)-\tWImag_{jik}(-\bk,\bl)\Big)\, d\bl   
\label{HST_CLMInPhysicalSpace_ww_fs}
\end{align}
and
\begin{align}
&
 \frac{\partial }{\partial t}\tWImag_{ijk}(\bk,\bl)           
-k_1\,\frac{\partial}{\partial k_2}\tWImag_{ijk}(\bk,\bl)
-l_1\,\frac{\partial}{\partial l_2}\tWImag_{ijk}(\bk,\bl)
\notag\\[4pt]
&     
+\delta_{i1}  \tWImag_{2jk}(\bk,\bl)
+\delta_{j1} \tWImag_{i2k}(\bk,\bl)
+\delta_{k1}\tWImag_{ij2}(\bk,\bl)   
-(k_l+l_l)\int_{\mathbb{R}^2}\tW_{iljk}(\bm,\bk,\bl)\,d\bm   
\notag\\
&     
+k_l\int_{\mathbb{R}^2}\tW_{jlik}(\bm,-\bk-\bl,\bl)\,d\bm
+l_l\int_{\mathbb{R}^2}\tW_{klij}(\bm,-\bk-\bl,\bk)\,d\bm      
\notag\\[4pt]
=&\,
 k_i\,\tQ_{jk}(\bk,\bl)
+l_i\,\tQ_{jk}(\bk,\bl)   
-k_j\,\tQ_{ik}(-\bk-\bl,\bl) 
-l_k\,\tQ_{ij}(-\bk-\bl,\bk)       
\notag\\[4pt]
&     
-2\,\big(|\bk|^2+|\bl|^2+\bk\s\cdot\s\bl\big)\,\tWImag_{ijk}(\bk,\bl) 
\label{HST_CLMInPhysicalSpace_www_fs}
\end{align}

\subsection{Primary Equations}
\ \ \ \
Equation \eqref{HST_DivergenceFreeInPhysicalSpace_qandw_fs} can be easily solved to obtain
\begin{align}
&
\tW_{11}(\bk)=:\soc(\bk)=\soc(-\bk)
,\quad
\tW_{ij}(\bk)=\bigg(\s-\frac{k_1}{k_2}\bigg)^{\s i+j-2}\,\soc(\bk)
\label{HST_DivergenceFreeInPhysicalSpace_wiwj_fs}
\end{align}
\begin{align}
&
\tW^{(I)}_{111}(\bk,\bl)=:\toc(\bk,\bl)=\toc(\bl,\bk)
=-\toc(-\bk,-\bl)
=\toc(-\bk-\bl,\bk)
,
\notag\\[8pt]
&
\tW^{(I)}_{ijk}(\bk,\bl)=\bigg(\s-\frac{k_1+l_1}{k_2+l_2}\bigg)^{\s i-1} \bigg(\s-\frac{k_1}{k_2}\bigg)^{\s j-1} \bigg(\s-\frac{l_1}{l_2}\bigg)^{\s k-1}\,\toc(\bk,\bl)
\label{HST_DivergenceFreeInPhysicalSpac_wiwjwk_fs}
\end{align}
and
\begin{align}
&
\tW_{1111}(\bk,\bl,\bm)=:\foc(\bk,\bl,\bm)
=\foc(\bk,\bm,\bl)
=\foc(\bm,\bl,\bk)
=\foc(\bl,\bk,\bm)
=\foc(-\bk,-\bl,-\bm)
\notag\\[4pt]
&
=\foc(-\bk-\bl-\bm,\bl,\bm)
,
\notag\\[8pt]
&
\tW_{ijkl}(\bk,\bl,\bm)=\bigg(\s-\frac{k_1+l_1+m_1}{k_2+l_2+m_2}\bigg)^{\s i-1} \bigg(\s-\frac{k_1}{k_2}\bigg)^{\s j-1} \bigg(\s-\frac{l_1}{l_2}\bigg)^{\s k-1} \bigg(\s-\frac{m_1}{m_2}\bigg)^{\s l-1}\,\foc(\bk,\bl,\bm)
\label{HST_DivergenceFreeInPhysicalSpace_wiwjwkwl_fs}
\end{align}
That is, $\soc$, $\toc$ and $\foc$ are, respectively, the primary components for the second, the third and the four order correlations.
  Next, the consistency between \eqref{HST_CLMInPhysicalSpace_ww_fs} and \eqref{HST_DivergenceFreeInPhysicalSpace_wiwj_fs} requires the existence of single equation of evolution for $\soc(\bk)$ and the consistency between \eqref{HST_CLMInPhysicalSpace_www_fs} and \eqref{HST_DivergenceFreeInPhysicalSpac_wiwjwk_fs} also demands single equation of evolution for $\toc(\bk,\bl)$.
 Both can be checked directly by the respective
 substitutions of \eqref{HST_DivergenceFreeInPhysicalSpace_wiwj_fs} into \eqref{HST_CLMInPhysicalSpace_ww_fs} and \eqref{HST_DivergenceFreeInPhysicalSpac_wiwjwk_fs} into \eqref{HST_CLMInPhysicalSpace_www_fs} and so on; straightforward but lengthy operations give
\begin{align}
&
 \bigg(\frac{\partial}{\partial t}- k_1\,\frac{\partial}{\partial k_2}\bigg)
     \bigg\{\frac{|\bk|^4}{(k_2)^2}\,\exp\s\Big[2 H(0,\bk)\Big]
\,\soc(\bk)\bigg\}  
\notag\\[4pt]
= &\,
2\,|\bk|^2 \,k_2\,\exp\s\Big[2 H(0,\bk)\Big]\s
 \int_{\mathbb{R}^2}\s \bigg[
     \frac{l_1}{l_2}
      + \frac{k_1+l_1}{k_2+l_2}\,\frac{k_1}{k_2}\,\frac{l_1}{l_2}
         -\frac{k_1}{k_2}
        -  \frac{k_1+l_1}{k_2+l_2}\,\bigg(\frac{k_1}{k_2}\bigg)^2
\bigg]\, \toc(\bk,\bl) \, d\bl
\label{HST_CLMInPhysicalSpace_w1w1_fs}
\end{align}
and
\begin{align}
&
\bigg(\frac{\partial }{\partial t}-k_1\,\frac{\partial}{\partial k_2}- l_1\,\frac{\partial}{\partial l_2} \bigg)\s
     \bigg\{\frac{|\bk|^2\,|\bl|^2\,|\bk+\bl|^2}{k_2\,l_2\,(k_2+l_2)}
                  \exp\s\Big[H(0,\bk)\s+\s H(0,\bl)\s+\s H(0,\bk+\bl)\Big]\, \toc(\bk,\bl)\bigg\}
\notag\\[4pt]
=&\,
\frac{|\bk|^2\,|\bl|^2\,|\bk+\bl|^2}{k_2\,l_2\,(k_2+l_2)}
                  \exp\s\Big[H(0,\bk)\s+\s H(0,\bl)\s+\s H(0,\bk+\bl)\Big]\notag\\
&\hskip 3mm
\times\s\bigg[
 (k_1+l_1)\,\frac{(k_2+l_2)^2}{|\bk+\bl|^2}\,\focI(\bk,\bl)
+(k_2+l_2)\,\bigg(1 -\frac{2(k_1+l_1)^2}{|\bk+\bl|^2} \bigg)\,\focII(\bk,\bl) 
\notag\\[4pt]
&\hskip 10mm
-k_1\,\frac{(k_2)^2}{|\bk|^2}\,\focI(-\bk-\bl,\bl)
-k_2\,\bigg(1-\frac{2(k_1)^2}{|\bk|^2}\bigg)\,\focII(-\bk-\bl,\bl)
\notag\\[4pt]
&\hskip 10mm
-l_1\,\frac{(l_2)^2}{|\bl|^2} \,\focI(-\bk-\bl,\bk)
-l_2\,\bigg(1-\frac{2(l_1)^2}{|\bl|^2}\bigg) \, \focII(-\bk-\bl,\bk)
\bigg]      
\label{HST_CLMInPhysicalSpace_w1w1w1_fs}
\end{align}
Here,
\begin{align}
&
H(\sigma',\bk):=-\frac{k_2}{k_1}\Big(\sigma'+(k_1)^2+\frac{1}{3}(k_2)^2\Big),\quad
\focI(\bk,\bl)
:= \int_{\mathbb{R}^2}\s \bigg(1-\frac{m_1+k_1+l_1}{m_2+k_2+l_2}\,\frac{m_1}{m_2}\bigg)\foc(\bm,\bk,\bl)\,d\bm,
\notag\\[4pt]&
\focII(\bk,\bl)
:=-\int_{\mathbb{R}^2}\frac{m_1}{m_2} \, \foc(\bm,\bk,\bl)\,d\bm
\label{HST_DivergenceFreeInPhysicalSpace_1s2s_fs_Asymp}
\end{align}
 Equations \eqref{HST_CLMInPhysicalSpace_w1w1_fs} and \eqref{HST_CLMInPhysicalSpace_w1w1w1_fs} are the two primary equations for $\soc(\bk)$ and $\toc(\bk,\bl)$, respectively;
 $\foc(\bm,\bk,\bl)$ is to be determined.
 The equations above have the linear structures involving $\soc(\bk)$, $\toc(\bk,\bl)$ and $\foc(\bm,\bk,\bl)$; the non-linearity comes into play through the nonlinear constraints of inequality to be discussed below.

\subsection{Constraints of Inequality}\label{Subsec:ConstraintsofInequality}
\ \ \ \
It is straightforward to check that the symmetries of the second and third order correlations listed in \eqref{HST_DivergenceFreeInPhysicalSpace_wiwj_fs} and \eqref{HST_DivergenceFreeInPhysicalSpac_wiwjwk_fs} are guaranteed by the structures of \eqref{HST_CLMInPhysicalSpace_w1w1_fs}, \eqref{HST_CLMInPhysicalSpace_w1w1w1_fs} and the symmetries of $\foc(\bk,\bl,\bm)$ in \eqref{HST_DivergenceFreeInPhysicalSpace_wiwjwkwl_fs} that are to be implemented.

There are constraints of inequality for the second, third and fourth order correlations from various considerations.
 Firstly, there are constraints of inequality for $\W_{ij}(\br)$ as discussed in PART I which will, in turn, result in a set of inequality constraints for $\tW_{ij}(\bk)$ and $\soc(\bk)$, (the summations are replaced with the corresponding integrations here due to the infinite domain of flow).
\begin{enumerate}
\item
The two-point correlations $\overline{w_i(\bx)\, w_j(\by)}$ in the physical space are supposed to be finite at any finite instant, and the finiteness supposedly holds also for the corresponding correlations in the wave number space. That is,
\begin{align}
\W_{ij}(\br),\ 
\tW_{ij}(\bk)\ \ \text{finite at any finite}\ t
\label{FiniteCorrelations_Wij}
\end{align}
\item
 We take $\soc(\bk)$ as non-negative,
\begin{align}
\soc(\bk)\geq 0
\label{NonNegativityofSoc11}
\end{align}
It guarantees the non-negativity of the energy spectrum distribution whose consequence or necessity will be demonstrated below.
 The constraint may also be justified if one starts from the Fourier transform of $w_1(\bx)$ and then applies the homogeneity to the resultant correlation of $\overline{w_1(\bx)\, w_1(\by)}$. We should mention that \eqref{NonNegativityofSoc11} is the only constraint formulated directly in the wave number space, we will not enforce similar inequalities for $\tW_{ijk}(\bk,\bl)$ and $\tW_{ijkl}(\bk,\bl,\bm)$ derived from the application of the Cauchy-Schwarz inequality to $\tW_{ijk}(\bk,\bl)$ and $\tW_{ijkl}(\bk,\bl,\bm)$, since  the involvement of the Dirac delta complicates the formulation. 
 The above adoption of the homogeneity before the Fourier transforms intends to avoid such complications. 
\item
The constraints of inequality from the positive semi-definiteness of the single-point correlations $\barwiwj$, $\overline{\wi,_k \wj,_k}$ and $\overline{\wk,_i \wk,_j}$ are satisfied automatically under  \eqref{HST_DivergenceFreeInPhysicalSpace_wiwj_fs} and \eqref{NonNegativityofSoc11}. For instance, in the case of
\begin{align*}
\Big(\overline{w_1,_k(\bx)\, w_2,_k(\bx)}\Big)^2
\leq \overline{w_1,_k(\bx)\, w_1,_k(\bx)}\,\,\overline{w_2,_l(\bx)\, w_2,_l(\bx)}
\end{align*}
 we have, with the help of the Cauchy-Schwarz inequality,
\begin{align*}
&
\bigg|\int_{\mathbb{R}^2} |\bk|^2\,\tW_{12}(\bk)\, d\bk\bigg|
=
 \bigg|\int_{\mathbb{R}^2} |\bk|^2\,\frac{k_1}{k_2}\,\soc(\bk)\, d\bk\bigg|
\leq
 \int_{\mathbb{R}^2} |\bk|^2\,\Big|\frac{k_1}{k_2}\Big|\,\soc(\bk)\, d\bk
\notag\\[4pt]
&
=
 \int_{\mathbb{R}^2} \Big(|\bk|\,\sqrt{\soc(\bk)}\Big)\,|\bk|\,\Big|\frac{k_1}{k_2}\Big|\,\sqrt{\soc(\bk)}\, d\bk
\leq
 \sqrt{\int_{\mathbb{R}^2} |\bk|^2\,\soc(\bk)\, d\bk\,\,
      \int_{\mathbb{R}^2} |\bk|^2\,\Big|\frac{k_1}{k_2}\Big|^2\,\soc(\bk)\, d\bk}
\notag\\[4pt]
&
=
 \sqrt{\int_{\mathbb{R}^2} |\bk|^2\,\tW_{11}(\bk)\, d\bk}\,\,
      \sqrt{\int_{\mathbb{R}^2} |\bk|^2\,\tW_{22}(\bk)\, d\bk}
\end{align*}
\item
We apply the Cauchy-Schwarz inequality to the two-point correlations of
\begin{align*}
\Big\{&
\overline{\wi(\bx)\,\wj(\by)},\ 
\overline{\wi(\bx)\,\wj,_m\s(\by)},\
\overline{\wi,_m\s(\bx)\,\wj,_n\s(\by)},\
\overline{\big(w_2,_1\s(\bx)-w_1,_2\s(\bx)\big)\big(w_2,_1\s(\by)-w_1,_2\s(\by)\big)},
\notag\\ &\qquad
\overline{\w_i(\bx)\big(w_2,_1\s(\by)-w_1,_2\s(\by)\big)}
         \Big\}
\end{align*}
to obtain a set of constraints of inequality for $\soc(\bk)$, and these constraints are also satisfied automatically under \eqref{NonNegativityofSoc11}. For example, in the case of 
\begin{align*}
\Big(\W_{12}(\br)\Big)^2=\Big(\overline{w_1(\bx)\, w_2(\by)}\Big)^2
\leq \overline{w_1(\bx)\, w_1(\bx)}\,\,\overline{w_2(\by)\, w_2(\by)}
=\W_{11}(\mathbf{0})\,\W_{22}(\mathbf{0})
\end{align*}
we have
\begin{align*}
&
\bigg|\int_{\mathbb{R}^2} \tW_{12}(\bk)\, \cos(\bk\s\cdot\s\br)\, d\bk\bigg|
\leq \int_{\mathbb{R}^2} \Big|\tW_{12}(\bk)\Big|\, d\bk
=\int_{\mathbb{R}^2}\Big|\frac{k_1}{k_2}\Big|\,\soc(\bk) \, d\bk
\notag\\[4pt]&
=\int_{\mathbb{R}^2}\sqrt{\soc(\bk)}\,\,\,\Big|\frac{k_1}{k_2}\Big|\,\sqrt{\soc(\bk)} \, d\bk
\leq
 \sqrt{\int_{\mathbb{R}^2} \soc(\bk) \, d\bk\,
         \int_{\mathbb{R}^2}\Big|\frac{k_1}{k_2}\Big|^2\,\soc(\bk) \, d\bk}
\notag\\[4pt]&
=\sqrt{\int_{\mathbb{R}^2} \tW_{11}(\bk) \, d\bk}\,
         \sqrt{\int_{\mathbb{R}^2}\tW_{22}(\bk) \, d\bk}
\end{align*}
with the help of \eqref{FourierTransform}, \eqref{Homogeneity_Symmetry_Inversion_fs}, \eqref{HST_DivergenceFreeInPhysicalSpace_wiwj_fs}, \eqref{NonNegativityofSoc11} and the Cauchy-Schwarz inequality to the functions in the wave number space.
\end{enumerate}

Next, we consider the multi-point correlations in the physical space involving the higher orders.
\begin{enumerate}
\item
It is expected that
\begin{align}
\W_{ijk}(\br,\bs),\ \W_{ijkl}(\br,\bs,\bs')\ \ \text{finite at any finite}\ t
\label{FiniteCorrelations}
\end{align}
which indicates the finiteness of the corresponding correlations at any finite time in the wave number space.
\item
The expected $\overline{w_{\underlinei}(\bx) w_{\underlinei}(\bx) w_{\underlinej}(\by) w_{\underlinej}(\by)}\geq 0$ and
  $\overline{w_{\underlinei}(\bx) w_{\underlinei}(\bx) w_{\underlinej},_{\underlinel}(\by) w_{\underlinej},_{\underlinel}(\by)}\geq 0$
   for all $\bx$ and $\by$ and $i$, $j$ and $l$ requires that 
\begin{align}
\W_{\underlinei\underlinei\underlinej\underlinej}(\mathbf{0},\br,\br)\geq 0,\quad
\frac{\partial}{\partial r_{\underlinel}}\frac{\partial}{\partial s_{\underlinel}}\W_{\underlinei\underlinei\underlinej\underlinej}(\mathbf{0},\br,\bs)\bigg|_{\bs\,=\,\br}\geq 0,
\quad
i\leq j
\label{wiwiwjwj_Nonnegativity}
\end{align}
Hereafter, the summation rule is suspended for underlined subscripts, following the convention.
 More such inequalities can be formulated for different combinations of partial derivatives of various orders.
\item
We can obtain constraints of inequality among $\tW_{ij}(\bk)$, $\tW^{(I)}_{ijk}(\bk,\bl)$ and $\tW_{ijkl}(\bk,\bl,\bm)$ by applying the Cauchy-Schwarz inequality to the correlations of  
$$\overline{\wi(\bx) \wj(\by) \wk(\bz)},\ \ \overline{\wi(\bx) \wj(\by) \wk(\bz) \wl(\bz')},\ \ \overline{q(\bx)q(\by)},\ \ \overline{q(\bx) \wi(\by)},\ \ \overline{q(\bx)\wi(\by)\wj(\bz)}$$
 as well as their spatial derivatives.
\begin{enumerate}
\item
Consider $\overline{\wi(\bx) \wj(\by) \wk(\bz)}$. The Cauchy-Schwarz inequality requires that
\begin{align*}
\Big(\overline{\wi(\bx) \wj(\by) \wk(\bz)}\Big)^2\leq
\min\s\Big(
&
 \overline{\w_{\underlinei}(\bx) \w_{\underlinei}(\bx)}\,\,\overline{\w_{\underlinej}(\by) \w_{\underlinej}(\by) \w_{\underlinek}(\bz) \w_{\underlinek}(\bz)},
\notag\\[4pt]
&\quad
\overline{\w_{\underlinej}(\by) \w_{\underlinej}(\by)}\,\,\overline{\w_{\underlinei}(\bx) \w_{\underlinei}(\bx) \w_{\underlinek}(\bz) \w_{\underlinek}(\bz)},
\notag\\[4pt]
&\quad
\overline{\w_{\underlinek}(\bz) \w_{\underlinek}(\bz)}\,\,\overline{\w_{\underlinei}(\bx) \w_{\underlinei}(\bx) \w_{\underlinej}(\by) \w_{\underlinej}(\by)}
\Big)
\end{align*}
That is,
\begin{align}
\big(\W_{ijk}(\br,\bs)\big)^2\leq
\min\s\Big(
&
 \W_{\underlinei\underlinei}(\mathbf{0})\,\,\W_{\underlinej\underlinej\underlinek\underlinek}(\mathbf{0},\bs-\br,\bs-\br),\ 
 \W_{\underlinej\underlinej}(\mathbf{0})\,\,\W_{\underlinei\underlinei\underlinek\underlinek}(\mathbf{0},\bs,\bs),
\notag\\[4pt]
&\quad
 \W_{\underlinek\underlinek}(\mathbf{0})\,\,\W_{\underlinei\underlinei\underlinej\underlinej}(\mathbf{0},\br,\br)
\Big),
\quad
i\leq j\leq k
\label{CS_Inequality_ps_01}
\end{align}
\item
The application to $\overline{\wi(\bx) \wj(\by) \wk(\bz) \wl(\bz')}$ results in 
\begin{align}
\big(\W_{ijkl}(\br,\bs,\bs')\big)^2\leq 
\min\s\Big(
&
 \W_{\underlinei\underlinei\underlinej\underlinej}(\mathbf{0},\br,\br)\,\,\W_{\underlinek\underlinek\underlinel\underlinel}(\mathbf{0},\bs'-\bs,\bs'-\bs),
\notag\\[4pt]
&\quad
 \W_{\underlinei\underlinei\underlinek\underlinek}(\mathbf{0},\bs,\bs)\,\,\W_{\underlinej\underlinej\underlinel\underlinel}(\mathbf{0},\bs'-\br,\bs'-\br), 
\notag\\[4pt]
&\quad
 \W_{\underlinei\underlinei\underlinel\underlinel}(\mathbf{0},\bs',\bs')\,\,\W_{\underlinej\underlinej\underlinek\underlinek}(\mathbf{0},\bs-\br,\bs-\br) 
\Big),
\quad
i\leq j\leq k\leq l
\label{CS_Inequality_wwww_01}
\end{align}

\item
$\overline{q(\bx)\, \q(\by)}$ leads to
\begin{align}
\big(\Q(\br)\big)^2\leq \big(\Q(\mathbf{0})\big)^2,\quad
 \Q(\mathbf{0})\geq 0
\label{CS_Inequality_qq_01}
\end{align}

\item
$\overline{q(\bx) \wi(\by)}$ and $\overline{q(\bx)\wi(\by)\wj(\bz)}$ give, respectively,
\begin{align}
&
\big(\Q_{i}(\br)\big)^2\leq \Q(\mathbf{0})\,\W_{\underlinei \underlinei}(\mathbf{0}),\quad 
\big(\Q_{ij}(\br,\bs)\big)^2\leq \Q(\mathbf{0})\,\W_{\underlinei \underlinei \underlinej \underlinej}(\mathbf{0},\bs-\br,\bs-\br),
\quad
i\leq j
\label{CS_Inequality_qws_01}
\end{align}
\end{enumerate}

\item
It is interesting to evaluate the average deviation of $w_{i}(\bx)\, w_{j}(\by)$ from $\overline{w_{i}(\bx) w_{j}(\by)}$ through
\begin{align*}
 \overline{\Big(w_{i}(\bx) w_{j}(\by)-\overline{w_{i}(\bx) w_{j}(\by)}\Big)^2}\geq 0
\end{align*}
or
\begin{align}
\W_{\underlinei\underlinei\underlinej\underlinej}(\mathbf{0},\br,\br)\geq 
\big(\W_{i j}(\br)\big)^2,
\quad
i\leq j
\label{wiwj_Deviation}
\end{align}
This inequality has certain similarity to the quasi-Normal approximation, and it has a significant implication to the asymptotic state solutions to be discussed. Similar inequalities involving $w_{i}(\bx)$ and $\w_{j,k}(\by)$ can also be formulated, such as
\begin{align}
&
 \overline{w_{\underlinei}(\bx)w_{\underlinei}(\bx)  w_{\underlinej},_{\underlinek}(\by)w_{\underlinej},_{\underlinek}(\by)}
-\Big(\overline{w_{i}(\bx) w_{j},_k(\by)}\Big)^2\geq 0,
\notag\\[4pt]&
 \overline{w_{\underlinei},_{\underlinek}(\bx)w_{\underlinei},_{\underlinek}(\bx)  w_{\underlinej},_{\underlinel}(\by) w_{\underlinej},_{\underlinel}(\by)}
-\Big(\overline{w_{i},_k(\bx) w_{j},_l(\by)}\Big)^2\geq 0
\label{wiwj,k_Deviation}
\end{align}
and so on.
\end{enumerate}

\subsection{Objective Function}
\ \ \ \
In PART I, we have restricted our treatment to the case of bounded flow domains so as to avoid the complication of a functional formulation of probability density. Therefore, we need to modify the objective function for the homogeneous shear turbulence in the unbounded flow domain of $\mathbb{R}^2$. We have established in PART I the proportional relationship between $\K$ and the total fluctuation kinetic energy possessed in a turbulent flow, and consequently, we will redefine here the objective as the fluctuation energy per unit area or equivalently
\begin{align}
\K^{\text{hom}}=\W_{kk}(\mathbf{0})
=\int_{\mathbb{R}^2} \tW_{kk}(\bk)\,d\bk
=\int_{\mathbb{R}^2} \frac{|\bk|^2}{(k_2)^2}\, \soc(\bk)\,d\bk
\label{ObjectiveFunction_01}
\end{align}
It is preferable to employ $\K$ as the alternative objective to be maximized which has a mathematically simple linear structure and a physically clear meaning, compared with the other invariants of the covariance matrix $\overline{\w_i(\bx) \w_j(\by)}$. We need to examine how the alternative affects the uniqueness of solutions and other issues.

It is clear that the mathematical problem of \eqref{HST_DivergenceFreeInPhysicalSpace_wiwj_fs} through \eqref{ObjectiveFunction_01}, together with \eqref{HST_PressureInPhysicalSpace_qq_fs} through \eqref{HST_PressureInPhysicalSpace_qww_fs}, is an optimal control problem of an infinite dimensional system governed by two integro-partial differential equations with $\soc$ and $\toc$ as the state variables and $\foc$ as the control variable (\cite{Barbu1993}, \cite{Lasiecka2002}). This link implies that we should solve the problem with the help of the relevant tools from optimal control theory and develop further analysis if required.

\section{Formal Solutions Without Enforcing Constraints}\label{Sec:FormalSolutionsWithoutEnforcingConstraints}
\ \ \ \
Equations \eqref{HST_CLMInPhysicalSpace_w1w1_fs} and \eqref{HST_CLMInPhysicalSpace_w1w1w1_fs} are of first order and linear forms, which can be solved formally with the help of the method of characteristics and the separation of variables under appropriate initial conditions. We explore the properties of the equations, without enforcing the maximization of objective and the constraints of inequality listed above.

\subsection{Transient States}\label{Subsec:TransientStates}
\ \ \ \
Under rather general initial conditions, we can find the formal solutions of \eqref{HST_CLMInPhysicalSpace_w1w1_fs} and \eqref{HST_CLMInPhysicalSpace_w1w1w1_fs} with the aid of the method of characteristics, which are presented below.
\begin{align}
&
\soc(t,\bk)
\notag\\
=\,&
 \frac{|\bk^{\prime\prime}|^4}{|\bk|^4}\,\frac{(k_2)^2}{(k^{\prime\prime}_2)^2}\,
     \exp\s\Big[2 \big(H(0,\bk^{\prime\prime})-H(0,\bk)\big)\Big]\,\soc_0(\bk^{\prime\prime})
\notag\\ &
+\frac{2(k_2)^2}{|\bk|^4} \int_0^t dt'   
   \,|\bk^{\prime}|^2 \,\exp\s\Big[2 \big(H\big(0,\bk^{\prime}\big)- H(0,\bk)\big)\Big]\s
 \int_{\mathbb{R}^2}\s d\bl\, |\bl|^2
   \,(k_1\,l_2-k'_2\,l_1)\,
 \frac{\toc\big(t',\bk^{\prime},\bl\big)}{k'_2\,l_2\,(k'_2+l_2)}
\label{HST_CLMInPhysicalSpace_w1w1_fs_Solution}
\end{align}
and 
\begin{align}
&
 \toc(t,\bk,\bl)
\notag\\
=\,&
\frac{|\bk^{\prime\prime}|^2\,|\bl^{\prime\prime}|^2\,|\bk^{\prime\prime}+\bl^{\prime\prime}|^2}{|\bk|^2\,|\bl|^2\,|\bk+\bl|^2}\,
     \frac{k_2\,l_2\,(k_2+l_2)}{k^{\prime\prime}_2\,l^{\prime\prime}_2\,(k^{\prime\prime}_2+l^{\prime\prime}_2)}
\notag\\[2pt]&\hskip 4mm \times\s
  \exp\s\Big[H(0,\bk^{\prime\prime})-H(0,\bk)+ H(0,\bl^{\prime\prime})- H(0,\bl)
\notag\\[2pt] &\hskip 18mm
+ H(0,\bk''\s+\s\bl'')-H(0,\bk\s+\s\bl)
             \Big] \toc_0(\bk'',\bl'')
\notag\\[2pt]
&
+\frac{k_2\,l_2\,(k_2+l_2)}{|\bk|^2\,|\bl|^2\,|\bk+\bl|^2}
\notag\\[2pt]
&\hskip 4mm\times\s
 \int_0^{t} dt'  
\, \frac{|\bk^{\prime}|^2\,|\bl^{\prime}|^2\,|\bk^{\prime}+\bl^{\prime}|^2}{k^{\prime}_2\,l^{\prime}_2\,(k^{\prime}_2+l^{\prime}_2)}
                  \exp\s\Big[H(0,\bk^{\prime})-H(0,\bk)+ H(0,\bl^{\prime})- H(0,\bl)
\notag\\[2pt]
&\hskip 60mm
                             + H(0,\bk^{\prime}+\bl^{\prime})- H(0,\bk+\bl)\Big]
\notag\\
&\hskip 15mm
\times\s\bigg[
 (k_1+l_1)\,\frac{(k^{\prime}_2+l^{\prime}_2)^2}{|\bk^{\prime}+\bl^{\prime}|^2}\,\focI(t',\bk^{\prime},\bl^{\prime})
+(k^{\prime}_2+l^{\prime}_2) \bigg(1 -\frac{2(k_1+l_1)^2}{|\bk^{\prime}+\bl^{\prime}|^2} \bigg)\,\focII(t',\bk^{\prime},\bl^{\prime}) 
\notag\\
&\hskip 22mm
-k_1\,\frac{(k_2^{\prime})^2}{|\bk^{\prime}|^2}\,\focI(t',-\bk^{\prime}-\bl^{\prime},\bl^{\prime})
-k_2^{\prime}\,\bigg(1-\frac{2(k_1)^2}{|\bk^{\prime}|^2}\bigg)\,\focII(t',-\bk^{\prime}-\bl^{\prime},\bl^{\prime})
\notag\\
&\hskip 22mm
-l_1\,\frac{(l_2^{\prime})^2}{|\bl^{\prime}|^2} \,\focI(t',-\bk^{\prime}-\bl^{\prime},\bk^{\prime})
-l^{\prime}_2\,\bigg(1-\frac{2(l_1)^2}{|\bl^{\prime}|^2}\bigg)  \focII(t',-\bk^{\prime}-\bl^{\prime},\bk^{\prime})
\bigg]      
\label{HST_CLMInPhysicalSpace_w1w1w1_fs_Solution}
\end{align}
Here, $\soc_0(\bk)=\soc(0,\bk)$ and $\toc_0(\bk,\bl)=\toc(0,\bk,\bl)$ are, respectively, the initial conditions of $\soc$ and $\toc$, and
\begin{align}
&
\bk''=(k_1,k_2+k_1\,t),\quad
\bl''=(l_1,l_2+l_1\,t),\quad
\bk'=\l(k_1,k_2+k_1(t-t')\r),
\notag\\[3pt] &
\bl'=\l(l_1,l_2+l_1(t-t')\r)
\end{align}
\begin{align*}
H(0,\bk'')-H(0,\bk)
=&
-t \bigg[(k_1)^2+\frac{1}{6}\,\Big(\big(k_2+k^{\prime\prime}_2\big)^2+(k_2)^2+\big(k^{\prime\prime}_2\big)^2 \Big) \bigg],
\notag\\
H(0,\bk')-H(0,\bk)
=&-(t-t') \bigg[(k_1)^2+\frac{1}{6}\,\Big(\big(k_2+k^{\prime}_2\big)^2+(k_2)^2+\big(k^{\prime}_2\big)^2 \Big) \bigg],
\quad \text{etc.}
\end{align*}

In the derivation of \eqref{HST_CLMInPhysicalSpace_w1w1_fs_Solution}, we have used
\begin{align}
\int_{\mathbb{R}^2}\s d\bl\, 
(k'_2\,l_1-k_1\,l_2)
\, \frac{\toc\big(t',\bk^{\prime},\bl\big)}{k'_2\,l_2\,(k'_2+l_2)}
=0,\quad
 \int_{\mathbb{R}^2}\s d\bl\, (\bk'+\bl)\s\cdot\s\bl\,(k'_2\,l_1-k_1\,l_2)\,
       \frac{\toc(t',\bk',\bl)}{k'_2\,l_2\,(k'_2+l_2)}\,             
=0
\label{HST_CLMInPhysicalSpace_w1w1_fs_Equalities}
\end{align}
which can be verified directly on the basis of $\toc(\bk',\bl)=\toc(\bk',-\bk'-\bl)$ from \eqref{HST_DivergenceFreeInPhysicalSpac_wiwjwk_fs}.

One prominent feature of the formal solutions \eqref{HST_CLMInPhysicalSpace_w1w1_fs_Solution} and \eqref{HST_CLMInPhysicalSpace_w1w1w1_fs_Solution} is the presence of the mixed modes of time and wave numbers such as $k_2+k_1t$, $l_2+l_1 t$, $k_2+k_1(t-t')$ and $l_2+l_1(t-t')$, which characterize the turbulent energy transfer among various wave numbers as time proceeds, as to be demonstrated below.

\subsubsection{Behaviors of $\soc(t,\mathbf{0})$ and $\toc(t,\mathbf{0},\bl)$}\label{Subsec:Behaviorsgsocandgtoc}
\ \ \ \
There is a singularity at $k_1=0$ contained in $\exp\s\big[2 H(0,\bk)\big]$ and $ k_1\,\partial/\partial k_2$ of \eqref{HST_CLMInPhysicalSpace_w1w1_fs}, and there are singularities at $k_1 l_1 (k_1+l_1)=0$ contained in $\exp\s\big[H(0,\bk)+ H(0,\bl) + H(0,\bk+\bl)\big]$ and $k_1\,\partial/\partial k_2+l_1\,\partial/\partial l_2$ of \eqref{HST_CLMInPhysicalSpace_w1w1w1_fs}. We may understand their consequences in \eqref{HST_CLMInPhysicalSpace_w1w1_fs_Solution} and \eqref{HST_CLMInPhysicalSpace_w1w1w1_fs_Solution} through the limit of $\bk\rightarrow\mathbf{0}$.

We can approach $\bk=\mathbf{0}$ in $\mathbb{R}^2$ from different directions. To simplify the analysis, we focus on the limits of
\begin{align*}
\lim_{k_1\rightarrow 0}\lim_{k_2\rightarrow 0}\big\{\soc(t,\bk),\ \toc(t,\bk,\bl)\big\},
\quad
\lim_{k_2\rightarrow 0}\lim_{k_1\rightarrow 0}\big\{\soc(t,\bk),\ \toc(t,\bk,\bl)\big\}
\end{align*}

We set first $k_2=0$ and $k_1\not=0$ in \eqref{HST_CLMInPhysicalSpace_w1w1_fs_Solution} to obtain 
\begin{align*}
\soc(t,\bk)=0
\end{align*}
and we then have
\begin{align*}
\lim_{k_1\rightarrow 0}\lim_{k_2\rightarrow 0}\soc(t,\bk)=0
\end{align*}
Alternatively, under a fixed $k_2\not=0$, taking $k_1\rightarrow 0$ in \eqref{HST_CLMInPhysicalSpace_w1w1_fs_Solution} gives
\begin{align}
\soc(t,\bk)
=\,&
 \exp\s\Big[-2\,(k_2)^2\,t\Big]\,\soc_0((0,k_2))
\notag\\ &
-2\,k_2 \int_0^t dt'   
   \,\exp\s\Big[-2\,(k_2)^2\,(t-t')\Big]\s
    \int_{\mathbb{R}^2}\s d\bl\,|\bl|^2\,l_1\,
           \frac{\toc\big(t',(0,k_2),\bl\big)}{k_2\,l_2\,(k_2+l_2)}
\label{HST_CLMInPhysicalSpace_w1w1_fs_Solution_k1=0}
\end{align}
Consequently,
\begin{align*}
\lim_{k_2\rightarrow 0}\lim_{k_1\rightarrow 0}\soc(t,\bk)=\soc_0(\mathbf{0})
\end{align*}
due to the expectantly bounded $\toc\big(t',\mathbf{0},\bl\big)$ and integral in \eqref{HST_CLMInPhysicalSpace_w1w1_fs_Solution_k1=0} at any finite time, (see also \eqref{SingularityRemovalTransform} below). It follows from the equality of the two limits that
\begin{align}
\soc_0(\mathbf{0})=\soc(t,\mathbf{0})=0
\label{Constraint_soc_bk=b0}
\end{align}

Similarly, we consider the case of $\toc(t,\mathbf{0},\bl)$. We have from \eqref{HST_CLMInPhysicalSpace_w1w1w1_fs_Solution}, under fixed $\bl\not=\mathbf{0}$,
\begin{align*}
\lim_{k_1\rightarrow 0}\lim_{k_2\rightarrow 0}\toc(t,\bk,\bl)=0
\end{align*}
and 
\begin{align*}
\lim_{k_2\rightarrow 0}\lim_{k_1\rightarrow 0} \toc(t,\bk,\bl)
=
\frac{|\bl^{\prime\prime}|^4}{|\bl|^4}\,
     \frac{(l_2)^2}{(l^{\prime\prime}_2)^2}\,
  \exp\s\Big[2 \big(H(0,\bl^{\prime\prime})- H(0,\bl)\big)\Big] \toc_0(\mathbf{0},\bl^{\prime\prime})
\end{align*}
These two limits should be the same, and thus, we have
\begin{align}
\toc_0(\mathbf{0},\bl)=\toc(t,\mathbf{0},\bl)=0
\label{Constraint_toc_bk=b0}
\end{align}

\subsubsection{Effects of Initial Conditions $\soc_0(\bk)$ and $\toc_0(\bk,\bl)$}\label{Subsec:Theeffectsofgsoc0andgtoc0}
\ \ \ \
We have some observations on the restrictions and effects of the initial conditions $\soc_0(\bk)$ and $\toc_0(\bk,\bl)$ as follows. 
\begin{enumerate}
\item
The $\soc_0(\bk)$ related term in \eqref{HST_CLMInPhysicalSpace_w1w1_fs_Solution} contains a possible singularity at $k''_2=k_2+k_1t=0$ under $k_1 k_2<0$, or at $t=-k_2/k_1\, (>0)$, which needs to be removed by the distribution of $\soc_0(\bk)$.
Similarly, the $\toc_0(\bk,\bl)$ related term in \eqref{HST_CLMInPhysicalSpace_w1w1w1_fs_Solution} contains possible singularities at $k_2+k_1t=0$, $l_2+l_1t=0$ or $k_2+l_2+(k_1+l_1)\, t=0$ under $k_1 k_2<0$, $l_1 l_2<0$ or $(k_1+l_1) (k_2+l_2)<0$, at certain $t$'s, which need to be removed by the adequate distribution of $\toc_0(\bk,\bl)$. Therefore, we impose the constraints that 
\begin{align}
\lim_{k_2\,\rightarrow\, 0}\frac{\soc_0(\bk)}{(k_2)^2}\ \ \ \text{and}\ \
\lim_{k_2\,\rightarrow\, 0\ \text{or}\ l_2\,\rightarrow\, 0\ \text{or}\ k_2+l_2\,\rightarrow\, 0}\frac{\toc_0(\bk,\bl)}{k_2\,l_2\,(k_2+l_2)}\ \ \text{exist}
\label{Constraint_soc0_toc0_k2l2=0}
\end{align}
or
\begin{align}
\lim_{k_2\,\rightarrow\, 0}\frac{\soc(t,\bk)}{(k_2)^2} \ \ \text{and}\ \
\lim_{k_2\,\rightarrow\, 0\ \text{or}\ l_2\,\rightarrow\, 0\ \text{or}\ k_2+l_2\,\rightarrow\, 0}\frac{\toc(t,\bk,\bl)}{k_2\,l_2\,(k_2+l_2)}\ \ \text{exist}
\label{Constraint_soc_toc_k2l2=0}
\end{align}
under the expected invariance of time translation. Otherwise, say, the limits of \eqref{Constraint_soc_toc_k2l2=0} did not exist at some $t_0>0$, we could then take $t=t_0$ as an initial instant and infer the validity of \eqref{Constraint_soc_toc_k2l2=0} at $t_0$ from the application of \eqref{Constraint_soc0_toc0_k2l2=0} to the new setting, a contradiction.

The constraints above suggest the transformations of
\begin{align}
&
\soc(\bk)=(k_2)^2\,\dsoc(\bk),\quad
\toc(\bk,\bl)=k_2\,l_2\,(k_2+l_2)\,\dtoc(\bk,\bl),
\notag\\[4pt]
&
\foc(\bk,\bl,\bm)=k_2\,l_2\,m_2\,(k_2+l_2+m_2)\,\dfoc(\bk,\bl,\bm)
\label{SingularityRemovalTransform}
\end{align}
with
\begin{align}
&
\dsoc(\bk)=\dsoc(-\bk),\quad
\dtoc(\bk,\bl)=\dtoc(\bl,\bk)=\dtoc(-\bk-\bl,\bl)=\dtoc(-\bk-\bl,\bk)=\dtoc(-\bk,-\bl),
\notag\\[4pt]
&
\dfoc(\bk,\bl,\bm)=\dfoc(-\bk,-\bl,-\bm)
=\dfoc(\bk,\bm,\bl)=\dfoc(\bm,\bl,\bk)=\dfoc(\bl,\bk,\bm)
\notag\\[4pt]
&
=\dfoc(-\bk-\bl-\bm,\bl,\bm)
=\dfoc(-\bk-\bl-\bm,\bk,\bm)
=\dfoc(-\bk-\bl-\bm,\bk,\bl)
\label{HST_DivergenceFreeInPhysicalSpace_wiwjwkwl_fs_Transf}
\end{align}
following from \eqref{HST_DivergenceFreeInPhysicalSpace_wiwj_fs} through \eqref{HST_DivergenceFreeInPhysicalSpace_wiwjwkwl_fs}.
 These transformations are compatible with \eqref{HST_DivergenceFreeInPhysicalSpace_qandw_fs} in the limit of $k_2\rightarrow 0$
  and the forms of \eqref{HST_DivergenceFreeInPhysicalSpace_wiwj_fs} through \eqref{HST_DivergenceFreeInPhysicalSpace_wiwjwkwl_fs},
 and they also make \eqref{Constraint_soc_bk=b0} and \eqref{Constraint_toc_bk=b0} satisfied automatically.  

If we substitute \eqref{SingularityRemovalTransform} into \eqref{HST_CLMInPhysicalSpace_w1w1_fs_Solution} and \eqref{HST_CLMInPhysicalSpace_w1w1w1_fs_Solution} and we require that 
\begin{align*}
\lim_{\bk\rightarrow\mathbf{0}}\dsoc(t,\bk)\ \ \text{and}\ \ \lim_{\bk\rightarrow\mathbf{0}}\dtoc(t,\bk,\bl)\ \text{exist},
\end{align*}
we get the constraints of
\begin{align}
\dsoc_0(\mathbf{0})=\dsoc(t,\mathbf{0})=0,\quad 
\dtoc_0(\mathbf{0},\bl)=\dtoc(t,\mathbf{0},\bl)=0,
\quad\dfocI(t,-\bl,\bl)=\dfocII(t,-\bl,\bl)=0
\label{Constraint_gtoc_bk=b0}
\end{align}

\item
 Under fixed $k_1\not= 0$, the first term on the right-hand side of \eqref{HST_CLMInPhysicalSpace_w1w1_fs_Solution} tends to
\begin{align}
\frac{(k_1 k_2)^2}{|\bk|^4}\, t^2 \exp\s\Big[-\frac{2}{3}\,(k_1)^2 t^3\Big]\,\soc_0(\bk'')\quad \text{at large}\ t
\label{HST_CLMInPhysicalSpace_w1w1_fs_Solution_larget}
\end{align}
The constraints of $\soc_0(\bk)\geq 0$ and $\int_{\mathbb{R}^2}\soc_0(\bk)\,d\bk = \W_{11}(0,\mathbf{0}) < \infty$ imply that $\soc_0(\bk)$ is bounded for all the wave numbers and is negligible at large $|\bk|$. Therefore, $\soc_0(\bk^{\prime\prime})$ will have negligible effects on $\soc(t,\bk)$, $k_1\not=0$, at large time.

In the case of $k_1=0$ and $k_2\not=0$, \eqref{HST_CLMInPhysicalSpace_w1w1_fs_Solution_k1=0} indicates that $\soc_0(\bk^{\prime\prime})$ will have negligible effects on $\soc(t,\bk)$ at large time. In the case of $\bk=\mathbf{0}$, \eqref{Constraint_soc_bk=b0} says that $\soc_0(\bk^{\prime\prime})=0$. Consequently, $\soc_0(\bk^{\prime\prime})$ will have negligible effects on $\soc(t,\bk)$ at large time.

\item
Under fixed $k_1$ and $l_1$ with $k_1 l_1 (k_1+l_1)\not=0$, the first term on the right-hand side of \eqref{HST_CLMInPhysicalSpace_w1w1w1_fs_Solution} has the asymptote of
\begin{align}
\frac{k_1\,l_1 (k_1+l_1)\,k_2\,l_2 (k_2+l_2)}{|\bk|^2\,|\bl|^2\,|\bk+\bl|^2}\, t^3 \exp\s\Big[-\frac{1}{3}\,\Big((k_1)^2+(l_1)^2+(k_1+l_1)^2 \Big) t^3\Big]
       \toc_0(\bk^{\prime\prime},\bl^{\prime\prime})\ \ \text{at large}\ t
\label{HST_CLMInPhysicalSpace_w1w1_fs_Solution_larget_02}
\end{align}
We expect that $\toc_0(\bk,\bl)$ is bounded under supposedly bounded $\W_{ijk}(0, \br, \bs)$ with 
$$\lim_{|\br|\rightarrow\infty}\W_{ijk}(0, \br, \bs)=\lim_{|\bs|\rightarrow\infty}\W_{ijk}(0, \br, \bs)=0$$
As a consequence, we conclude that the effect of $\toc_0(\bk,\bl)$ on $\toc(t,\bk,\bl)$ will become negligible at large time.

\item
The effect of  $\toc_0(\bk,\bl)$ on $\soc(t,\bk)$ is described by the term of
\begin{align}
&
 \frac{2(k_2)^2}{|\bk|^4} \int_0^t dt'   
   \,\exp\s\bigg[-2\bigg((k_1)^2+\frac{1}{3}\Big((k'_2)^2+k'_2\,k_2+(k_2)^2\Big)\bigg)(t-t')\bigg]
\notag\\[4pt] &\hskip 18mm \times\s
 \int_{\mathbb{R}^2}\s d\bl\, |\bl|^2
   \,(k_1\,l_2-k'_2\,l_1)\,
 \frac{\toc_0(\bk'',(l_1,l_2+l_1t'))}{k''_2\,(l_2+l_1t')\,(k''_2+l_2+l_1t')}
\notag\\[4pt] &\hskip 28mm \times\s
\frac{|\bk''|^2\,[(l_1)^2+(l_2+l_1t')^2]\,[(k_1+l_1)^2+(k''_2+l_2+l_1t')^2]}{|\bl|^2\,|\bk'+\bl|^2}\,
\notag\\[4pt]&\hskip 28mm \times\s
  \exp\s\bigg[                 
-t' \Big((k_1)^2+(l_1)^2+(k_1+l_1)^2\Big)                  
-\frac{t'}{3}\Big(                               
\big(k^{\prime}_2\big)^2                               
+k^{\prime}_2\,k^{\prime\prime}_2                               
+\big(k^{\prime\prime}_2\big)^2                            
\Big)\notag\\[4pt] &\hskip 41mm                 
 -\frac{t'}{3} \Big(                                    
(l_2)^2                                     
+l_2\,\big(l_2+l_1\,t'\big)                                     
\s+\s\big(l_2+l_1\,t'\big)^2                                    
\Big)\notag\\[4pt] &\hskip 41mm                 
-\frac{t'}{3}\Big(\s                           
\big(k^{\prime}_2+l_2\big)^2                           
\s+\s \big(k^{\prime}_2+l_2\big)\big(k^{\prime\prime}_2+l_2+l_1\,t'\big)                            
\s+\s \big(k^{\prime\prime}_2+l_2+l_1\,t'\big)^2                       
\Big)              
\bigg]
\label{INsOfw1w1w1Onw1w1}
\end{align}
from \eqref{HST_CLMInPhysicalSpace_w1w1_fs_Solution} and \eqref{HST_CLMInPhysicalSpace_w1w1w1_fs_Solution}.  
  The constraint of \eqref{Constraint_soc0_toc0_k2l2=0} makes $\dtoc_0\big(\bk'',(l_1,l_2+l_1t')\big)$ finite.
  Furthermore, under expectantly bounded $\W_{ijk}(0, \br, \bs)$, $\toc_0(\bk,\bl)$ is bounded and goes to zero rapidly in the limits of
 $|\bk|\rightarrow +\infty$ or $|\bl|\rightarrow +\infty$.
 Therefore, under $k_1\not=0$, $\toc_0\big(\bk'',(l_1,l_2+l_1t')\big)$ rapidly approaches zero at large $t$. Also, the exponential functions contained in the integrand of \eqref{INsOfw1w1w1Onw1w1} approaches zero at large $t$ under $k_1\not=0$. Consequently, under $k_1\not=0$, \eqref{INsOfw1w1w1Onw1w1} is expected to be very small and $\toc_0(\bk,\bl)$ has a negligible effect on $\soc(t,\bk)$ at large $t$.
 Similarly, we can argue that, in the case of $k_1=0$ and $k_2\not=0$, $\toc_0(\bk,\bl)$ has a negligible effect on $\soc(t,\bk)$, which is also guaranteed by the adoption of \eqref{HST_CLMInPhysicalSpace_w1w1_w1w1w1_ET_k1l1=0} below. The case of $\bk=\mathbf{0}$ is trivial due to \eqref{Constraint_gtoc_bk=b0}.

\end{enumerate}

 The above conclusion of negligible effects is drawn based solely on the formal transient solutions for $\soc(t,\bk)$ and $\toc(t,\bk,\bl)$ without the enforcement of the constraints of inequality and the maximization of the objective function. Therefore, it does not exclude the impacts of the initial distributions on $\soc(t,\bk)$ and $\toc(t,\bk,\bl)$ at large time via the constraints and the maximization which shape the optimal control starting at $t=0$ with $\soc_0(\bk)$, $\toc_0(\bk,\bl)$ and $\foc_0(\bn,\bk,\bl)$. For example, the existence of asymptotic state solutions of various exponential time rates to be discussed may be viewed as the evidence bearing such impacts. The negligible effects discussed above are more relevant to the possibility that two different sets of initial conditions for $\{\soc_0(\bk)$, $\toc_0(\bk,\bl)$, $\foc_0(\bn,\bk,\bl)\}$ may evolve into the same asymptotic solution
 of $\{\soc(t,\bk)$, $\toc(t,\bk,\bl)$, $\foc(t,\bn,\bk,\bl)\}$ at great $t$.

The discussion above has used the implicit assumptions that, at large time, the $\soc_0(\bk)$-term in \eqref{HST_CLMInPhysicalSpace_w1w1_fs_Solution} is much smaller than the integral term and the $\toc_0(\bk,\bl)$-term in \eqref{HST_CLMInPhysicalSpace_w1w1w1_fs_Solution} much smaller than the other integral term.
   These assumptions seemingly hold if both the integral terms evolve at large time according to $\exp(\sigma\,t)$ with $\sigma$ being constant of any value, given the presence of $-t^3$ in the two exponential functions in \eqref{HST_CLMInPhysicalSpace_w1w1_fs_Solution_larget} and \eqref{HST_CLMInPhysicalSpace_w1w1_fs_Solution_larget_02}.
   However, the complication caused by the dependence of the two exponential functions on the wave numbers needs to examined.
    Some scenarios can occur, for example, the initial distribution terms may be greater than or have the same order of magnitude as the integral terms, 
  such as under the condition of certain small turbulent fluctuations and so on.
   In this case, the solutions may be dominated or significantly modified by the initial distribution terms and decay in a rather complicated fashion as indicated by the initial condition related terms in \eqref{HST_CLMInPhysicalSpace_w1w1_fs_Solution}
  and \eqref{HST_CLMInPhysicalSpace_w1w1w1_fs_Solution},
    in contrast to the constant exponential time rates of the asymptotic state solutions to be discussed. 
 This point may also have certain relevance to the issue of stability analysis to be considered later.

\subsubsection{Intrinsic Equalities}\label{Subsec:IntrinsicEqualities}
\ \ \ \
There are certain intrinsic equalities associated with \eqref{HST_CLMInPhysicalSpace_w1w1_fs_Solution} which can be established as follows. To this end, we first introduce
\begin{align}
L(\bk,t,t'):=
 \int_{\mathbb{R}^2}\s d\bl\, |\bl|^2
   \,(k_1\,l_2-k'_2\,l_1)\,
 \frac{\toc\big(t',\bk',\bl\big)}{k'_2\,l_2\,(k'_2+l_2)},\quad
\bk'=(k_1,k_2+k_1(t-t'))
\end{align}
Resorting to the symmetry $\toc(\bk',\bl)=\toc(-\bk'-\bl,\bl)$ of \eqref{HST_DivergenceFreeInPhysicalSpac_wiwjwk_fs} and the transformation of $k_1\rightarrow -k_1-l_1$ and $k_2\rightarrow -k_2-l_2+l_1\,(t-t')$ which gives $k'_2\rightarrow -k'_2-l_2$,
 we can show that
\begin{align}
 \int_{\mathbb{R}^2}\s d\bk\,L(\bk,t,t')
=
0,\quad \forall t'\in [0,t]
\label{IntrinsicEquality_Wholek1}
\end{align}
Moreover, we have
\begin{align}
 \int_{-\infty}^0\s dk_1\,
 \int_{\mathbb{R}}\s dk_2\,L(\bk,t,t')
= \int^{+\infty}_0\s dk_1\,
 \int_{\mathbb{R}}\s dk_2\,L(\bk,t,t')
=
0,\quad \forall t'\in [0,t]
\label{IntrinsicEquality_Halfk1}
\end{align}
which can be proved by using
\begin{align*}
L(-\bk,t,t')=L(\bk,t,t')
\end{align*}
on the basis of the transformation of $l_1\rightarrow -l_1$ and $l_2\rightarrow -l_2$ and $\toc(-\bk',-\bl)=-\toc(\bk',\bl)$ from \eqref{HST_DivergenceFreeInPhysicalSpac_wiwjwk_fs}, and then,
\begin{align*}
&
2\, \int_{-\infty}^0\s dk_1\,
 \int_{\mathbb{R}}\s dk_2\,L(\bk,t,t')
=
\int_{-\infty}^0\s dk_1\,
 \int_{\mathbb{R}}\s dk_2\, L(\bk,t,t')
+\int_{-\infty}^0\s dk_1\,
 \int_{\mathbb{R}}\s dk_2\, L(\bk,t,t')
\notag\\ 
=\,&
\int_{-\infty}^0\s dk_1\,
 \int_{\mathbb{R}}\s dk_2\, L(\bk,t,t')
-\int_{+\infty}^0\s dk_1\,
 \int_{\mathbb{R}}\s dk_2\, L((-k_1,k_2),t,t')
\notag\\ 
=\,&
\int_{-\infty}^0\s dk_1\,
 \int_{\mathbb{R}}\s dk_2\, L(\bk,t,t')
-\int_{+\infty}^0\s dk_1\,
 \int_{\mathbb{R}}\s dk_2\, L((-k_1,-k_2),t,t')
\notag\\ 
=\,&
\int_{-\infty}^0\s dk_1\,
 \int_{\mathbb{R}}\s dk_2\, L(\bk,t,t')
+\int^{+\infty}_0\s dk_1\,
 \int_{\mathbb{R}}\s dk_2\, L(\bk,t,t')
=\int_{\mathbb{R}^2}\s d\bk\, L(\bk,t,t')=0
\end{align*}

Next, we define
\begin{align}
L(\bk',t'):=
 \int_{\mathbb{R}^2}\s d\bl\, |\bl|^2
   \,(k_1\,l_2-k'_2\,l_1)\,
 \frac{\toc\big(t',\bk',\bl\big)}{k'_2\,l_2\,(k'_2+l_2)},\quad
\bk'=(k_1,k'_2)
\end{align}
and using arguments similar to the ones above, we can also show that
\begin{align}
 \int_{\mathbb{R}^2}\s d\bk'\,L(\bk',t')
=
 \int_{-\infty}^0\s dk_1\,
 \int_{\mathbb{R}}\s dk'_2\,L(\bk',t')
= \int^{+\infty}_0\s dk_1\,
 \int_{\mathbb{R}}\s dk'_2\,L(\bk',t')
=0,\quad \forall t'\in [0,t]
\label{IntrinsicEquality_bkp}
\end{align}

The significance of \eqref{IntrinsicEquality_Wholek1}, \eqref{IntrinsicEquality_Halfk1} and \eqref{IntrinsicEquality_bkp} may be understood by recasting \eqref{HST_CLMInPhysicalSpace_w1w1_fs_Solution} in the form of
\begin{align}
\soc(t,\bk)
=\,&
 \frac{|\bk^{\prime\prime}|^4}{|\bk|^4}\,\frac{(k_2)^2}{(k^{\prime\prime}_2)^2}\,
     \exp\s\Big[2 \big(H(0,\bk^{\prime\prime})-H(0,\bk)\big)\Big]\,\soc_0(\bk^{\prime\prime})
\notag\\ &\s\s\s
+\frac{2(k_2)^2}{|\bk|^4} \int_0^t dt'   
   \,|\bk^{\prime}|^2 \,\exp\s\Big[2 \big(H\big(0,\bk^{\prime}\big)- H(0,\bk)\big)\Big]
      \,L(\bk,t,t')
\end{align}
The non-negativity of $|\bk^{\prime}|^2 \,\exp\s\Big[2 \big(H\big(0,\bk^{\prime}\big)- H(0,\bk)\big)\Big]$ and the negative values of $L(\bk,t,t')$ in certain spatial regions resulting from the zero sum balance of $L(\bk,t,t')$ with respect to $\bk\in\mathbb{R}^2$ and $L(\bk,t,t')=L(\bk',t')$ with respect to $\bk'\in\mathbb{R}^2$ are expected to restrict the structures of $\toc(t,\bk,\bl)$ and $\foc(t,\bm,\bk,\bl)$ so as to satisfy the non-negativity of $\soc(t,\bk)\geq 0$.

\subsection{Asymptotic States}\label{Subsec:AsymptoticStates}
\ \ \ \
To solve \eqref{HST_CLMInPhysicalSpace_w1w1_fs} and \eqref{HST_CLMInPhysicalSpace_w1w1w1_fs} with the separation of variables, we take 
\begin{align}
&
\soc(t,\bk)=\socAsy(\bk)\,\soc'(t),\quad
\toc(t,\bk,\bl)=\tocAsy(\bk,\bl)\,\toc'(t),\quad
\foc(t,\bk,\bl,\bm)=\focAsy(\bk,\bl,\bm)\,\foc'(t),
\notag\\
&
\tQ(t,\bk)=\tQAsy(\bk)\,\Q'(t),\quad
\tQ^{(I)}_j(t,\bk)=\tQ^{(Ia)}_j(\bk)\,\Q'_j(t),\quad
\tQ_{jk}(t,\bk,\bl)=\tQAsy_{\underlinej\underlinek}(\bk,\bl)\,\Q'_{\underlinej\underlinek}(t)
\label{SeparationOfVariables_Sol}
\end{align}
where the quantities with the superscript \text{\small $(a)$} are independent of $t$.
 Substitution of the above expressions into \eqref{HST_PressureInPhysicalSpace_qq_fs} through \eqref{HST_PressureInPhysicalSpace_qww_fs} results in
\begin{align}
\Q'_{jk}(t)=\Q'_j(t)=\Q'(t)=\foc'(t)=\toc'(t)=\soc'(t)
\end{align}
It then follows from \eqref{HST_CLMInPhysicalSpace_w1w1_fs} and \eqref{HST_CLMInPhysicalSpace_w1w1w1_fs} that
\begin{align}
\soc'(t)=\exp(2 \sigma t)
\label{SeparationOfVariables_Sol_t}
\end{align}
where $\sigma$ is a constant,
\begin{align}
&
 \frac{\partial}{\partial k_2}\bigg\{\frac{|\bk|^4}{(k_2)^2}\,\exp\s\Big[2H(\sigma,\bk)\Big]\,\socAsy(\bk)\bigg\}   
\notag\\[4pt]
= &\,
-\frac{2 |\bk|^2 k_2}{k_1}\,\exp\s\Big[2H(\sigma,\bk)\Big]
\s \int_{\mathbb{R}^2}\s \bigg[
     \frac{l_1}{l_2}
      + \frac{k_1+l_1}{k_2+l_2}\,\frac{k_1}{k_2}\,\frac{l_1}{l_2}
         -\frac{k_1}{k_2}
        -  \frac{k_1+l_1}{k_2+l_2}\,\bigg(\frac{k_1}{k_2}\bigg)^2
\bigg]\, \tocAsy(\bk,\bl) \, d\bl
\label{HST_CLMInPhysicalSpace_w1w1_fs_Asymp}
\end{align}
and
\begin{align}
&
 \bigg(k_1\,\frac{\partial}{\partial k_2}
       +l_1\,\frac{\partial}{\partial l_2}\bigg)
         \bigg\{\frac{|\bk|^2\,|\bl|^2\,|\bk+\bl|^2}{k_2\,l_2\,(k_2+l_2)}
                  \exp\s\Big[H(\sigma_1,\bk)+H(\sigma_2,\bl)+H(\sigma_3,\bk+\bl)\Big] \,\tocAsy(\bk,\bl)\bigg\}
\notag\\[4pt]
=&\,
-\frac{|\bk|^2\,|\bl|^2\,|\bk+\bl|^2}{k_2\,l_2\,(k_2+l_2)}
\exp\s\Big[H(\sigma_1,\bk)+H(\sigma_2,\bl)+H(\sigma_3,\bk+\bl)\Big]
\notag\\
&
\times\s
\bigg[
 (k_1+l_1)\,\frac{(k_2+l_2)^2}{|\bk+\bl|^2}\,\focIAsy(\bk,\bl)
+(k_2+l_2)\,\bigg(1 -\frac{2(k_1+l_1)^2}{|\bk+\bl|^2} \bigg)\,\focIIAsy(\bk,\bl) 
\notag\\[4pt]
&\hskip 8mm
-k_1\,\frac{(k_2)^2}{|\bk|^2}\,\focIAsy(-\bk-\bl,\bl) 
-k_2\,\bigg(1-\frac{2(k_1)^2}{|\bk|^2}\bigg)\,\focIIAsy(-\bk-\bl,\bl)
\notag\\[4pt]
&\hskip 8mm
-l_1\,\frac{(l_2)^2}{|\bl|^2} \,\focIAsy(-\bk-\bl,\bk)
-l_2\,\bigg(1-\frac{2(l_1)^2}{|\bl|^2}\bigg) \, \focIIAsy(-\bk-\bl,\bk)\bigg]     
\label{HST_CLMInPhysicalSpace_w1w1w1_fs_Asymp}
\end{align}
with $\sigma_1=\sigma_2=\sigma_3=2\sigma/3$.

Integrations of \eqref{HST_CLMInPhysicalSpace_w1w1_fs_Asymp} and \eqref{HST_CLMInPhysicalSpace_w1w1w1_fs_Asymp} under the conditions of 
\begin{align}
\lim_{k_2\rightarrow\pm\infty}\socAsy(\bk)=0,\quad
\lim_{k_2\rightarrow\pm\infty}\tocAsy(\bk,\bl)=0
\end{align}
 respectively, result in
\begin{align}
\socAsy(\bk)
=\,&
-\frac{2\,(k_2)^2}{k_1\, |\bk|^4}
  \int^{k_2}_{-\infty} d k'_2\,|\bk'|^2 
   \exp\s\Big[2 \Big(H\big(\sigma,\bk^{\prime}\big)- H(\sigma,\bk)\Big)\Big]
\notag\\[4pt] & \hskip 28mm \times\s
 \int_{\mathbb{R}^2}\s d\bl\, |\bl|^2\,(k_1\,l_2-k'_2\,l_1)\,
     \frac{\tocAsy\big(\bk^{\prime},\bl\big)}{k'_2\,l_2\,(k'_2+l_2)}          
\label{HST_CLMInPhysicalSpace_w1w1_fs_Asymp_Sol}
\end{align}
where
\begin{align*}
k_1<0,\quad \bkp=(k_1, \kp_2)
\end{align*}
\begin{align*}
H(\sigma,\bk^{\prime})-H(\sigma,\bk)
=\frac{k_2-k'_2}{k_1}\,\bigg[\sigma+(k_1)^2+\frac{1}{6}\,\Big(\big(k_2+k^{\prime}_2\big)^2+(k_2)^2+\big(k^{\prime}_2\big)^2 \Big) \bigg]
\end{align*}
and 
\begin{align}
 \tocAsy(\bk,\bl)
=\,&
 -\int^{k_2}_{-\infty} d k''_2  \,\frac{k_2 l_2 (k_2+l_2)}{k_1 |\bk|^2 |\bl|^2 |\bk+\bl|^2}\,
   \frac{|\bk''|^2 |\bl''|^2 |\bk''+\bl''|^2}{k''_2 l''_2 (k''_2+l''_2)}\,
                  \exp\s\Big[\Sigma(\sigma,\bk''+\bl'',\bk'',\bl'';\bk+\bl,\bk,\bl)\Big]
\notag\\[4pt]
&\hskip 15mm
\times\s\bigg[
 (k_1+l_1)\,\frac{(k''_2+l''_2)^2}{|\bk''+\bl''|^2}\,\focIAsy(\bk'',\bl'')
+(k''_2+l''_2) \bigg(1 -\frac{2(k_1+l_1)^2}{|\bk''+\bl''|^2} \bigg)\,\focIIAsy(\bk'',\bl'') 
\notag\\[4pt] &\hskip 22mm
-k_1\,\frac{(k_2'')^2}{|\bk''|^2}\,\focIAsy(-\bk''-\bl'',\bl'')
-k_2''\,\bigg(1-\frac{2(k_1)^2}{|\bk''|^2}\bigg)\,\focIIAsy(-\bk''-\bl'',\bl'')
\notag\\[4pt]
&\hskip 22mm
-l_1\,\frac{(l_2'')^2}{|\bl''|^2} \,\focIAsy(-\bk''-\bl'',\bk'')
-l''_2\,\bigg(1-\frac{2(l_1)^2}{|\bl''|^2}\bigg)  \focIIAsy(-\bk''-\bl'',\bk'')
\bigg]     
\label{HST_CLMInPhysicalSpace_w1w1w1_fs_Asymp_Sol}
\end{align}
where
\begin{align*}
 k_1 <0,\quad
\bk''=(k_1, k''_2), \quad
\bl''=(l_1, l''_2), \quad
l''_2=l_2+\frac{l_1}{k_1}(k''_2-k_2)
\end{align*}
\begin{align*}
&
\Sigma(\sigma,\bk''+\bl'',\bk'',\bl'';\bk+\bl,\bk,\bl)
\notag\\[4pt]
=&\,
\frac{k_2-k''_2}{k_1}\bigg[
 \sigma
+(k_1+l_1)^2+(k_1)^2+(l_1)^2
\notag\\[4pt]
&\hskip 17mm
+\frac{1}{6}\Big(\big(k_2+l_2+k''_2+l''_2\big)^2
                  +(k_2+l_2)^2
                  +\big(k''_2+l''_2\big)^2
\notag\\[4pt]
&\hskip 28mm
                  +\big(k_2+k''_2\big)^2 
                  +(k_2)^2
                  +\big(k''_2\big)^2 
                  +\big(l_2+l''_2\big)^2
                  +(l_2)^2+\big(l''_2\big)^2
            \Big) \bigg]
\end{align*}
In the derivation of \eqref{HST_CLMInPhysicalSpace_w1w1_fs_Asymp_Sol} we have used
\begin{align}
\int_{\mathbb{R}^2} d\bl\, (k'_2\,l_1-k_1\,l_2)\, \frac{\tocAsy(\bk',\bl)}{k'_2\,l_2\,(k'_2+l_2)}=0,\quad
 \int_{\mathbb{R}^2}\s d\bl\, (\bk'+\bl)\s\cdot\s\bl\,\,   (l_1\,k'_2-l_2\,k_1)\, \frac{\tocAsy\big(\bk',\bl\big)}{k'_2\,l_2\,(k'_2+l_2)}=0
\end{align}
which can be shown in a fashion similar to that of \eqref{HST_CLMInPhysicalSpace_w1w1_fs_Equalities}.

The structure of the expression for $l''_2$ in \eqref{HST_CLMInPhysicalSpace_w1w1w1_fs_Asymp_Sol} is essential for the symmetry property of $\tocAsy(\bk,\bl)$ $=$ $\tocAsy(\bl,\bk)$. In the case of $k_1>0$, we need to replace $-\infty$ with $+\infty$ in the limits of the integrals above. The separate treatments are required by the structures of the exponential function parts contained in the integrands of \eqref{HST_CLMInPhysicalSpace_w1w1_fs_Asymp_Sol} and \eqref{HST_CLMInPhysicalSpace_w1w1w1_fs_Asymp_Sol} in that  $(k_2-k''_2)/k_1 \leq 0$ should hold for the sake of integrability, especially under $k_1\rightarrow 0$.
  We may also get the solutions for $\socAsy(\bk)$ and $\tocAsy(\bk,\bl)$ under $k_1>0$ from \eqref{HST_CLMInPhysicalSpace_w1w1_fs_Asymp_Sol} and \eqref{HST_CLMInPhysicalSpace_w1w1w1_fs_Asymp_Sol} with the help of $\socAsy(\bk)=\socAsy(-\bk)$ from \eqref{HST_DivergenceFreeInPhysicalSpace_wiwj_fs} and $\tocAsy(\bk,\bl)=-\tocAsy(-\bk,-\bl)$ from \eqref{HST_DivergenceFreeInPhysicalSpac_wiwjwk_fs}, as to be pursued below.  

Equations \eqref{SeparationOfVariables_Sol} through \eqref{SeparationOfVariables_Sol_t}
      require the special initial conditions for the correlations of $\W_{ij}(t,\br)$, $\W_{ijk}(t,\br,\bs)$
       and $\W_{ijkl}(t,\br,\bs,\bs')$, etc.,
    satisfying \eqref{HST_CLMInPhysicalSpace_w1w1_fs_Asymp_Sol}, \eqref{HST_CLMInPhysicalSpace_w1w1w1_fs_Asymp_Sol} and the constraints,  
   if we intend to solve for the correlations from the time-dependent equations of \eqref{HST_CLMInPhysicalSpace_w1w1_fs} and \eqref{HST_CLMInPhysicalSpace_w1w1w1_fs}.

\subsubsection{Intrinsic Equalities}
\ \ \ \
In the case of the asymptotic solution, we also have a set of intrinsic equalities of zero sum balance corresponding to \eqref{IntrinsicEquality_bkp},
\begin{align}
 \int_{\mathbb{R}^2}\s d\bk'\,\LAsy(\bk')
= \int_{-\infty}^0\s dk_1\,
 \int_{\mathbb{R}}\s dk'_2\,\LAsy(\bk')
= \int^{+\infty}_0\s dk_1\,
 \int_{\mathbb{R}}\s dk'_2\,\LAsy(\bk')
=
0
\label{Integrand_ZeroBalanceHalfPlane}
\end{align}
where
\begin{align}
\LAsy(\bk'):=\int_{\mathbb{R}^2}\s d\bl\, |\bl|^2\,(k_1\,l_2-k'_2\,l_1)\,\frac{\tocAsy(\bk',\bl)}{k'_2\,l_2\,(k'_2+l_2)},
\quad \bk'=(k_1,k'_2),\quad
\LAsy(-\bk')=\LAsy(\bk')
\label{LAsy_Defn}
\end{align}
Moreover, we can apply \eqref{HST_CLMInPhysicalSpace_2p_r} to the case of asymptotic states to obtain
the known intrinsic equality of
\begin{align}
 \sigma\,\WAsy_{jj}(\mathbf{0})
+\WAsy_{12}(\mathbf{0})
-\frac{\partial^2}{\partial r_k\partial  r_k}\WAsy_{jj}(\br)\bigg|_{\br\,=\,\mathbf{0}}
=0
\label{IntrinsicRelationAsymptoticStateForsigma}
\end{align}
Here, we have used \eqref{Homogeneity_Symmetry},\eqref{Homogeneity_Inversion} and \eqref{HST_DivergenceFreeInPhysicalSpace_2p_3p_rs} in the derivation.
     In the Fourier wave number space, the equality takes the form of
\begin{align}
\int_{-\infty}^0dk_1\, \int_{\mathbb{R}} dk_2\,
    \Big[\big(\sigma+|\bk|^2\big)\,\tWAsy_{kk}(\bk)+\tWAsy_{12}(\bk)\Big]
=0
\label{IntrinsicEqualityForWij_FourierWaveNumberSpace}
\end{align}
We can verify directly that, with the help of \eqref{HST_DivergenceFreeInPhysicalSpace_wiwj_fs}, \eqref{HST_CLMInPhysicalSpace_w1w1_fs_Asymp} and \eqref{Integrand_ZeroBalanceHalfPlane}, the above equality is redundant.
However, it can provide some interesting estimates in a rather simple manner.
   The consequences of \eqref{Integrand_ZeroBalanceHalfPlane}
 and \eqref{IntrinsicEqualityForWij_FourierWaveNumberSpace} will be explored in Section \ref{sec:AsymptoticStateSolution}.

\subsubsection{The Existence of Certain Limits}\label{TheExistenceofCertainLimits}
\ \ \ \
To help model $\LAsy(\bk')$, $\tocAsy(\bk',\bl)$ or $\focAsy(\bk'',\bl,\bm)$ appropriately, we need to examine their asymptotic behaviors under certain limits.
 For this purpose, we recast \eqref{HST_CLMInPhysicalSpace_w1w1_fs_Asymp_Sol} and
   \eqref{HST_CLMInPhysicalSpace_w1w1w1_fs_Asymp_Sol} in the form of
\begin{align*}
\socAsy(\bk)= \int^{k_2}_{-\infty}\s dk'_2\,\rho_{\soc}(\bk;k'_2),\quad
\tocAsy(\bk,\bl)= \int^{k_2}_{-\infty}\s dk''_2\,\rho_{\toc}(\bk,\bl;k''_2)
\end{align*}
where
\begin{align}
\rho_{\soc}(\bk;k'_2):=
\frac{2\,(k_2)^2\,|\bk'|^2}{|k_1|\,|\bk|^4}\, 
   \exp\s\Big[2 \Big(H\big(\sigma,\bk^{\prime}\big)- H(\sigma,\bk)\Big)\Big]
         \,\LAsy(\bk')
\label{IntegrandOfbeta}
\end{align}
and
\begin{align}
\rho_{\toc}(\bk,\bl;k''_2)
:=\,&
 \frac{k_2 l_2 (k_2+l_2)}{|k_1|\, |\bk|^2 |\bl|^2 |\bk+\bl|^2}\,
   \frac{|\bk''|^2 |\bl''|^2 |\bk''+\bl''|^2}{k''_2 l''_2 (k''_2+l''_2)}\,
                  \exp\s\Big[\Sigma(\sigma,\bk''+\bl'',\bk'',\bl'';\bk+\bl,\bk,\bl)\Big]
\notag\\[4pt]
&
\times\s\bigg[
 (k_1+l_1)\,\frac{(k''_2+l''_2)^2}{|\bk''+\bl''|^2}\,\focIAsy(\bk'',\bl'')
+(k''_2+l''_2) \bigg(1 -\frac{2(k_1+l_1)^2}{|\bk''+\bl''|^2} \bigg)\,\focIIAsy(\bk'',\bl'') 
\notag\\[4pt] &\hskip 7mm
-k_1\,\frac{(k_2'')^2}{|\bk''|^2}\,\focIAsy(-\bk''-\bl'',\bl'')
-k_2''\,\bigg(1-\frac{2(k_1)^2}{|\bk''|^2}\bigg)\,\focIIAsy(-\bk''-\bl'',\bl'')
\notag\\[4pt]
&\hskip 7mm
-l_1\,\frac{(l_2'')^2}{|\bl''|^2} \,\focIAsy(-\bk''-\bl'',\bk'')
-l''_2\,\bigg(1-\frac{2(l_1)^2}{|\bl''|^2}\bigg)  \focIIAsy(-\bk''-\bl'',\bk'')
\bigg]     
\label{IntegrandOfgamma}
\end{align}

The structures of the integrands $\rho_{\soc}(\bk;k'_2)$ and $\rho_{\toc}(\bk,\bl;k''_2)$ indicate that
 $\bk=(0^-,k^0_2)$ with $k^0_2\not=0$ is a point of interest.
 The asymptotic behaviors of the integrands in a neighborhood of the point along certain directions are listed below.
  In the analysis we assume that $\socAsy(\bk)$, $\LAsy(\bk)$, $\tocAsy(\bk,\bl)$ and $\focAsy(\bk,\bl,\bm)$ are bounded
 and they tend to zero rapidly as $|\bk|$ goes to infinity.

 The existence of $\rho_{\soc}(\bk;k'_2)$ in the limiting procedure of 
   $k_1\rightarrow 0^-$ and $k'_2\rightarrow (k^0_2)^-$ requires that
\begin{align*}
\lim_{k_1\rightarrow 0^-}\lim_{k'_2\rightarrow (k^0_2)^-}\rho_{\soc}((k_1,k^0_2);k'_2)
=
\lim_{k'_2\rightarrow (k^0_2)^-}\lim_{k_1\rightarrow 0^-}\rho_{\soc}((k_1,k^0_2;k'_2)
\end{align*}
which results in
\begin{align}
\lim_{k_1\rightarrow 0^-}\frac{\LAsy(\bk)}{k_1}=0,\quad \bk=(k_1,k^0_2)
\label{CVL_Limits}
\end{align}
Based this relation and the definition of \eqref{LAsy_Defn}, we infer that
\begin{align}
\lim_{k_1\rightarrow 0^-}\frac{\tocAsy(\bk,\bl)}{k_1}=0,\quad \bk=(k_1,k^0_2)
\label{CVgamma_Limits}
\end{align}

Similarly, the existence of $\rho_{\toc}(\bk,\bl;k''_2)$ in the limiting procedure of 
   $k_1\rightarrow 0^-$ and $k''_2\rightarrow (k^0_2)^-$ requires that
\begin{align*}
\lim_{k_1\rightarrow 0^-}\lim_{k''_2\rightarrow (k^0_2)^-}\rho_{\toc}(\bk,\bl;k''_2)
=
\lim_{k''_2\rightarrow (k^0_2)^-}\lim_{k_1\rightarrow 0^-}\rho_{\toc}(\bk,\bl;k''_2)
\end{align*}
which can be satisfied sufficiently by
\begin{align}
\lim_{k_1\rightarrow 0^-} \frac{\focAsy(\bk,\bl,\bm)}{k_1}
=\lim_{k_1\rightarrow 0^-} \frac{\focAsy(-\bk-\bl,\bl,\bm)}{k_1}
=\lim_{k_1\rightarrow 0^-} \frac{\focAsy(-\bk-\bl,\bk,\bm)}{k_1}
=0,\quad \bk=(k_1,k^0_2)    
\label{CVdelta_Limits}
\end{align}
Here, we have restricted our derivation to the conditions of $l_1\,l_2\,(k^0_2+l_2)\not=0$. Other cases can be discussed in a similar fashion.

\subsection{Evolution and Energy Transfer}\label{Subsec:MechanismofEnergyTransfer}
\ \ \ \
Since the structures of \eqref{HST_CLMInPhysicalSpace_w1w1_fs_Solution} and \eqref{HST_CLMInPhysicalSpace_w1w1w1_fs_Solution} contain the mixed modes, we may explore the relationship between the transient solutions and the asymptotic solutions and understand the mechanism of turbulent energy transfer among the wave numbers. For the sake of simplicity, we will resort to the results above to neglect the initial distribution terms. That is, we focus on the flows where this neglect is appropriate at large time as mentioned in Section \ref{Subsec:TransientStates}. It then follows from \eqref{HST_CLMInPhysicalSpace_w1w1_fs_Solution} and \eqref{HST_CLMInPhysicalSpace_w1w1w1_fs_Solution} that at large $t$, under the change of variables $\tau=t-t'$,
\begin{align}
\soc(t,\bk)
=
\frac{2(k_2)^2}{|\bk|^4} \int_0^t d\tau  
   \,|\bk^{\prime}|^2 \,\exp\s\Big[2 \big(H\big(0,\bk^{\prime}\big)- H(0,\bk)\big)\Big]\s
 \int_{\mathbb{R}^2}\s d\bl\, |\bl|^2
   \,(k_1\,l_2-k'_2\,l_1)\,
 \frac{\toc\big(t-\tau,\bk^{\prime},\bl\big)}{k'_2\,l_2\,(k'_2+l_2)}
\label{HST_CLMInPhysicalSpace_w1w1_ET}
\end{align}
and 
\begin{align}
 \toc(t,\bk,\bl)
=\,&
 \int_0^{t} d\tau\,\frac{k_2\,l_2\,(k_2+l_2)}{|\bk|^2\,|\bl|^2\,|\bk+\bl|^2}\,
         \frac{|\bk^{\prime}|^2\,|\bl^{\prime}|^2\,|\bk^{\prime}+\bl^{\prime}|^2}{k^{\prime}_2\,l^{\prime}_2\,(k^{\prime}_2+l^{\prime}_2)}
                  \exp\s\Big[H(0,\bk^{\prime})\s-H(0,\bk)\s+\s H(0,\bl^{\prime})\s -\s H(0,\bl)
\notag\\[4pt]
&\hskip 80mm
                             + H(0,\bk^{\prime}+\bl^{\prime})-\s H(0,\bk+\bl)
\Big]
\notag\\[4pt]
&\hskip 8mm
\times\s\bigg[
 (k_1+l_1)\,\frac{(k^{\prime}_2+l^{\prime}_2)^2}{|\bk^{\prime}+\bl^{\prime}|^2}\,\focI(t-\tau,\bk^{\prime},\bl^{\prime})
\notag\\[4pt] &\hskip 14mm
+(k^{\prime}_2+l^{\prime}_2) \bigg(1 -\frac{2(k_1+l_1)^2}{|\bk^{\prime}+\bl^{\prime}|^2} \bigg)\,\focII(t-\tau,\bk^{\prime},\bl^{\prime}) 
\notag\\[4pt]
&\hskip 14mm
-k_1\,\frac{(k_2^{\prime})^2}{|\bk^{\prime}|^2}\,\focI(t-\tau,-\bk^{\prime}-\bl^{\prime},\bl^{\prime})
-k_2^{\prime}\,\bigg(1-\frac{2(k_1)^2}{|\bk^{\prime}|^2}\bigg)\,\focII(t-\tau,-\bk^{\prime}-\bl^{\prime},\bl^{\prime})
\notag\\[4pt]
&\hskip 14mm
-l_1\,\frac{(l_2^{\prime})^2}{|\bl^{\prime}|^2} \,\focI(t-\tau,-\bk^{\prime}-\bl^{\prime},\bk^{\prime})
-l^{\prime}_2\,\bigg(1-\frac{2(l_1)^2}{|\bl^{\prime}|^2}\bigg)  \focII(t-\tau,-\bk^{\prime}-\bl^{\prime},\bk^{\prime})
\bigg]     
\label{HST_CLMInPhysicalSpace_w1w1w1_ET}
\end{align}
with
\begin{align*}
\bk^{\prime}=(k_1,k_2+k_1\,\tau),\quad
\bl^{\prime}=(l_1,l_2+l_1\,\tau)
\end{align*}
\begin{align*}
H(0,\bk^{\prime})-H(0,\bk)
=
-\tau\, \bigg[(k_1)^2+\frac{1}{4}\,(k_2)^2+\frac{1}{3}\,\Big(\frac{3}{2}\,k_2+k_1\,\tau \Big)^2 \bigg],
\quad \text{etc.}
\end{align*}

Due to the peculiar structures of the exponential function parts, the integrands of $\int_0^t d\tau$ in \eqref{HST_CLMInPhysicalSpace_w1w1_ET} and \eqref{HST_CLMInPhysicalSpace_w1w1w1_ET} are extremely small under $\bk\not= \mathbf{0}$, $\tau \sim t$ and $t$ sufficiently great; the major contributions to the integrals are expected to come from the $\toc(t-\tau,\cdots)$ and $\foc(t-\tau,\cdots)$ of $\tau<< t$. Motivated by this observation and the separation of variables \eqref{SeparationOfVariables_Sol} through \eqref{SeparationOfVariables_Sol_t}, we take the approximation of
\begin{align}
&
\soc(t-\tau,\bk)\approx\socAsy(\bk)\exp\big(2 \sigma (t-\tau)\big),\quad
\toc(t-\tau,\bk,\bl)\approx\tocAsy(\bk,\bl)\exp\big(2 \sigma (t-\tau)\big),
\notag\\[4pt] &
\foc(t-\tau,\bk,\bl,\bm)\approx \focAsy(\bk,\bl,\bm)\exp\big(2 \sigma (t-\tau)\big)
\label{EmergingFieldDecomposition}
\end{align}
Here, $\sigma$ is taken as a constant;  
   the arguments of $\bk$, $\bl$ and $\bm$ can also be the mixed modes.
   The adequacy of such a treatment may be seen partly by the consistent emergence of the asymptotic solutions from the transient solutions to be established below.
   Next, substitution of the above approximations into \eqref{HST_CLMInPhysicalSpace_w1w1_ET} and \eqref{HST_CLMInPhysicalSpace_w1w1w1_ET} gives
\begin{align}
\socAsy(\bk)
=\,&
\frac{2(k_2)^2}{|\bk|^4} \int_0^t d\tau  
   \,|\bk^{\prime}|^2 \,\exp\s\Big[2 \big(H\big(0,\bk^{\prime}\big)- H(0,\bk)\big)-2 \sigma\tau\Big]\s
\notag\\[4pt] &\hskip 25mm \times\s
 \int_{\mathbb{R}^2}\s d\bl\, |\bl|^2
   \,(k_1\,l_2-k'_2\,l_1)\,
 \frac{\tocAsy\big(\bk^{\prime},\bl\big)}{k'_2\,l_2\,(k'_2+l_2)}
\label{HST_CLMInPhysicalSpace_w1w1_ET_Asy}
\end{align}
and 
\begin{align}
 \tocAsy(\bk,\bl)
=\,&
 \int_0^{t} d\tau  \,\frac{k_2\, l_2\, (k_2+l_2)}{|\bk|^2 |\bl|^2 |\bk+\bl|^2}\,
   \frac{|\bk^{\prime}|^2 |\bl^{\prime}|^2 |\bk^{\prime}+\bl^{\prime}|^2}{k^{\prime}_2 \, l^{\prime}_2\, (k^{\prime}_2+l^{\prime}_2)}
                  \exp\s\Big[H(0,\bk^{\prime})\s-H(0,\bk)\s+\s H(0,\bl^{\prime})\s-\s H(0,\bl)
\notag\\[4pt]
&\hskip 77mm
                             +\s H(0,\bk^{\prime}+\bl^{\prime})-\s H(0,\bk+\bl)-2 \sigma\tau
\Big]
\notag\\[4pt]
&\hskip 8mm
\times\s\bigg[
 (k_1+l_1)\,\frac{(k^{\prime}_2+l^{\prime}_2)^2}{|\bk^{\prime}+\bl^{\prime}|^2}\,\focIAsy(\bk^{\prime},\bl^{\prime})
+(k^{\prime}_2+l^{\prime}_2) \bigg(1 -\frac{2(k_1+l_1)^2}{|\bk^{\prime}+\bl^{\prime}|^2} \bigg)\,\focIIAsy(\bk^{\prime},\bl^{\prime}) 
\notag\\ &\hskip 14mm
-k_1\,\frac{(k_2^{\prime})^2}{|\bk^{\prime}|^2}\,\focIAsy(-\bk^{\prime}-\bl^{\prime},\bl^{\prime})
-k_2^{\prime}\,\bigg(1-\frac{2(k_1)^2}{|\bk^{\prime}|^2}\bigg)\,\focIIAsy(-\bk^{\prime}-\bl^{\prime},\bl^{\prime})
\notag\\[4pt]
&\hskip 14mm
-l_1\,\frac{(l_2^{\prime})^2}{|\bl^{\prime}|^2} \,\focIAsy(-\bk^{\prime}-\bl^{\prime},\bk^{\prime})
-l^{\prime}_2\,\bigg(1-\frac{2(l_1)^2}{|\bl^{\prime}|^2}\bigg)  \focIIAsy(-\bk^{\prime}-\bl^{\prime},\bk^{\prime})
\bigg]     
\label{HST_CLMInPhysicalSpace_w1w1w1_ET_Asy}
\end{align}
where
\begin{align*}
\bk^{\prime}=(k_1,k_2+k_1\,\tau),\quad
\bl^{\prime}=(l_1,l_2+l_1\,\tau)
\end{align*}
\begin{align*}
H(0,\bk^{\prime})-H(0,\bk)
=
-\tau \bigg[(k_1)^2+\frac{1}{4}\,(k_2)^2+\frac{1}{3}\,\Big(\frac{3}{2}\,k_2+k_1\,\tau \Big)^2 \bigg],
\quad \text{etc.}
\end{align*}

\subsubsection{$\soc(t,\bk)$ and $\toc(t,\bk,\bl)$ with $k_1 \not= 0$}
\ \ \ \
That $k_1\not= 0$ makes it possible to adopt the change of variables,
\begin{align*}
\tau=\frac{k'_2-k_2}{k_1},\quad
d\tau=\frac{dk'_2}{k_1},\quad
l'_2=l_2+\frac{l_1}{k_1}\,(k'_2-k_2)
\end{align*}
and consequently, \eqref{HST_CLMInPhysicalSpace_w1w1_ET_Asy} and \eqref{HST_CLMInPhysicalSpace_w1w1w1_ET_Asy} can be recast in the forms of
\begin{align}
\socAsy(\bk)
=\,&
\frac{2(k_2)^2}{k_1 |\bk|^4} \int_{k_2}^{k_2+k_1 t} dk'_2 
   \,|\bk^{\prime}|^2 \,\exp\s\Big[2 \Big(H\big(\sigma,\bk^{\prime}\big)- H(\sigma,\bk)\Big)\Big]\s
\notag\\[4pt] &\hskip 30mm \times\s
 \int_{\mathbb{R}^2}\s d\bl\, |\bl|^2
   \,(k_1\,l_2-k'_2\,l_1)\,
 \frac{\tocAsy\big(\bk^{\prime},\bl\big)}{k'_2\,l_2\,(k'_2+l_2)}
\label{HST_CLMInPhysicalSpace_w1w1_ET_Asy_01}
\end{align}
and 
\begin{align}
 \tocAsy(\bk,\bl)
=\,&
 \int_{k_2}^{k_2+k_1 t} d k'_2  \,\frac{k_2\, l_2\, (k_2+l_2)}{k_1 |\bk|^2 |\bl|^2 |\bk+\bl|^2}\,
   \frac{|\bk^{\prime}|^2 |\bl^{\prime}|^2 |\bk^{\prime}+\bl^{\prime}|^2}{k^{\prime}_2\, l^{\prime}_2\, (k^{\prime}_2+l^{\prime}_2)}\,
                  \exp\s\Big[\Sigma(\sigma,\bkp+\blp,\bkp,\blp;\bk+\bl,\bk,\bl)\Big]
\notag\\[4pt]
&\hskip 15mm
\times\s\bigg[
 (k_1+l_1)\,\frac{(k^{\prime}_2+l^{\prime}_2)^2}{|\bk^{\prime}+\bl^{\prime}|^2}\,\focIAsy(\bk^{\prime},\bl^{\prime})
+(k^{\prime}_2+l^{\prime}_2) \bigg(1 -\frac{2(k_1+l_1)^2}{|\bk^{\prime}+\bl^{\prime}|^2} \bigg)\,\focIIAsy(\bk^{\prime},\bl^{\prime}) 
\notag\\[4pt] &\hskip 22mm
-k_1\,\frac{(k_2^{\prime})^2}{|\bk^{\prime}|^2}\,\focIAsy(-\bk^{\prime}-\bl^{\prime},\bl^{\prime})
-k_2^{\prime}\,\bigg(1-\frac{2(k_1)^2}{|\bk^{\prime}|^2}\bigg)\,\focIIAsy(-\bk^{\prime}-\bl^{\prime},\bl^{\prime})
\notag\\[4pt]
&\hskip 22mm
-l_1\,\frac{(l_2^{\prime})^2}{|\bl^{\prime}|^2} \,\focIAsy(-\bk^{\prime}-\bl^{\prime},\bk^{\prime})
-l^{\prime}_2\,\bigg(1-\frac{2(l_1)^2}{|\bl^{\prime}|^2}\bigg)  \focIIAsy(-\bk^{\prime}-\bl^{\prime},\bk^{\prime})
\bigg]     
\label{HST_CLMInPhysicalSpace_w1w1w1_ET_Asy_01}
\end{align}
where
\begin{align*}
\bk^{\prime}=(k_1,k^{\prime}_2),\quad
\bl^{\prime}=(l_1,l^{\prime}_2),\quad
l'_2=l_2+\frac{l_1}{k_1}\,(k'_2-k_2)
\end{align*}
\begin{align*}
 H(\sigma,\bkp)- H(\sigma,\bk)
=\frac{k_2-k'_2}{k_1}\,
\bigg[\sigma+(k_1)^2+\frac{1}{6}\,\Big((k_2+k'_2)^2+(k_2)^2+(k'_2)^2\Big)\bigg]
\end{align*}
\begin{align*}
&
\Sigma(\sigma,\bkp+\blp,\bkp,\blp;\bk+\bl,\bk,\bl)
\notag\\[4pt]
=&\,
\frac{k_2-k'_2}{k_1}\bigg[
 \sigma
+(k_1+l_1)^2+(k_1)^2+(l_1)^2
\notag\\[4pt]
&\hskip 18mm
+\frac{1}{6}\Big(\big(k_2+l_2+k'_2+l'_2\big)^2
                  +(k_2+l_2)^2
                  +\big(k'_2+l'_2\big)^2
\notag\\[4pt]
&\hskip 18mm
                  +\big(k_2+k'_2\big)^2 
                  +(k_2)^2
                  +\big(k'_2\big)^2 
                  +\big(l_2+l'_2\big)^2
                  +(l_2)^2+\big(l'_2\big)^2
            \Big) \bigg]
\end{align*}
In the analysis below, we will take $k_1<0$ without loss of generality.
 Since the above relations supposedly hold at sufficiently great $t$, we may approximate $t$ with $+\infty$ or take the limit of $t\rightarrow +\infty$ to obtain
\begin{align}
\socAsy(\bk)
=\,&
-\frac{2(k_2)^2}{k_1 |\bk|^4} \int^{k_2}_{-\infty} dk'_2 
   \,|\bk^{\prime}|^2 \,\exp\s\Big[2 \Big(H\big(\sigma,\bk^{\prime}\big)- H(\sigma,\bk)\Big)\Big]\s
\notag\\[4pt] &\hskip 30mm \times\s
 \int_{\mathbb{R}^2}\s d\bl\, |\bl|^2
   \,(k_1\,l_2-k'_2\,l_1)\,
 \frac{\tocAsy\big(\bk^{\prime},\bl\big)}{k'_2\,l_2\,(k'_2+l_2)}
\label{HST_CLMInPhysicalSpace_w1w1_ET_Asy_00}
\end{align}
and 
\begin{align}
 \tocAsy(\bk,\bl)
=\,&
 -\int^{k_2}_{-\infty} \s d k'_2  \frac{k_2\, l_2\, (k_2+l_2)}{k_1 |\bk|^2 |\bl|^2 |\bk+\bl|^2}\,
   \frac{|\bk^{\prime}|^2 |\bl^{\prime}|^2 |\bk^{\prime}+\bl^{\prime}|^2}{k^{\prime}_2\, l^{\prime}_2\, (k^{\prime}_2+l^{\prime}_2)}\,
                  \exp\s\Big[\Sigma(\sigma,\bkp+\blp,\bkp,\blp;\bk+\bl,\bk,\bl)\Big]
\notag\\[4pt]
&\hskip 15mm
\times\s\bigg[
 (k_1+l_1)\,\frac{(k^{\prime}_2+l^{\prime}_2)^2}{|\bk^{\prime}+\bl^{\prime}|^2}\,\focIAsy(\bk^{\prime},\bl^{\prime})
+(k^{\prime}_2+l^{\prime}_2) \bigg(1 -\frac{2(k_1+l_1)^2}{|\bk^{\prime}+\bl^{\prime}|^2} \bigg)\,\focIIAsy(\bk^{\prime},\bl^{\prime}) 
\notag\\[4pt] &\hskip 22mm
-k_1\,\frac{(k_2^{\prime})^2}{|\bk^{\prime}|^2}\,\focIAsy(-\bk^{\prime}-\bl^{\prime},\bl^{\prime})
-k_2^{\prime}\,\bigg(1-\frac{2(k_1)^2}{|\bk^{\prime}|^2}\bigg)\,\focIIAsy(-\bk^{\prime}-\bl^{\prime},\bl^{\prime})
\notag\\[4pt]
&\hskip 22mm
-l_1\,\frac{(l_2^{\prime})^2}{|\bl^{\prime}|^2} \,\focIAsy(-\bk^{\prime}-\bl^{\prime},\bk^{\prime})
-l^{\prime}_2\,\bigg(1-\frac{2(l_1)^2}{|\bl^{\prime}|^2}\bigg)  \focIIAsy(-\bk^{\prime}-\bl^{\prime},\bk^{\prime})
\bigg]    
\label{HST_CLMInPhysicalSpace_w1w1w1_ET_Asy_00}
\end{align}
\begin{enumerate}
\item
Equations \eqref{HST_CLMInPhysicalSpace_w1w1_ET_Asy_00} and \eqref{HST_CLMInPhysicalSpace_w1w1w1_ET_Asy_00} have the same structures as the asymptotic state solutions of \eqref{HST_CLMInPhysicalSpace_w1w1_fs_Asymp_Sol} and \eqref{HST_CLMInPhysicalSpace_w1w1w1_fs_Asymp_Sol}, which implies that a transient solution evolves toward a corresponding asymptotic state solution. Mathematically, such an evolution is possible because of the presence of the mixed modes, $k^{\prime}_2=k_2+k_1(t-t')$, in the transient solution, which makes possible the transformation of integrals from the time domain to the wave number domain.
 
\item
The mixed modes may characterize a mechanism for the turbulent energy transfer and redistribution among various wave numbers, considering that, say, $|\bk'|^2=(k_1)^2+\big(k_2+k_1(t-t')\big)^2$ monotonically increases as $t$ increases under $k_1 k_2>0$; or it increases monotonically as $t-t'$ $(> |k_2|/|k_1|)$ increases under $k_1 k_2<0$. As time proceeds, the initial energy distribution associated with $\soc_0(\bk)$ and $\toc_0(\bk,\bl)$ is redistributed among the wave numbers via the mixed modes, dissipated by the viscous effect and modified by the continual energy supply from the fixed average shearing; its effect on the state of $\soc(t,\bk)$ and $\toc(t,\bk,\bl)$ at great $t$ becomes negligible in the sense discussed before. The eventual emergence of \eqref{EmergingFieldDecomposition} may be interpreted as that, for the concerned case of a fixed shearing in the present study, the energy possessed in each and every wave number is finally saturated such that the exponential time rate of change is synchronized to the same. 
\end{enumerate}

\subsubsection{$\soc(t,\bk)$ and $\toc(t,\bk,\bl)$ under $\bk=(0,k_2)\not=\mathbf{0}$}
\ \ \ \
We now deal with the case of $\bk=(0,k_2)\not=\mathbf{0}$, which is special due to the related occurrence of singularity in \eqref{HST_CLMInPhysicalSpace_w1w1_fs} and \eqref{HST_CLMInPhysicalSpace_w1w1w1_fs}. In this case, under the limit of $k_1\rightarrow 0$, \eqref{HST_CLMInPhysicalSpace_w1w1_ET} and \eqref{HST_CLMInPhysicalSpace_w1w1w1_ET} reduce to
\begin{align}
\soc(t,\bk)
=
\int_0^t d\tau \, 2\,k_2 \,\exp\s\Big[-2\,(k_2)^2\,\tau\Big]
 \int_{\mathbb{R}^2}\s d\bl\, \frac{l_1}{l_2} \toc\big(t-\tau,\bk,\bl\big)
\label{HST_CLMInPhysicalSpace_w1w1_ET_k1=0}
\end{align}
where (\ref{HST_CLMInPhysicalSpace_w1w1_fs_Equalities}$)_2$ has been used and 
\begin{align}
 \toc(t,\bk,\bl)
=\,&
 \int_0^{t} d\tau\,\frac{l_2\,(k_2+l_2)}{|\bl|^2\,|\bk+\bl|^2}\,
         \frac{|\bl^{\prime}|^2\,|\bk+\bl^{\prime}|^2}{l^{\prime}_2\,(k_2+l^{\prime}_2)}
                  \exp\s\Big[\Sigma(\bk+\bl',\bk,\bl';\bk+\bl,\bl)\Big]
\notag\\[4pt]
&\hskip 8mm
\times\s\bigg[
 l_1\,\frac{(k_2+l^{\prime}_2)^2}{|\bk+\bl^{\prime}|^2}\,\focI(t-\tau,\bk,\bl^{\prime})
+(k_2+l^{\prime}_2) \bigg(1 -\frac{2(l_1)^2}{|\bk+\bl^{\prime}|^2} \bigg)\,\focII(t-\tau,\bk,\bl^{\prime}) 
\notag\\[4pt]
&\hskip 14mm
-k_2\,\focII(t-\tau,-\bk-\bl^{\prime},\bl^{\prime})
\notag\\[4pt]
&\hskip 14mm
-l_1\,\frac{(l_2^{\prime})^2}{|\bl^{\prime}|^2} \,\focI(t-\tau,-\bk-\bl^{\prime},\bk)
-l^{\prime}_2\,\bigg(1-\frac{2(l_1)^2}{|\bl^{\prime}|^2}\bigg)  \focII(t-\tau,-\bk-\bl^{\prime},\bk)
\bigg]     
\label{HST_CLMInPhysicalSpace_w1w1w1_ET_k1=0}
\end{align}
with
\begin{align*}
\bl^{\prime}=(l_1,l_2+l_1\,\tau)
\end{align*}
\begin{align*}
\Sigma(\bk+\bl',\bk,\bl';\bk+\bl,\bl)
=
-\bigg[\,& (k_2)^2
           +2\,(l_1)^2+\frac{1}{4}\,(l_2)^2+\frac{1}{3}\,\Big(\frac{3}{2}\,l_2+l_1\,\tau \Big)^2
\notag\\[4pt]
&
+\frac{1}{4}\,(k_2+l_2)^2+\frac{1}{3}\,\Big(\frac{3}{2}\,(k_2+l_2)+l_1\,\tau \Big)^2 \bigg]\tau
\end{align*}

One special feature of these expressions is the missing  mixed mode $k_2+k_1\tau$. If we consider that the mixed modes are the dominant means for the energy transfer among different wave numbers during the transient period, we may neglect the contributions of $\soc(t,\bk)$ and $\toc(t,\bk,\bl)$ under $k_1=0$ by taking
\begin{align}
\soc(t,\bk)=\toc(t,\bk,\bl)=\foc(t,\bk,\bl,\bm)=0,\ \ k_1=0;\quad
\foc(t,\bk,\bl,\bm)=0,\ \ k_1+l_1=0,\ \text{etc.}
\label{HST_CLMInPhysicalSpace_w1w1_w1w1w1_ET_k1l1=0}
\end{align}
Moreover, this constraint is consistent with \eqref{CVL_Limits} through  \eqref{CVdelta_Limits}.

\subsection{Relevance to Stability Analysis}
\ \ \ \
Given that the basic flow field  of \eqref{HST_AverageVelocityField} is taken mathematically fixed in the present analysis and the behavior of disturbances is investigated from the perspective of statistical averaging, it is interesting to notice that the existence of the above transient and asymptotic state solutions has close relevance to the issue of stability of this basic flow.
 Whether the correlations decay or not with time under a set of initial conditions $\{\soc_0(\bk)$,  $\toc_0(\bk,\bl)$, $\foc_0(\bk,\bl,\bm)\}$ indicates, in a statistical sense, whether the basic flow is stable or unstable under the disturbances characterized by the set. This analysis is of statistical nature, in contrast to the conventional stability theory (\cite{DrazinReid2004}, \cite{Joseph1976a}).
 
As a possible extension, we may apply the present formulation of turbulence modeling as optimal control to study the stability of a concerned basic flow field, outlined and explained as follows.
 (i) The basic field is solved from the conservation of mass and the Navier-Stokes equations under appropriate initial and boundary conditions, as done conventionally. Mathematically, the basic field is also a solution of the averaged continuity equation and the averaged Navier-Stokes equations with the Reynolds stress related terms ignored. 
 (ii) The averaged continuity equation and the averaged Navier-Stokes equations are left out, and the basic field is held fixed in the equations of evolution for the second and third order correlations.
 These evolution equations will be solved under the basic field, the constraints of equality and inequality and the maximization of the turbulent energy contained in the domain. 
 (iii) The treatment differs from the conventional stability theory in that the equations for the perturbation field are recast in the forms of statistical correlations, which may partially reflect the nature of disturbances with some degree of randomness and data uncertain and incomplete.
 (iv) The treatment is consistent with the purpose of stability analysis of the concerned flow in the sense that
 the basic flow field is fixed, and it allows only a one-way impact of the basic flow to the correlations. Consequently, the analysis is much less complicated than the full formulation of turbulence modeling as optimal control.
   
Here, motivated by the specific problem of two-dimensional homogeneous shear turbulence, we attempt to establish a link
 between flow stability analysis and optimal control and optimization; and more are to be explored to test its adequacy and consequence.

\section{\label{sec:AsymptoticStateSolution}Asymptotic State Solution}
\ \ \ \
To understand better the mathematical issues involved in the optimal control problem, we explore the special asymptotic state solutions of \eqref{HST_CLMInPhysicalSpace_w1w1_fs} and \eqref{HST_CLMInPhysicalSpace_w1w1w1_fs} with the help of the separation of variables as initiated in Subsection \ref{Subsec:AsymptoticStates}. That is, we seek the asymptotic form solutions of
\begin{align}
\psi=\psiAsy\,\exp(2\sigma t)
\label{AsymptoticForm_01}
\end{align}
where $\psi$ represents any of $\{\soc$, $\toc$, $\foc$,
     $\tW_{ij}$, $\tW^{(I)}_{ijk}$, $\tW_{ijkl}$, $\tQ$, $\tQ^{(I)}_i$, $\tQ_{ij}$,
     $\W_{ij}$, $\W_{ijk}$, $\W_{ijkl}$, $\Q$, $\Q_i$, $\Q_{ij}\}$,  $\psiAsy$ is the time-independent part of $\psi$, and $\sigma$ a constant fixed. 

 We have obtained the formal solution for $\socAsy(\bk)$ and $\tocAsy(\bk,\bl)$ in \eqref{HST_CLMInPhysicalSpace_w1w1_fs_Asymp_Sol} and \eqref{HST_CLMInPhysicalSpace_w1w1w1_fs_Asymp_Sol}, without implementing the constraints of inequality and the optimization. The task now is how to determine $\focAsy(\bk,\bl,\bm)$ with the help of the optimization under the constraints.

Substitution of \eqref{AsymptoticForm_01} into \eqref{HST_DivergenceFreeInPhysicalSpace_wiwjwkwl_fs}, \eqref{NonNegativityofSoc11} and \eqref{wiwiwjwj_Nonnegativity} through \eqref{wiwj,k_Deviation} results in 
\begin{align}
&
 \focAsy(\bk,\bl,\bm)
=\focAsy(\bk,\bm,\bl)
=\focAsy(\bm,\bl,\bk)
=\focAsy(\bl,\bk,\bm)
=\focAsy(-\bk,-\bl,-\bm)
\notag\\[4pt] &
=\focAsy(-\bk-\bl-\bm,\bl,\bm)
\label{HST_DivergenceFreeInPhysicalSpace_wiwjwkwl_fs_Asymp}
\end{align}
\begin{align}
\socAsy(\bk)\geq 0
\label{NonNegativityofSoc11_Asymp}
\end{align}
\begin{align}
\WAsy_{\underlinei\,\underlinei\underlinej\underlinej}(\mathbf{0},\br,\br)\geq 0,\quad
\frac{\partial}{\partial r_{\underlinel}}\frac{\partial}{\partial s_{\underlinel}}\WAsy_{\underlinei\,\underlinei\underlinej\underlinej}(\mathbf{0},\br,\bs)\bigg|_{\bs\,=\,\br}\geq 0,
\quad
i\leq j
\label{wiwiwjwj_Nonnegativity_Asymp}
\end{align}
\begin{align}
\l|\QAsy(\br)\r|\leq \QAsy(\mathbf{0}),\quad
 \QAsy(\mathbf{0})\geq 0
\label{CS_Inequality_qq_01_Asymp}
\end{align}
\begin{align}
\Big(\WAsy_{ijk}(\br,\bs)\Big)^2\leq
\min\s\Big(
&
 \WAsy_{\underlinei\,\underlinei}(\mathbf{0})\,\,\WAsy_{\underlinej\underlinej\underlinek\underlinek}(\mathbf{0},\bs-\br,\bs-\br),\ 
 \WAsy_{\underlinej\underlinej}(\mathbf{0})\,\,\WAsy_{\underlinei\,\underlinei\underlinek\underlinek}(\mathbf{0},\bs,\bs),
\notag\\
&\quad
 \WAsy_{\underlinek\underlinek}(\mathbf{0})\,\,\WAsy_{\underlinei\,\underlinei\underlinej\underlinej}(\mathbf{0},\br,\br)\Big),
\quad i\leq j\leq k
\label{CS_Inequality_ps_01_Asymp}
\end{align}
\begin{align}
\Big(\WAsy_{ijkl}(\br,\bs,\bs')\Big)^2\leq 
\min\s\Big(
&
 \WAsy_{\underlinei\underlinei\underlinej\underlinej}(\mathbf{0},\br,\br)\,\,\WAsy_{\underlinek\underlinek\underlinel\underlinel}(\mathbf{0},\bs'-\bs,\bs'-\bs),
\notag\\
&\quad
 \WAsy_{\underlinei\underlinei\underlinek\underlinek}(\mathbf{0},\bs,\bs)\,\,\WAsy_{\underlinej\underlinej\underlinel\underlinel}(\mathbf{0},\bs'-\br,\bs'-\br), 
\notag\\
&\quad
 \WAsy_{\underlinei\underlinei\underlinel\underlinel}(\mathbf{0},\bs',\bs')\,\,\WAsy_{\underlinej\underlinej\underlinek\underlinek}(\mathbf{0},\bs-\br,\bs-\br) 
\Big),
\quad
i\leq j\leq k \leq l
\label{CS_Inequality_wwww_01_Asymp}
\end{align}
\begin{align}
\Big(\QAsy_{i}(\br)\Big)^2\leq \QAsy(\mathbf{0})\,\WAsy_{\underlinei \underlinei}(\mathbf{0}),
\quad
\Big(\QAsy_{ij}(\br,\bs)\Big)^2\leq \QAsy(\mathbf{0})\,\WAsy_{\underlinei \underlinei \underlinej \underlinej}(\mathbf{0},\bs-\br,\bs-\br),
\quad
i\leq j
\label{CS_Inequality_qws_01_Asymp}
\end{align}
\begin{align}
\WAsy_{\underlinei\underlinei\underlinej\underlinej}(\mathbf{0},\br,\br)\geq 
\big(\WAsy_{i j}(\br)\big)^2\,\exp(2 \sigma t),
\quad
i\leq j
\label{wiwj_Deviation_Asymp}
\end{align}
and
\begin{align}
&
 \overline{w_{\underlinei}(\bx)w_{\underlinei}(\bx)  w_{\underlinej},_{\underlinek}(\by)w_{\underlinej},_{\underlinek}(\by)}^{(a)}
\geq\Big(\overline{w_{i}(\bx) w_{j},_k(\by)}^{(a)}\Big)^2\,\exp(2 \sigma t),
\notag\\[4pt]&
 \overline{w_{\underlinei},_{\underlinek}(\bx)w_{\underlinei},_{\underlinek}(\bx)  w_{\underlinej},_{\underlinel}(\by) w_{\underlinej},_{\underlinel}(\by)}^{(a)}
\geq\Big(\overline{w_{i},_k(\bx) w_{j},_l(\by)}^{(a)}\Big)^2\,\exp(2 \sigma t)
\label{wiwj,k_Deviation_Asymp}
\end{align}
It is interesting to notice that the constraints of \eqref{wiwj_Deviation_Asymp} and \eqref{wiwj,k_Deviation_Asymp} demand that
\begin{align}
 \sigma\leq 0
\label{MaxsigmaFromDeviationConstraint}
\end{align}
That is, the asymptotic solution form cannot hold at great time under $\sigma>0$, and if growing, the turbulent energy will tend toward a constant state of $\sigma=0$.

The limits of \eqref{CVL_Limits} through \eqref{CVdelta_Limits} require that
\begin{align}
&
\lim_{k_1\rightarrow 0^-}\frac{\LAsy(\bk)}{k_1}=
\lim_{k_1\rightarrow 0^-}\frac{\tocAsy(\bk,\bl)}{k_1}=
\lim_{k_1\rightarrow 0^-} \frac{\focAsy(\bk,\bl,\bm)}{k_1}
=\lim_{k_1\rightarrow 0^-} \frac{\focAsy(-\bk-\bl,\bl,\bm)}{k_1}
\notag\\ &
=\lim_{k_1\rightarrow 0^-} \frac{\focAsy(-\bk-\bl,\bk,\bm)}{k_1}
=0,\quad k_2\not=0
\label{CVFOC_Limits}
\end{align}

Regarding the maximization of $\K^{\text{hom}}$ in \eqref{ObjectiveFunction_01},
\begin{align}
\K^{\text{hom}}
=\K^{\text{hom}(a)}(\sigma)\,\exp(2\sigma t),
\quad
\K^{\text{hom}(a)}(\sigma):=\int_{\mathbb{R}^2} \frac{|\bk|^2}{(k_2)^2}\, \socAsy(\bk)\,d\bk
\label{ObjectiveFunction_01_Asymp}
\end{align}
we will maximize $\K^{\text{hom}(a)}(\sigma)$ under a specific value of $\sigma$. Unlike the case of the transient solution in which we start from the prescribed initial condition of $\{\soc_0(\bk), \toc_0(\bk,\bl), \foc_0(\bk,\bl,\bm)\}$ and find the associated $\sigma$ as part of the optimal control solution, the initial condition of the asymptotic state solution here is yet to be solved from the optimal control, and we need to provide the value of $\sigma$ explicitly. 

We will set the bounds of $\focAsy(\bk,\bl,\bm)$ according to 
\begin{align}
|\focAsy(\bk,\bl,\bm)|\leq C
 \label{focBounds}
\end{align}
where $C$ is a positive constant. This constraint is required, since the mathematical structures of
  \eqref{HST_CLMInPhysicalSpace_w1w1_fs_Asymp_Sol}, \eqref{HST_CLMInPhysicalSpace_w1w1w1_fs_Asymp_Sol} and
   \eqref{HST_DivergenceFreeInPhysicalSpace_wiwjwkwl_fs_Asymp} through \eqref{ObjectiveFunction_01_Asymp}
  allow arbitrary linear scaling. This practice is consistent with the form of \eqref{AsymptoticForm_01}
   due to the lacking of the specific initial instant. Though the optimal control of the asymptotic state
   solution does not yield absolute values for $\socAsy(\bk)$, $\tocAsy(\bk,\bl)$
   and $\focAsy(\bk,\bl,\bm)$, it gives us the normalized distributions of 
    $\overline{w_i(\bx) w_j(\by)}/\overline{w_k(\mathbf{0}) w_k(\mathbf{0})}$, etc.

It follows that, under an adequate discretization of $\focAsy(\bk,\bl,\bm)$, we have a quadratically constrained programming problem: a linear objective function with linear constraints of equality and inequality and quadratic constraints of inequality.

Regarding the values of $\sigma$ in \eqref{HST_CLMInPhysicalSpace_w1w1_fs_Asymp_Sol},  \eqref{HST_CLMInPhysicalSpace_w1w1w1_fs_Asymp_Sol} and \eqref{AsymptoticForm_01}, we now focus on
\begin{align}
\sigma\geq 0
\label{PositivityOfsigma}
\end{align}
while aware of \eqref{MaxsigmaFromDeviationConstraint}.
 The inclusion of the positive values is mainly motivated mathematically,
 which is allowed in a model without the presence of the fourth order correlations to be discussed
 and which provides us better understanding about the multi-scale structure of the turbulent flow.
 The restriction to the non-negative is due to the analytical and computational simplicity it brings, as to become clear below.
   We notice that the structures of \eqref{HST_CLMInPhysicalSpace_w1w1_fs_Asymp_Sol} and \eqref{HST_CLMInPhysicalSpace_w1w1w1_fs_Asymp_Sol} display a critical difference between $\sigma\geq 0$ and $\sigma<0$;
 the exponential functions are unbounded under $\sigma<0$ in certain subdomains containing $k_1=0$ or $l_1=0$,
 which can be demonstrated by taking
\begin{align}
\sigma=\sigma_0<0,\quad
k_2=\delta_0 \in \Big(0,\sqrt{|\sigma_0|}\Big],\quad
k'_2=-\,a\,\l[\min(0.5,\delta_0)\r]^n,\quad
a\in(0,1],\quad
n\geq 1
\label{SpecialChoiceUnderNegsigma}
\end{align}
It results in, under $k_1<0$,
\begin{align*}
H(\sigma_0,\bk')- H(\sigma_0,\bk)>\frac{2\, \delta_0}{k_1} \bigg((k_1)^2-\frac{1}{3}\,(\delta_0)^2\bigg)
\end{align*}
and
\begin{align*}
\lim_{k_1\rightarrow 0^-}\Big[ H(\sigma_0,\bkp)- H(\sigma_0,\bk)\Big]
\geq \frac{2\, (\delta_0)^3}{3}\,\lim_{|k_1|\rightarrow 0^+}\frac{1}{|k_1|}=+\infty
\end{align*}
This unbounded $H(\sigma_0,\bk')-H(\sigma_0,\bk)$ results in an unbounded integrand of $\int^{k_2}_{-\infty} d k'_2$ in \eqref{HST_CLMInPhysicalSpace_w1w1_fs_Asymp_Sol}, if $\tocAsy(\bk^{\prime},\bl)$ does not have such an adequate decrease in the above limit in order to counter or control the unbounded increase of the exponential function part. 
 It then follows that $|\tocAsy(\bk',\bl)|$ should be adequately small in a neighborhood of $k_1=0$ and in a neighborhood of $l_1=0$ so as to guarantee the boundedness of $\tWAsy_{ij}(\bk)$ and $\WAsy_{ij}(\br)$.
   The issue should be looked into further from the perspective of all allowable values of $\sigma$, especially regarding the supports of $\tocAsy(\bk',\bl)$ affected by $\sigma$.

\subsection{Linearization and Convex Constraints}
\ \ \ \
The nonlinear constraints such as \eqref{CS_Inequality_ps_01_Asymp} through \eqref{wiwj_Deviation_Asymp} (under $\sigma=0$) apparently pose a challenge to the determination of the solutions.
 However,  \eqref{CS_Inequality_ps_01_Asymp} through \eqref{CS_Inequality_qws_01_Asymp} are of quadratic structures in the form of
\begin{align}
 (c)^2\leq a\,b,\quad
 a=A_j\,\delta_j\geq 0,\ b=B_j\,\delta_j\geq 0,\ c=C_j\,\delta_j
\label{QuadraticFormPeculiar}
\end{align}
Here, within the context of numerical simulation, we have supposedly adopted certain scheme to discretize the distribution of $\focAsy(\bk,\bl,\bm)$ with $(\delta_i)=\{\delta_i\}$ to denote its representative values, say at a finite collection of nodes. 
 The coefficients of $A_i$, $B_i$ and $C_i$ are constant for each inequality of specifically chosen $\br$ and $\bs$.
 It seems rather difficult to check directly whether the above constraint is convex or not with respect to $(\delta_i)$. We will show indirectly the constraint is convex through the linearization scheme to be discussed below.

The constraint of \eqref{QuadraticFormPeculiar} can be linearized through
\begin{align}
  2\,|c|
\leq 2\,\sqrt{a}\,\sqrt{b}
\leq \frac{a}{d}+d\,b,
\quad \forall d>0
\label{QuadraticFormLinearization_a}
\end{align}
or
\begin{align}
-\l(\frac{a}{d}+d\,b\r)
\leq
  2\,c
\leq \frac{a}{d}+d\,b,
\quad \forall d>0
\label{QuadraticFormLinearization}
\end{align}
Both \eqref{QuadraticFormPeculiar} and \eqref{QuadraticFormLinearization} are equivalent, which can be simply seen as follows. Consider an arbitrary point of $(a_0, b_0)$ on the surface of $2\sqrt{a}\,\sqrt{b}$.
 Then, there is $d_0=\sqrt{a_0/b_0}$ such that $a_0/d_0+d_0\, b_0$ $=$ $2\,\sqrt{a_0}\,\sqrt{b_0}$.
 That is, the envelope of the linear constraints is the quadratic constraint. 

To argue for the convexity of \eqref{QuadraticFormPeculiar}, let us fix $A_i$, $B_i$ and $C_i$ and introduce a finite discrete domain for $d$,
\begin{align}
 {\cal V}_n=
\l\{
d_k:\ d_k\in \frac{1}{2^n}\,\mathbb{N},\ d_k\leq 2^n
\r\},\ \ \
\forall n\in\mathbb{N}
\end{align}
It is easy to check that
\begin{align}
 {\cal V}_n\subset {\cal V}_{n+1},\ \ \
\forall n\in\mathbb{N}
\end{align}
Furthermore, we have
\begin{align}
 \l\{(\delta_i):\ |c|\leq \sqrt{a}\,\sqrt{b}\r\}
=
 \bigcap_{n\,\in\,\mathbb{N}}
\,\bigcap_{d\,\in\,{\cal V}_n}
\l\{(\delta_i):\ -\l(\frac{a}{d}+d\,b\r)\leq 2\,c\leq \frac{a}{d}+d\,b
\r\}
\label{QuadraticConvex}
\end{align}
whose proof is sketched below. 
\begin{enumerate}
 \item
\eqref{QuadraticFormLinearization_a} implies that
\begin{align*}
 \l\{(\delta_i):\ |c|\leq \sqrt{a}\,\sqrt{b}\r\}
\subseteq
\l\{(\delta_i):\ -\l(\frac{a}{d}+d\,b\r)\leq 2\,c\leq \frac{a}{d}+d\,b
\r\},\ \ \
\forall d\in{\cal V}_n,\ \forall n\in\mathbb{N}
\end{align*}

\item
Fix $(\delta^0_i)$, $a^0=A_i\,\delta^0_i>0$, $b^0=B_i\,\delta^0_i>0$, $c^0=C_i\,\delta^0_i$, satisfying
\begin{align*}
-\l(\frac{a^0}{d}+d\,b^0\r)\leq 2\,c^0\leq \frac{a^0}{d}+d\,b^0,\ \ \
\forall d\in{\cal V}_n,\ \forall n\in\mathbb{N}
\end{align*}
Then, $d^0=\sqrt{a^0/b^0}>0$. $\exists N\in\mathbb{N}$ s.t. $d^0\leq 2^{N-1}$; and
\begin{align*}
 \forall n\geq N,\ \exists K_n\in\mathbb{N}\ \text{s.t.}\ 
d^0\in\l(\frac{K_n}{2^n},\frac{K_n+1}{2^n}\r],\ 
\frac{K_n}{2^n}\in {\cal V}_n,\ \frac{K_n+1}{2^n}\in{\cal V}_n
\end{align*}
It follows that
\begin{align*}
 2 \l|c^0\r|\leq \frac{a^0}{{K_n}/{2^n}}+\frac{K_n}{2^n}\,b^0,\ \ \
 \l|d^0-\frac{K_n}{2^n}\r|\leq\frac{1}{2^n},
\ \ \forall n\geq N
\end{align*}
which implies that
\begin{align*}
 \l|c^0\r|\leq \frac{1}{2}\l(\frac{a^0}{d^0}+d^0\,b^0\r)=\sqrt{a^0}\,\sqrt{b^0}
\end{align*}
\end{enumerate}
Equation \eqref{QuadraticConvex} says that the quadratic constraint of \eqref{QuadraticFormPeculiar} is convex.

In the case of \eqref{wiwj_Deviation_Asymp} (under $\sigma=0$), the constraint can be linearized in the form of
\begin{align}
\WAsy_{\underlinei\underlinei\underlinej\underlinej}(\mathbf{0},\br,\br)\geq 
\big(\WAsy_{ij}(\br)\big)^2
\geq d\,\big(2\,\WAsy_{ij}(\br)-d\big),\  \forall d\in\mathbb{R};\ \
  \lim_{d\rightarrow \,\WAsy_{ij}(\br)} d\,\big(2\,\WAsy_{ij}(\br)-d\big)=\big(\WAsy_{ij}(\br)\big)^2
\label{wiwj_Deviation_Linearization}
\end{align}
An argument like the above can be used to show the convexity of the constraint.
 Similarly, one can demonstrate the convexity of \eqref{wiwj,k_Deviation_Asymp} (under $\sigma=0$).

We have shown that the present nonlinear programming problem is a convex programming problem.
 Moreover, it is easy to infer from the above arguments that the corresponding constraints of time-dependence are convex,
 and in general, the constraints derived from the Cauchy-Schwarz inequality or the like are also convex.
 Therefore, we have a convex optimization problem here
 and we can resort to convex analysis and algorithms to analyze and solve the problem \cite{BoydVandenberghe2009}.

We can also linearize each and every quadratic constraints of \eqref{CS_Inequality_ps_01_Asymp}
 through \eqref{CS_Inequality_qws_01_Asymp} and \eqref{wiwj_Deviation_Asymp} (under $\sigma=0$)
 with $d$'s from finite discrete sets of ${\cal V}_d=\{d_n\}$ and convert the nonlinear programming
 to a corresponding linear programming problem.
 We can then resort to linear programming to solve the linearized optimization problem.

\subsection{Transformed Equations}
\ \ \ \
For the sake of convenience, we transform \eqref{HST_CLMInPhysicalSpace_w1w1_fs_Asymp_Sol} and \eqref{HST_CLMInPhysicalSpace_w1w1w1_fs_Asymp_Sol} by adopting \eqref{SingularityRemovalTransform},
\begin{align}
&
\socAsy(\bk)=(k_2)^2\,\dsocAsy(\bk),\quad
\tocAsy(\bk,\bl)=k_2\,l_2\,(k_2+l_2)\,\dtocAsy(\bk,\bl),
\notag\\[4pt]
&
\focAsy(\bk,\bl,\bm)=k_2\,l_2\,m_2\,(k_2+l_2+m_2)\,\dfocAsy(\bk,\bl,\bm)
\label{SingularityRemovalTransform_Asymp}
\end{align}
with
\begin{align}
&
\dsocAsy(\bk)=\dsocAsy(-\bk),\quad
\dtocAsy(\bk,\bl)=\dtocAsy(\bl,\bk)=\dtocAsy(-\bk-\bl,\bl)
=\dtocAsy(-\bk,-\bl),
\notag\\[4pt]
&
\dfocAsy(\bk,\bl,\bm)=\dfocAsy(-\bk,-\bl,-\bm)
=\dfocAsy(\bk,\bm,\bl)=\dfocAsy(\bm,\bl,\bk)=\dfocAsy(\bl,\bk,\bm)
\notag\\[4pt]
&
=\dfocAsy(-\bk-\bl-\bm,\bl,\bm)
=\dfocAsy(-\bk-\bl-\bm,\bk,\bm)
=\dfocAsy(-\bk-\bl-\bm,\bk,\bl)
\label{HST_DivergenceFreeInPhysicalSpace_wiwjwkwl_fs_Transf_Asymp}
\end{align}
      Consequently, we obtain
\begin{align}
\gsocAsy(\bk)
=&\,
 -\int^{k_2}_{-\infty}\s d k'_2 \,\frac{2\,|\bk'|^2}{k_1\,|\bk|^4}\, 
   \exp\s\bigg\{\frac{2(k_2-k'_2)}{k_1}\, \bigg[\sigma+(k_1)^2+\frac{1}{6}\,\Big(\big(k_2+k^{\prime}_2\big)^2+(k_2)^2+\big(k^{\prime}_2\big)^2 \Big) \bigg] \bigg\}\s
\notag\\[4pt] &\hskip 15mm \times\s
\int_{\mathbb{R}^2}\s d\bl\, |\bl|^2\,\big(k_1\,l_2-k'_2\,l_1\big)\, \gtocAsy(\bk',\bl)
\label{HST_CLMInPhysicalSpace_w1w1_fs_Transf_Asymp_Sol}
\end{align}
where $k_1<0$, $\bkp=(k_1, k'_2)$, and 
\begin{align}
\gtocAsy(\bk,\bl)  
=\,&
 -\int^{k_2}_{-\infty} d k''_2  \,\frac{|\bk''|^2 |\bl''|^2 |\bk''+\bl''|^2}{k_1 |\bk|^2 |\bl|^2 |\bk+\bl|^2}\,
                  \exp\s\Big[\Sigma(\sigma,\bk''+\bl'',\bk'',\bl'';\bk+\bl,\bk,\bl)\Big]
\notag\\
&\hskip 15mm
\times\s\bigg[
 \frac{(k_1+l_1)(k''_2+l''_2)}{|\bk''+\bl''|^2}\,\gfocIAsy(\bk'',\bl'')
+ \bigg(1 -\frac{2(k_1+l_1)^2}{|\bk''+\bl''|^2} \bigg)\,
           \gfocIIAsy(\bk'',\bl'') 
\notag\\ &\hskip 22mm
+\frac{k_1 k''_2}{|\bk''|^2}\,
           \gfocIAsy(-\bk''-\bl'',\bl'')  
+\bigg(1-\frac{2(k_1)^2}{|\bk''|^2}\bigg)\,
           \gfocIIAsy(-\bk''-\bl'',\bl'')  
\notag\\
&\hskip 22mm
+\frac{l_1 l''_2}{|\bl''|^2} \,
           \gfocIAsy(-\bk''-\bl'',\bk'') 
+\bigg(1-\frac{2(l_1)^2}{|\bl''|^2}\bigg) 
           \gfocIIAsy(-\bk''-\bl'',\bk'') 
\bigg]     
\label{HST_CLMInPhysicalSpace_w1w1w1_fs_Transf_Asymp_Sol}
\end{align}
where
\begin{align*}
k_1 <0, \quad
\bk''=(k_1, k''_2), \quad
\bl''=(l_1, l''_2), \quad
l''_2=l_2-l_1\,\frac{k_2-k''_2}{k_1}
\end{align*}
\begin{align*}
&
\Sigma(\sigma,\bk''+\bl'',\bk'',\bl'';\bk+\bl,\bk,\bl)
\notag\\
=&\,
\frac{k_2-k''_2}{k_1}\, \bigg[
 \sigma
+(k_1+l_1)^2+(k_1)^2+(l_1)^2
\notag\\
&\hskip 17mm
+\frac{1}{6}\Big(\big(k_2+l_2+k''_2+l''_2\big)^2
                  +(k_2+l_2)^2
                  +\big(k''_2+l''_2\big)^2
\notag\\
&\hskip 28mm
                  +\big(k_2+k''_2\big)^2 
                  +(k_2)^2
                  +\big(k''_2\big)^2 
                  +\big(l_2+l''_2\big)^2
                  +(l_2)^2+\big(l''_2\big)^2
            \Big) \bigg]
\end{align*}
\begin{align*}
\gfocIAsy(\bk,\bl)=\int_{\mathbb{R}^2}\s \bigg(1-\frac{m_1+k_1+l_1}{m_2+k_2+l_2}\,\frac{m_1}{m_2}\bigg)(k_2+l_2+m_2)\,m_2\,\dfocAsy(\bk,\bl,\bm)\,d\bm
\end{align*}
\begin{align*}
\gfocIIAsy(\bk,\bl)=- \int_{\mathbb{R}^2}\frac{m_1}{m_2} \, (k_2+l_2+m_2)\,m_2\,\dfocAsy(\bk,\bl,\bm)\,d\bm
\end{align*}

The limiting constraints of \eqref{CVL_Limitsk2zero_Reduced} and \eqref{CVL_Limitsk2k1_Reduced} should be extended to $\dfocAsy(\bk,\bl,\bm)$ and implemented.

The objective function of \eqref{ObjectiveFunction_01_Asymp} is transformed into
\begin{align}
\K^{\text{hom}(a)}(\sigma)=2\int_{-\infty}^0\s dk_1\s \int_{\mathbb{R}}\s dk_2\, |\bk|^2\, \dsocAsy(\bk) \end{align}
which is to be maximized under \eqref{HST_DivergenceFreeInPhysicalSpace_wiwjwkwl_fs_Transf_Asymp}
   through  \eqref{HST_CLMInPhysicalSpace_w1w1w1_fs_Transf_Asymp_Sol},  
    all the constraints of  \eqref{HST_DivergenceFreeInPhysicalSpace_wiwjwkwl_fs_Asymp}
    through \eqref{CVFOC_Limits}
   and the extended \eqref{CVL_Limitsk2zero_Reduced} and \eqref{CVL_Limitsk2k1_Reduced}, with $\dfocAsy(\bk,\bl,\bm)$ as the control variable.

\subsection{Reduced Model up to Third Order Correlation}
\ \ \ \
The task now is to determine $\dfocAsy(\bk,\bl,\bm)$ through optimization. 
 We have mentioned that this is essentially a convex programming problem. For the sake of simplicity and exploration, we consider first the reduced model consisting only of the averaged velocity, the averaged pressure, the second and the third order correlations, by simply dropping all the relations involving the fourth order correlation; The third order correlation then becomes the control variable in the reduced optimal control or optimization problem. Such a reduced formulation has a much simpler mathematical structure of linear programming, which offers an advantage for a more detailed analysis and understanding of the issues involved in the optimal control theory of turbulence modeling. 
   The reduced model, however, cannot resolve the pressure fluctuation correlation and cannot guarantee \eqref{MaxsigmaFromDeviationConstraint}. 

For the asymptotic state solutions within the reduced formulation, we now have
\begin{align}
&
\gsocAsy(\bk)=
\notag\\
\,&
 -\int^{k_2}_{-\infty}\s d k'_2 \,\frac{2\,|\bk'|^2}{k_1\,|\bk|^4}\, 
   \exp\s\bigg\{\frac{2(k_2-k'_2)}{k_1}\, \bigg[\sigma+(k_1)^2+\frac{1}{6}\,\Big(\big(k_2+k^{\prime}_2\big)^2+(k_2)^2+\big(k^{\prime}_2\big)^2 \Big) \bigg] \bigg\}\,
       \LAsy(\bk')
\label{HST_CLMInPhysicalSpace_w1w1_fs_Transf_Asymp_Sol_Reduced}
\end{align}
where $k_1<0$, $\bk'=(k_1, k'_2)$,
\begin{align}
\LAsy(\bk')=\int_{\mathbb{R}^2}\s d\bl\, |\bl|^2\,(k_1\,l_2-k'_2\,l_1)\,\gtocAsy(\bk',\bl)
\label{DefiningNkp}
\end{align} 
The constraints of equality and inequality are
\begin{align}
 \dAtocAsy(\bk,\bl)
=\dAtocAsy(\bl,\bk)
=\dAtocAsy(-\bk-\bl,\bl)
=\dAtocAsy(-\bk,-\bl)
\label{HST_DivergenceFreeInPhysicalSpace_wiwjwk_fs_Transf_Asymp}
\end{align}
\begin{align}
\dsocAsy(\bk)\geq 0
\label{NonNegativityofSoc11_Reduced}
\end{align}
\begin{align}
\l|\dAtocAsy(\bk,\bl)\r|\leq C
\label{tocBounds}
\end{align}
\begin{align}
\lim_{k_1\rightarrow 0^-}\frac{\LAsy(\bk)}{k_1}=
\lim_{k_1\rightarrow 0^-}\frac{\dAtocAsy(\bk,\bl)}{k_1}
=0,\quad k_2\not=0
\label{CVLgamma_Limits_Reduced}
\end{align}
The quantity of $\gtocAsy(\bk',\bl)$ in \eqref{DefiningNkp} may be replaced with $(\gtocAsy(\bk',\bl)-\gtocAsy(\bk',-\bl))/2$ due to the antisymmetry of $|\bl|^2\,(k_1\,l_2-k'_2\,l_1)$ under $\bl \rightarrow -\bl$.
 Consequently, the reduced model will yield at most an estimate to $(\gtocAsy(\bk',\bl)-\gtocAsy(\bk',-\bl))$.
 $\gtocAsy(\bk',\bl)$ is employed here since its model has the advantage of straight-forwardness in its capacity to satisfy the limits of \eqref{CVLgamma_Limits_Reduced} and the ones established below.
 The imposition of \eqref{tocBounds} has the justification similar to that underlying \eqref{focBounds}. Additional limit constraints are to be derived in the next subsection.
 The objective is
\begin{align}
\K^{\text{hom}(a)}(\sigma)=2\int_{-\infty}^0\s dk_1\s \int_{\mathbb{R}}\s dk_2\, |\bk|^2\, \dsocAsy(\bk)\ \ \text{to be maximized under fixed $\sigma\geq 0$}
\label{ObjectiveFunction_01_Asymp_Reduced}
\end{align}

 Equations \eqref{HST_CLMInPhysicalSpace_w1w1_fs_Transf_Asymp_Sol_Reduced} through \eqref{ObjectiveFunction_01_Asymp_Reduced} may be simplified further by taking $\LAsy(\bk)$ as the control variable satisfying
\begin{align}
 \LAsy(-\bk)=\LAsy(\bk),\quad
 \LAsy(0,k_2)=0,\quad
\int_{-\infty}^0\s dk_1\, \int_{\mathbb{R}}\s dk_2\,\LAsy(\bk)=0,\quad
\l|\LAsy(\bk)\r|\leq C
\label{N_ControlVariable}
\end{align}
where we have employed $\LAsy(k_1,k_2)=\LAsy((k_1,k_2))$. This is the simplest model within the present framework.
  One advantage of  $\LAsy(\bk)$ as the control variable
 is the reduction of the wave number space dimensions involved in the optimization procedure.
 However, it does not provide any detailed information about the third order correlations $\tW^{(I)}_{ijk}(\bk,\bl)$.
 We will study both cases of $\LAsy(\bk)$ and $\dAtocAsy(\bk,\bl)$
 as control variables in order to understand more about the modeling issue.

\subsubsection{Consequence of \eqref{IntrinsicRelationAsymptoticStateForsigma}}
\ \ \ \
The intrinsic constraint of \eqref{IntrinsicRelationAsymptoticStateForsigma} can be recast in the form of
\begin{align}
\sigma 
=\,&\frac{1}{\WAsy_{kk}(\mathbf{0})}\,\bigg(\big(-\WAsy_{12}(\mathbf{0})\big)-\overline{\frac{\partial \w_k(\bx)}{\partial x_j}\,\frac{\partial \w_k(\bx)}{\partial x_j}}^{(a)}\,\bigg)
\label{IntrinsicRelationForsigma}
\end{align}
\begin{enumerate}
 \item 
Under $\sigma\geq 0$, the relation requires that 
\begin{align}
-\WAsy_{12}(\bo)>\sigma\,\WAsy_{kk}(\bo)\geq 0
\label{AnisotropicTensor12_Negative} 
\end{align}
or
\begin{align}
\int_{-\infty}^0\s dk_1\, \int_{\mathbb{R}} dk_2\, |k_1|\,k_2\,\gsocAsy(\bk)
<-\sigma\,\WAsy_{kk}(\bo)\leq 0
\end{align}
implying that $\gsocAsy(\bk)$ has its predominant values in the region of $k_2<0$.
 In the case of $\sigma=0$,
\begin{align}
 \int_{-\infty}^0dk_1 \int_{\mathbb{R}}dk'_2\,\Big(|k_1|\,k'_2+|\bk'|^4\Big)\,\dsocAsy(\bk')=0
\label{IntrinsicRelationForsigma=0}
\end{align}
which implies the necessity for $\dsocAsy(\bk')$ to be significantly positive under $k'_2<0$.

\item
The positive semi-definiteness of $\WAsy_{ij}(\mathbf{0})$ gives
\begin{align}
 -\WAsy_{12}(\bo) 
\leq \sqrt{\WAsy_{11}(\bo)}\,\sqrt{\WAsy_{22}(\bo)}
\leq \frac{1}{2}\l(\WAsy_{11}(\bo)+\WAsy_{22}(\bo)\r)
\label{InequalityFromWijPSD}
\end{align}
and thus, \eqref{IntrinsicRelationForsigma} becomes
\begin{align}
\sigma 
\leq
 \frac{1}{2}
-\frac{1}{\WAsy_{kk}(\bo)}\overline{\frac{\partial \w_k(\bx)}{\partial x_j}\,\frac{\partial \w_k(\bx)}{\partial x_j}}^{(a)}
\label{IntrinsicUpperBoundForsigma}
\end{align}
implying the existence of an upper bound of
\begin{align}
\sigma \leq\sigma_{\max}\leq\frac{1}{2} 
\label{AbsoluteUpperBoundForsigma}
\end{align}
which is compatible with \eqref{MaxsigmaFromDeviationConstraint}.

Since the constraint of \eqref{wiwj_Deviation_Asymp} is not explicitly present in the reduced model,
 $\sigma_{\max}$ is allowed mathematically to be positive.
 Also, the exploration of such values provides us more insights to the turbulent structures in the wave number space.
 The exact value of $\sigma_{\max}$ allowed in the reduced model is yet to be determined, and it may occur that $\sigma<\sigma_{\max}$ in the possible case of $\sigma_{\max}=0.5$,
 since this possibility implies that the support for $\gsocAsy(\bk)$ may be as small as possible and shrinks to the point
 of $\bk=\mathbf{0}$ under the limit of $\sigma\rightarrow \sigma_{\max}=0.5$.
 If we examine \eqref{InequalityFromWijPSD} in detail, we find out the two inequality signs there can be strict: First, it is expected that $\W^{(a)\prime}_{11}(\bo)>0$ and $\W^{(a)\prime}_{22}(\bo)>0$ in any Cartesian coordinate system rotated with respect to the present one, and thus,
\begin{align*}
 -\WAsy_{12}(\bo)< \sqrt{\WAsy_{11}(\bo)}\,\sqrt{\WAsy_{22}(\bo)}
\end{align*}
Next, the average flow field of \eqref{HST_AverageVelocityField} and the data from the corresponding three-dimensional shear homogeneous turbulence indicate that
\begin{align*}
 \bigg(\sqrt{\WAsy_{11}(\bo)}-\sqrt{\WAsy_{22}(\bo)}\bigg)^2>0
\end{align*}

 The issue regarding the determination of $\sigma_{\max}$ will be discussed in more detail later.

We can also cast \eqref{IntrinsicUpperBoundForsigma} in the form of
\begin{align}
 \int_{-\infty}^0dk_1 \int_{\mathbb{R}}dk'_2\,\Big(1-2\,\sigma-2\,|\bk'|^2\Big)\,|\bk'|^2\,\dsocAsy(\bk')\geq 0
\label{IntrinsicUpperBoundForsigma=0}
\end{align}
which implies that $|\bk'|^2\,\dsocAsy(\bk')$ should be predominant inside the domain of
\begin{align}
 |\bk'|^2<\frac{1-2\,\sigma}{2},\quad
 |k_1|<\sqrt{\frac{1-2\,\sigma}{2}},\quad
 |k'_2|<\sqrt{\frac{1-2\,\sigma}{2}}
\label{SupportEstimateFromsigma=0}
\end{align}
\end{enumerate}

\subsubsection{Existence of Certain Limits}
\ \ \ \
To help model $\LAsy(\bk')$ and $\dAtocAsy(\bk',\bl)$ appropriately, we need to examine their asymptotic behaviors under $k_1\rightarrow 0^-$ and so on. For this purpose, we recast \eqref{HST_CLMInPhysicalSpace_w1w1_fs_Transf_Asymp_Sol_Reduced} in the form of
\begin{align*}
\gsocAsy(\bk)=
 \int^{k_2}_{-\infty}\s dk'_2\,\rho_{\soc}(\bk;k'_2)
\end{align*}
where
\begin{align}
\rho_{\soc}(\bk;k'_2):=
\frac{2\,|\bk'|^2}{|k_1|\,|\bk|^4}\, 
   \exp\s\bigg\{\frac{2(k_2-k'_2)}{k_1}\, \bigg[\sigma+(k_1)^2+\frac{1}{6}\,\Big(\big(k_2+k^{\prime}_2\big)^2+(k_2)^2+\big(k^{\prime}_2\big)^2 \Big) \bigg] \bigg\}\,
       \LAsy(\bk')
\end{align}

Within a similar context in Subsection \ref{TheExistenceofCertainLimits},
   we have discussed the point of $\bk=(0^-,k_2)$ with $k_2\not=0$ and $\bk'\rightarrow\bk$; the results of \eqref{CVL_Limits} through \eqref{CVgamma_Limits} or \eqref{CVFOC_Limits} are converted to those of \eqref{CVLgamma_Limits_Reduced}.

Now, we focus at the point of $\bk=(0^-,0)$.
 To calculate the value of $\rho_{\soc}(\bk;k'_2)$ at $\bk'=\bk=(0^-,0)$, we can take different paths.
 The value supposedly unique  provides us the basis to determine the asymptotic behavior of $\rho_{\soc}(\bk;k'_2)$ 
in a small neighborhood of $\bk'=\bk=(0^-,0)$. In the analysis below, $\dsocAsy(\bk)$, $\LAsy(\bk)$ and $\dAtocAsy(\bk,\bl)$
 are treated as bounded.

First, we take $\bk=(k_1,0)$, $k_1<0$ to evaluate $\rho_{\soc}(\bk;k'_2)$. The requirement of
\begin{align*}
\lim_{k_1\rightarrow 0^-}\lim_{k'_2\rightarrow 0^-}\rho_{\soc}(\bk;k'_2)=
\lim_{k'_2\rightarrow 0^-}\lim_{k_1\rightarrow 0^-}\rho_{\soc}(\bk;k'_2)
\end{align*}
\begin{align}
\lim_{k_1\rightarrow 0^-}\frac{\LAsy(k_1,0)}{(k_1)^3}=0
\label{CVL_Limitsk2zero_Reduced}
\end{align}

Next, we set $\bk=(k_1,k_1)$, $k_1<0$ to obtain 
\begin{align*}
\lim_{k_1\rightarrow 0^-}\rho_{\soc}(\bk;k'_2)=0,\quad
\lim_{k'_2\rightarrow 0^-}\lim_{k_1\rightarrow 0^-}\rho_{\soc}(\bk;k'_2)=0
\end{align*}
and
\begin{align*}
 \lim_{k'_2\rightarrow k_1^-}\rho_{\soc}(\bk;k'_2)=\frac{\LAsy(k_1,k_1)}{|k_1|^3},\quad
 \lim_{k_1\rightarrow 0^-}\lim_{k'_2\rightarrow k_1^-}\rho_{\soc}(\bk;k'_2)
=\lim_{k_1\rightarrow 0^-}\frac{\LAsy(k_1,k_1)}{|k_1|^3}
\end{align*}
The supposed equality of the two limit values gives
\begin{align}
\lim_{k_1\rightarrow 0^-}\frac{\LAsy(k_1,k_1)}{(k_1)^3}=0
\label{CVL_Limitsk2k1_Reduced}
\end{align}

\subsubsection{Effect of the Exponential Function Part}
Along with the zero sum balance \eqref{Integrand_ZeroBalanceHalfPlane}, the exponential function part contained in \eqref{HST_CLMInPhysicalSpace_w1w1_fs_Transf_Asymp_Sol_Reduced} also plays a crucial role for a structure of $\dAtocAsy(\bk,\bl)$ to satisfy \eqref{NonNegativityofSoc11_Reduced}. To appreciate this role, we recast \eqref{HST_CLMInPhysicalSpace_w1w1_fs_Transf_Asymp_Sol_Reduced} in the form of
\begin{align}
\gsocAsy(\bk)
=
\frac{2}{|k_1|\,|\bk|^4}\s\int^{k_2}_{-\infty}\s dk'_2\,M(\bk;k'_2)\,\LAsy(\bk')
\label{HST_CLMInPhysicalSpace_w1w1_fs_Transf_Asymp_Sol_Reduced_Recast}
\end{align}
where $k_1<0$, $\bk'=(k_1, k'_2)$ and
\begin{align}
M(\bk;k'_2)=
 |\bk'|^2 \exp\s\bigg\{-\frac{2(k_2-k'_2)}{|k_1|}\bigg[\sigma+|k_1|^2+\frac{1}{6}\Big((k_2+k'_2)^2+(k_2)^2+(k'_2)^2 \Big)\bigg]\s \bigg\}\geq 0
\label{MVsk2}       
\end{align}
Here, $M(\bk;k'_2)$ contains the exponential time rate of growth $2\sigma$ as a parameter which provides a basis for the existence of an upper bound on the value of $\sigma$ as to be established below.

We can estimate the asymptotic behavior of $\gsocAsy(\bk)$ at large $|k_1|$ or $|k_2|$ by evaluating $M(\bk;k'_2)$ approximately. To this end, we fix two bounds for $k_1$ and $k_2$, respectively, as $k_{1c}>0$ and $k_{2c}>0$, whose approximate values will be estimated below.
\begin{enumerate}
\item
In the case of $k_1\leq -k_{1c}$,
\begin{align*}
\frac{2(k_2-k'_2)}{|k_1|}\bigg[\sigma+|k_1|^2+\frac{1}{6}\Big((k_2+k'_2)^2+(k_2)^2+(k'_2)^2 \Big)\bigg]
\geq 2\,|k_1|\,(k_2-k'_2)
\geq 2\,k_{1c}\,(k_2-k'_2)
\end{align*}
\item
In the case of $k_1\in[-k_{1c},0)$ and $|k_2|\geq k_{2c}$, 
\begin{align*}
&
\frac{2(k_2-k'_2)}{|k_1|}\bigg[\sigma+|k_1|^2+\frac{1}{6}\Big((k_2+k'_2)^2+(k_2)^2+(k'_2)^2 \Big)\bigg]
\notag\\[4pt] 
\geq\,& \frac{2(k_2-k'_2)}{3|k_1|}\Big((k_2)^2+k_2 k'_2+(k'_2)^2 \Big)
\geq \frac{(k_2)^2\,(k_2-k'_2)}{2\,|k_1|}
\geq \frac{(k_{2c})^2\,(k_2-k'_2)}{2\,|k_{1c}|}
\end{align*}
\end{enumerate}
Introduce two specific bounds of, say, $k_{1c}=10$ and $k_{2c}=2\,k_{1c}=20$. It can be seen that, under $k_2-k'_2\geq 1$, $M(\bk;k'_2)$ is negligible wherever either $\{k_1\leq -k_{1c}$, $k_2\in\mathbb{R}\}$ or $\{k_1\in[-k_{1c},0)$, $|k_2|\geq k_{2c}\}$. Therefore, if $\LAsy(\bk')$ does not vary drastically like that of $M(\bk;k_2)/M(\bk;k'_2)$, we can approximate the integral of \eqref{HST_CLMInPhysicalSpace_w1w1_fs_Transf_Asymp_Sol_Reduced_Recast} so as to obtain
\begin{align}
\gsocAsy(\bk)
\approx\,&
 \frac{2\,\LAsy(\bk)}{|k_1|\,|\bk|^2}\s
       \int^{k_2}_{-\infty}\s dk'_2\,
           \exp\s\bigg\{-\frac{2(k_2-k'_2)}{|k_1|}\bigg[\sigma+|k_1|^2+\frac{1}{6}\Big((k_2+k'_2)^2+(k_2)^2+(k'_2)^2 \Big)\bigg]\s \bigg\}
\notag\\[4pt]
              \geq\,& 0
\label{gsocAsy_BehaviorAtLargeks}
\end{align}
Here, $\bk$ is in the exterior or complement of the domain defined by
\begin{align}
{\cal S}:=\l\{\bk:\,k_1\in (-k_{1c},0), \ |k_2|<k_{2c}\r\}
\label{DomainOfNegativeNk}
\end{align}
 Equation \eqref{gsocAsy_BehaviorAtLargeks} together with \eqref{NonNegativityofSoc11_Reduced} and 
  \eqref{Integrand_ZeroBalanceHalfPlane} implies that there is $\LAsy(\bk)<0$ occurring inside and only inside ${\cal S}$, which imposes a constraint on the structure of $\gAtocAsy(\bk,\bl)$ required for non-trivial solutions. 

Next, we discuss the role of the exponential function coefficient $M(\bk;k'_2)$ in restricting the values of $\sigma$ and the support of $\gAtocAsy(\bk,\bl)$. To have a non-trivial solution of $\gsocAsy(\bk)$, there is a certain requirement on the behavior of $M(\bk;k'_2)$ over ${\cal S}$. Specifically, we need to exclude the possibility of
\begin{align}
\frac{\partial M(\bk;k'_2)}{\partial k'_2}\geq 0,\ \forall k_1\in(-k_{1c}, 0),\, \forall k_2\in \mathbb{R},\, \forall k'_2\in (-\infty, k_2] 
\label{SufficientCondition_TrivialSolution}
\end{align}  
This result may be understood on the basis of \eqref{Integrand_ZeroBalanceHalfPlane}, \eqref{NonNegativityofSoc11_Reduced} and \eqref{gsocAsy_BehaviorAtLargeks}; the proof is sketched below. 
\begin{enumerate}
\item
It follows from \eqref{Integrand_ZeroBalanceHalfPlane} and the above remark regarding \eqref{gsocAsy_BehaviorAtLargeks} and \eqref{DomainOfNegativeNk} that there exist $k^0_1 \in (-k_{1c}, 0)$ with 
\begin{align}
\int_{\mathbb{R}}\s dk'_2\,\LAsy(k^0_1,k'_2)\leq 0,\quad
\LAsy(k^0_1,k'_2)\not=0
\label{SufficientCondition_TrivialSolution_Argument00}
\end{align}
It then follows that $\LAsy(k^0_1,k'_2)<0$ in some non-empty open interval of $k'_2$, say $(a, b) \subseteq (-k_{2c}, k_{2c})$.
 We have $\LAsy(k^0_1,k'_2)\geq 0$, if $|k'_2| \geq k_{2c}$, from \eqref{gsocAsy_BehaviorAtLargeks}. 

\item
The open interval can be chosen or enlarged in such a fashion that $\LAsy(k^0_1,k'_2)< 0$, $k'_2 \in (a, b)$; and $\LAsy(k^0_1,a)$ $=$ $\LAsy(k^0_1,b)=0$, due to the continuity of $\LAsy(\bk')$. In general, there may be a number of such non-intersecting intervals denoted as
\begin{align*}
(a_{n}, b_{n}) \subseteq (-k_{2c}, k_{2c}),\ n=1,2,\cdots,n_0
\end{align*} 
in which $\LAsy(k^0_1,k'_2)<0$, and $\LAsy(k^0_1,a_{n})$ $=$ $\LAsy(k^0_1,b_{n})=0$; This number of $n_0$ is taken as finite since $\LAsy(\bk')$ is continuous, supposedly well-behaved and all the open intervals are subsets of $(-k_{2c}, k_{2c})$. We can order the intervals with
 $a_{1}<b_{1}\leq a_{2}<b_{2}\leq \cdots \leq a_{n_0-1}<b_{n_0-1}\leq a_{n_0}<b_{n_0}$. 
  Equation \eqref{SufficientCondition_TrivialSolution_Argument00} can now be recast as
\begin{align*}
 \int_{-\infty}^{b_{n_0}}\s dk'_2\,\LAsy(k^0_1,k'_2)
+\int_{b_{n_0}}^{+\infty}\s dk'_2\,\LAsy(k^0_1,k'_2)
\leq 0
\end{align*}
which gives, along with \eqref{gsocAsy_BehaviorAtLargeks},
\begin{align}
 \int_{-\infty}^{b_{n_0}}\s dk'_2\,\LAsy(k^0_1,k'_2)\leq 0
\label{SufficientCondition_TrivialSolution_Argument02}
\end{align}

\item
We can now show inductively that equation \eqref{SufficientCondition_TrivialSolution_Argument02} is incompatible with \eqref{NonNegativityofSoc11_Reduced}, if the assumed condition of \eqref{SufficientCondition_TrivialSolution} holds.

  The condition of \eqref{SufficientCondition_TrivialSolution} and the specific structure of \eqref{MVsk2} imply that $M(k^0_1,k_2;k'_2)$ is a positive, monotonically increasing function of $k'_2$ at fixed $k_2$ and $\sigma$. Therefore, $\forall n$,
\begin{align*}
\int_{b_{n-1}}^{a_{n}}\s dk'_2\,M(k^0_1,b_{n_0};k'_2)\,|\LAsy(k^0_1,k'_2)| \leq M(k^0_1,b_{n_0};a_{n}) \int_{b_{n-1}}^{a_{n}}\s dk'_2\,|\LAsy(k^0_1,k'_2)|
\end{align*}
\begin{align*}
\int^{b_{n}}_{a_{n}}\s dk'_2\,M(k^0_1,b_{n_0};k'_2)\,|\LAsy(k^0_1,k'_2)| > M(k^0_1,b_{n_0};a_{n}) \int^{b_{n}}_{a_{n}}\s dk'_2\,|\LAsy(k^0_1,k'_2)|
\end{align*}

Consider first the case of $\gsocAsy(k^0_1,b_1)\geq 0$,
\begin{align*}
0 \leq 
&\, \int^{b_{1}}_{-\infty}\s dk'_2\,M(k^0_1,b_{1};k'_2)\,\LAsy(k^0_1,k'_2)
\notag\\[4pt]
=&\,
 \int_{-\infty}^{a_{1}}\s dk'_2\,M(k^0_1,b_{1};k'_2)\,|\LAsy(k^0_1,k'_2)|
-\int_{a_{1}}^{b_{1}}\s dk'_2\,M(k^0_1,b_{1};k'_2)\,|\LAsy(k^0_1,k'_2)|
\notag\\[4pt]
<\,&
 M(k^0_1,b_{1};a_1) \int_{-\infty}^{a_{1}}\s dk'_2\,|\LAsy(k^0_1,k'_2)|
-M(k^0_1,b_{1};a_{1}) \int^{b_{1}}_{a_{1}}\s dk'_2\,|\LAsy(k^0_1,k'_2)|
\notag\\[4pt]
=\,&
 M(k^0_1,b_{1};a_1) \int_{-\infty}^{b_{1}}\s dk'_2\,\LAsy(k^0_1,k'_2)
\end{align*}
That is,
\begin{align}
 \int_{-\infty}^{b_{1}}\s dk'_2\,\LAsy(k^0_1,k'_2) >0
\label{SufficientCondition_TrivialSolution_ArgumentB01}
\end{align}
Next, in the case of $\gsocAsy(k^0_1,b_2)\geq 0$,
\begin{align*}
0 \leq 
&\, \int^{b_{2}}_{-\infty}\s dk'_2\,M(k^0_1,b_{2};k'_2)\,\LAsy(k^0_1,k'_2)
\notag\\[4pt]
=&\,
 \int_{-\infty}^{a_{1}}\s dk'_2\,M(k^0_1,b_{2};k'_2)\,|\LAsy(k^0_1,k'_2)|
-\int_{a_{1}}^{b_{1}}\s dk'_2\,M(k^0_1,b_{2};k'_2)\,|\LAsy(k^0_1,k'_2)|
\notag\\[4pt] &
+\int_{b_{1}}^{a_{2}}\s dk'_2\,M(k^0_1,b_{2};k'_2)\,|\LAsy(k^0_1,k'_2)|
-\int_{a_{2}}^{b_{2}}\s dk'_2\,M(k^0_1,b_{2};k'_2)\,|\LAsy(k^0_1,k'_2)|
\notag\\[4pt]
<\,&
 M(k^0_1,b_{2};a_1) \int_{-\infty}^{a_{1}}\s dk'_2\,|\LAsy(k^0_1,k'_2)|
-M(k^0_1,b_{2};a_1) \int_{a_{1}}^{b_{1}}\s dk'_2\,|\LAsy(k^0_1,k'_2)|
\notag\\[4pt] &
+M(k^0_1,b_{2};a_2) \int_{b_{1}}^{a_{2}}\s dk'_2\,|\LAsy(k^0_1,k'_2)|
-M(k^0_1,b_{2};a_2) \int_{a_{2}}^{b_{2}}\s dk'_2\,|\LAsy(k^0_1,k'_2)|
\notag\\[4pt]
=\,&
 M(k^0_1,b_{2};a_1) \int_{-\infty}^{b_{1}}\s dk'_2\,\LAsy(k^0_1,k'_2)
+M(k^0_1,b_{2};a_2) \int_{b_{1}}^{b_{2}}\s dk'_2\,\LAsy(k^0_1,k'_2)
\end{align*}
Using \eqref{SufficientCondition_TrivialSolution_ArgumentB01} and $M(k^0_1,b_{2};a_1)<M(k^0_1,b_{2};a_2)$, we get
\begin{align}
 \int_{-\infty}^{b_{2}}\s dk'_2\,\LAsy(k^0_1,k'_2) >0
\label{SufficientCondition_TrivialSolution_ArgumentB02}
\end{align}
It follows that, inductively, $\gsocAsy(k^0_1,b_n)\geq 0$, $n=1$, $2$, $\cdots$, $n_0$, results in
\begin{align}
 \int_{-\infty}^{b_{n}}\s dk'_2\,\LAsy(k^0_1,k'_2) >0,\ \ n=1, 2, \cdots, n_0
\label{SufficientCondition_TrivialSolution_ArgumentB0n}
\end{align}
which contradicts \eqref{SufficientCondition_TrivialSolution_Argument02}.
\end{enumerate}

The above result implies that to find non-trivial solutions of $\gsocAsy(\bk)\geq 0$ the support of $\LAsy(\bk')$ needs to contain a subdomain in which the following condition holds,
\begin{align}
\frac{\partial M(\bk;k'_2)}{\partial k'_2}< 0 
\label{NecessaryCondition_NontrivialSolution}
\end{align}  
which can be easily evaluated, with the help of \eqref{MVsk2}, to obtain
\begin{align}
k_1\,k'_2 > |\bk'|^2\,\Big(\sigma+|\bk'|^2\Big), \ \ k_1 \in (-k'_{1c},0), \ \ \sigma \geq 0
\label{NecessaryCondition_NontrivialSolution_Specific}
\end{align}
where the value of $k'_{1c}$ $(<k_{1c})$ is to be fixed.
 A prominent feature of the above relation is the absence of $k_2$.  
   One immediate consequence of \eqref{NecessaryCondition_NontrivialSolution_Specific} is that
\begin{align}
k'_2<0
\label{M_Extreme_Consequence_kp2}
\end{align}
Therefore, equation \eqref{NecessaryCondition_NontrivialSolution_Specific} restricts the arguments $(k_1,k'_2)$ of $M(\bk;k'_2)$ satisfying \eqref{NecessaryCondition_NontrivialSolution} to a region in the third quadrant of the plane coordinate system $(k_1,k'_2)$.
  To extract more information, we exploit the symmetric structure of \eqref{NecessaryCondition_NontrivialSolution_Specific} with respect to $k_1$ and $k'_2$ by introducing
\begin{align}
k_1:=|\bk'|\,\cos\theta,\quad
k'_2:=|\bk'|\,\sin\theta,\ \ \theta\in \Big(\pi, \frac{3\pi}{2}\Big)
\label{M_Extreme_Consequence_Definingtheta}
\end{align}
Substitution of the defined into \eqref{NecessaryCondition_NontrivialSolution_Specific} gives
\begin{align}
\sin(2\,\theta) > 2\l(\sigma+|\bk'|^2\r),\ \ \theta\in \Big(\pi, \frac{3\pi}{2}\Big)
\label{M_Extreme_Consequence_theta}
\end{align}
 This representation can help us visualize the solutions of $k'_2$ from \eqref{NecessaryCondition_NontrivialSolution_Specific} under given $k_1$ and $\sigma$; Its specifics  and certain consequences are discussed below.
\begin{enumerate}
\item
Since $k_1<0$ underlies \eqref{M_Extreme_Consequence_theta}, we have
\begin{align}
2\,\sigma < 2 \l(\sigma+|\bk'|^2\r) < 1
\label{M_Extreme_Consequence_sigmakp}
\end{align}
or
\begin{align}
2\,\sigma < 1,\quad
|\bk'| < \sqrt{\frac{1-2\,\sigma}{2}},\quad
 |k_1|<\sqrt{\frac{1-2\,\sigma}{2}},\quad
 |k'_2|<\sqrt{\frac{1-2\,\sigma}{2}}
\label{M_Extreme_Consequence_sigma_UpperBound}
\end{align}
which is consistent with that of \eqref{AbsoluteUpperBoundForsigma} and \eqref{SupportEstimateFromsigma=0}.
  These results also hold for a formulation involving higher order correlations,
 since we have obtained them based on the generally held \eqref{HST_CLMInPhysicalSpace_w1w1_fs_Transf_Asymp_Sol_Reduced}, \eqref{HST_DivergenceFreeInPhysicalSpace_wiwjwk_fs_Transf_Asymp} and \eqref{NonNegativityofSoc11_Reduced},
    without resorting to any approximations to $\dAtocAsy(\bk',\bl)$.

\item
If $\sigma<0$ is considered, \eqref{M_Extreme_Consequence_theta} (with a modified range of $\theta$) implies that
\begin{align*}
 -\sigma-\frac{1}{2} < |\bk'|^2 < -\sigma+\frac{1}{2},\quad
\max\bigg(0,\ |\sigma|-\frac{1}{2}\bigg) < |\bk'|^2 < |\sigma|+\frac{1}{2}
\end{align*}
 whose consequence is not explored here.

\item
By including $\theta=\pi$ and $\theta=3\pi/2$, equation \eqref{M_Extreme_Consequence_theta} represents a closed domain, denoted as $\MExtremeDomain$, in the third quadrant of the Cartesian coordinate system of $(k_1, k'_2)$.  
 Its boundary $\partial\MExtremeDomain$ is determined by
\begin{align}
\sin(2\,\theta) = 2\l(\sigma+|\bk'|^2\r),\ \ \theta\in\Big[\pi,\frac{3\pi}{2}\Big]
\label{M_Extreme_Consequence_Boundary}
\end{align}
Under given $\sigma$ and $|\bk'|$ satisfying \eqref{M_Extreme_Consequence_sigma_UpperBound}, if $\theta' \in [\pi,5\pi/4]$ is a solution of \eqref{M_Extreme_Consequence_Boundary}, $5\pi/4+(5\pi/4-\theta')$ is a solution too, which can be verified directly. Therefore, $\partial\MExtremeDomain$ is a closed loop in the polar coordinate system of $(|\bk'|,\theta)$ with $\theta\in[\pi,3\pi/2]$, symmetric with respect to the line of $\theta=5\pi/4$, as sketched in Fig. \ref{NGD}. The loop shrinks in the ranges of both $\theta$ and $|\bk'|$, as $\sigma$ increases from zero toward $0.5$. It degenerates to a single point of $|\bk'|=0$, ($\theta=5\pi/4$), at $\sigma=0.5$, following from \eqref{M_Extreme_Consequence_sigma_UpperBound} or \eqref{M_Extreme_Consequence_Boundary};
 Of course, this degeneration will not occur due to the upper bound of \eqref{AbsoluteUpperBoundForsigma}. 
\begin{figure}[t]
\centering
\psfrag{k1}{$k_1$}
\psfrag{k2}{$k'_2$}
\psfrag{0d0}{$\sigma=0$}
\psfrag{0d2}{$\sigma=0.2$}
\psfrag{0d4}{$\sigma=0.4$}
\psfrag{k1min}{$\koneminzero$}
\psfrag{k2p}{$\MPeak$}
\psfrag{k2v}{$\MValley$}
\includegraphics[width = 4.5in]{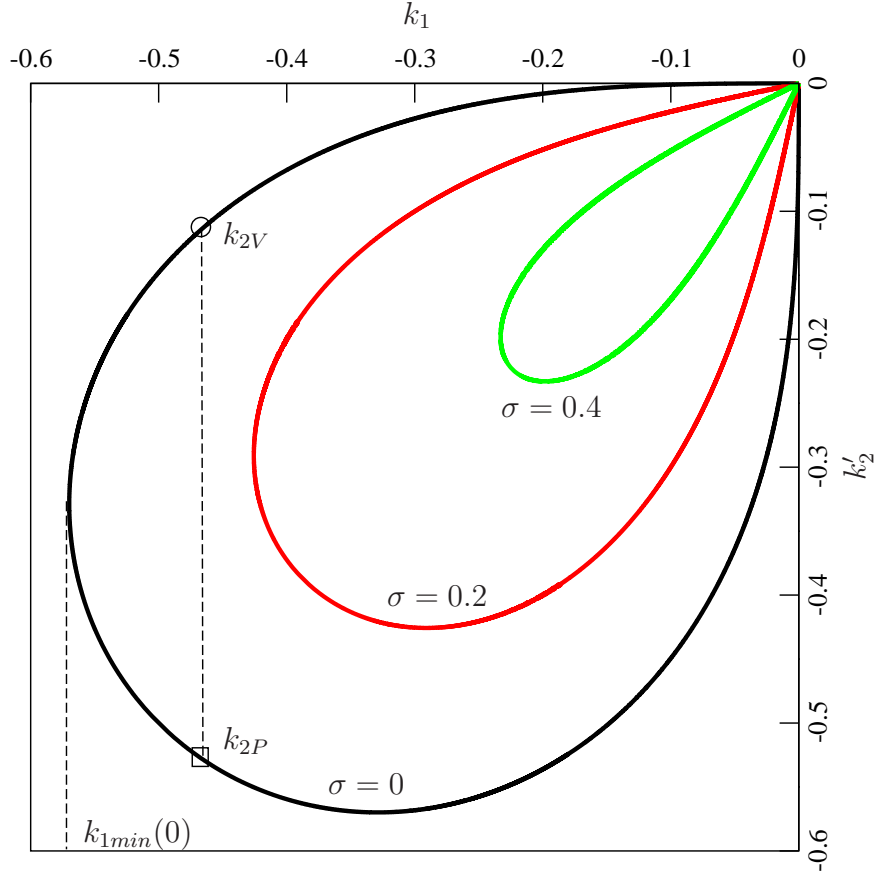}
\vskip -3mm
\caption{The geometry of $\partial\MExtremeDomain$ at various values of $\sigma$. The cases of $\sigma=0.2$, $0.4$ are plotted for the sake of comparison.} 
\label{NGD}
\end{figure}

It is easy to check that
\begin{align}
\frac{\partial M(\bk;k'_2)}{\partial k'_2}\,
\l\{
\begin{array}{ll}
<0, & \bk'\in (\MExtremeDomain)^o;\\[4pt]
=0, & \bk'\in \partial\MExtremeDomain;\\[4pt] 
>0, & \bk'\in (\MExtremeDomain)^c
\end{array}
\r.
\end{align}
Here, $(\MExtremeDomain)^o$ and $(\MExtremeDomain)^c$ denote, respectively, the interior and the compliment of $\MExtremeDomain$.

\item
Fix $\sigma \in [0,\sigma_{\max}]$.
 Fig. \ref{NGD} indicates the existence of the minimum $k_1$ of the loop $\partial\MExtremeDomain$, denoted as $\konemin$, which is negative and can be obtained by taking $\konemin=-|k_1|$ with $|k_1|$ to be solved from
\begin{align}
|k_1|\,|k'_2|=\Big(|k_1|^2+|k'_2|^2\Big)\,\Big(\sigma+|k_1|^2+|k'_2|^2\Big),\quad
\frac{\partial |k_1|}{\partial |k'_2|}=0
\label{k1min}
\end{align}
One can verify that $k'_{1c}=|\konemin|\leq |\koneminzero|<0.56988$.
   The specific $k'_2$ $(<0)$ on the loop as a function of $k_1$ can be solved from (\ref{k1min}$)_1$ under $k_1$ $\in (\konemin,\, 0)$; There are two distinct solutions, denoted by $\MPeak(k_1,\sigma)$ and $\MValley(k_1,\sigma)$, respectively, a special case of which is illustrated in Fig. \ref{MCurve}.
\begin{figure}[t]
\centering
\psfrag{M}{$M(\bk;k'_2)$}
\psfrag{k2}{$k'_2$}
\psfrag{a}{$\MValley'$}
\psfrag{b}{$\MPeak$}
\psfrag{c}{$\MValley$}
\psfrag{d}{$\MPeak'$}
\includegraphics[width = 5in]{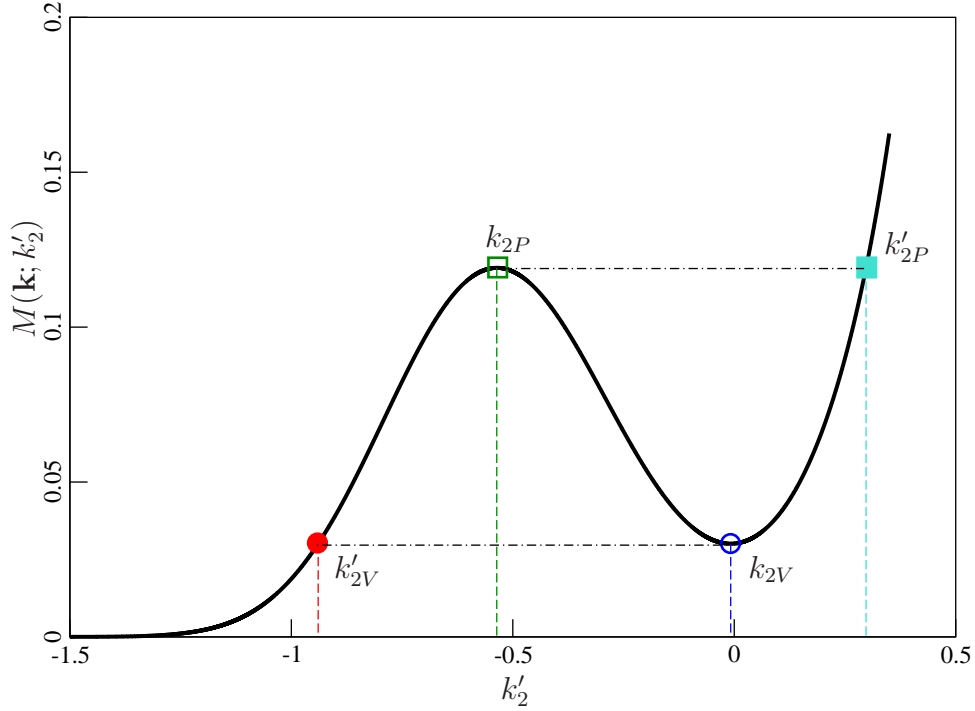}
\vskip -3mm
\caption{The variation of $M(\bk;k'_2)$ versus $k'_2$ under $\sigma=0$ and $\bk=(-0.2,0.35)$; the relevant definitions are indicated.}
\label{MCurve}
\end{figure}
Here, $\MPeak(k_1,\sigma)$ represents the lower branch and $\MValley(k_1,\sigma)$ the upper branch of the loop displayed in Fig. \ref{NGD}, $\konemin<\MPeak<\MValley <0$, and
\begin{align}
\frac{\partial M(\bk;k'_2)}{\partial k'_2}\,
\l\{
\begin{array}{ll}
<0, & k'_2 \in (\MPeak,\, \MValley);\\[4pt]
=0, & k'_2=\MPeak, \, \MValley; \\ [4pt]
>0, & k'_2\in (-\infty,\,\MPeak)\cup(\MValley,\,+\infty)
\end{array}
\r.
\end{align}
$\MExtremeDomain$ and  $(\MPeak,\,\MValley)$ are related through
\begin{align}
(\MExtremeDomain)^o=\cup_{k_1\in(\konemin,\, 0)}\Big\{\{k_1\} \s\times\s (\MPeak,\,\MValley)\Big\},\quad
\sigma\in [0,\sigma_{\max}]
\end{align}

\item
Recall \eqref{SufficientCondition_TrivialSolution_Argument00} that there exist $k^0_1 \in (-k_{1c}, 0)$ such that 
\begin{align*}
\int_{\mathbb{R}}\s dk'_2\,\LAsy(k^0_1,k'_2)\leq 0,\quad
\LAsy(k^0_1,k'_2)\not=0
\end{align*}
Constraint \eqref{NonNegativityofSoc11_Reduced} requires that
\begin{align*}
\int^{k_2}_{-\infty}\s dk'_2\,M((k^0_1,k_2);k'_2)\,\LAsy(k^0_1,k'_2)\geq 0,\ \forall k_2
\end{align*}
The two relations above and the behavior of $M(\bk;k'_2)$ illustrated in Figs. \ref{NGD} and \ref{MCurve} imply that
 $k^0_1 \in (\konemin,0)$ and
  $k'_2$ should lie preferably in a small neighborhood of $\MValley(k^0_1,\sigma)$ when $\LAsy(k^0_1,k'_2)$ is predominantly negative
  so as to satisfy the constraint of \eqref{NonNegativityofSoc11_Reduced}.
  This observation offers us a ground to estimate the support of $\LAsy(\bk')$, $\MaxSupportNbkp$, as follows.
  
As illustrated by the specific curve of Fig. \ref{MCurve}, there is a unique $\MValley'$ such that $\MValley'(k_1,\sigma)<\MPeak(k_1,\sigma)$ and $M(\bk;\MValley')=M(\bk;\MValley)$, i.e.,
\begin{align}
\frac{(k_1)^2+(\MValley')^2}{(k_1)^2+(\MValley)^2}
=\exp\s\bigg[
 \frac{2}{3\,|k_1|}\,\Big(
3\l(\sigma+|k_1|^2\r)(\MValley-\MValley')
+(\MValley)^3
-(\MValley')^3
\Big)\bigg]
\label{LowerBoundForNkPositive}
\end{align}
We may take this $\MValley'(k_1,\sigma)$ as a lower bound for the set $\{k'_2:\, \LAsy(\bk')\not=0\}$ under fixed $k_1\in(\konemin,0)$, considering \eqref{NonNegativityofSoc11_Reduced}, \eqref{ObjectiveFunction_01_Asymp_Reduced} and the behavior of $M(\bk;k'_2)$.
 Furthermore, there is a unique $\MPeak'(k_1,\sigma)>\MValley(k_1,\sigma)$ with $M(\bk;\MPeak')=M(\bk;\MPeak)$ or
\begin{align}
\frac{(k_1)^2+(\MPeak')^2}{(k_1)^2+(\MPeak)^2}
=\exp\s\bigg[
 \frac{2}{3\,|k_1|}\,\Big(
3\l(\sigma+|k_1|^2\r)(\MPeak-\MPeak')
+(\MPeak)^3
-(\MPeak')^3
\Big)\bigg]<1
\label{UpperBoundForNkNegative}
\end{align}
This $\MPeak'(k_1,\sigma)$ may be treated as a upper bound for the set $\{k'_2:\, \LAsy(\bk')\not=0\}$, $k_1\in(\konemin,0)$. 
 
  With the help of $\MValley'$ and $\MPeak'$, we may have an estimate for the support of $\LAsy(\bk')$
\begin{align}
\MaxSupportNbkp =\MaxSupportNbkpNeg \cup \MaxSupportNbkpPos
\label{PossibleSupportOfNkp}
\end{align}
where
\begin{align}
\MaxSupportNbkpNeg:=\cup_{k_1\in[\konemin,\, 0]} \Big\{\{k_1\}\s\times\s\big[\MValley',\,\MPeak'\big]\Big\},\quad
\MaxSupportNbkpPos:=\Big\{-\bk':\ \bk'\in \MaxSupportNbkpNeg\Big\}
\label{PossibleSupportOfNkp_Subdomains}
\end{align}
The boundaries of $\MaxSupportNbkpNegVoid(\sigma)$ with $\sigma=0$, $0.4$ are, respectively, sketched in Fig. \ref{NSupportMax}.
\begin{figure}[t]
\centering
\psfrag{k1}{$k_1$}
\psfrag{k2}{$k'_2$}
\psfrag{NGD0}{$\partial\MExtremeDomainVoid(0)$}
\psfrag{NSNeg0}{$\partial\MaxSupportNbkpNegVoid(0)$}
\psfrag{NGD04}{$\partial\MExtremeDomainVoid(0.4)$}
\psfrag{NSNeg04}{$\partial\MaxSupportNbkpNegVoid(0.4)$}
\includegraphics[width = 5.4in]{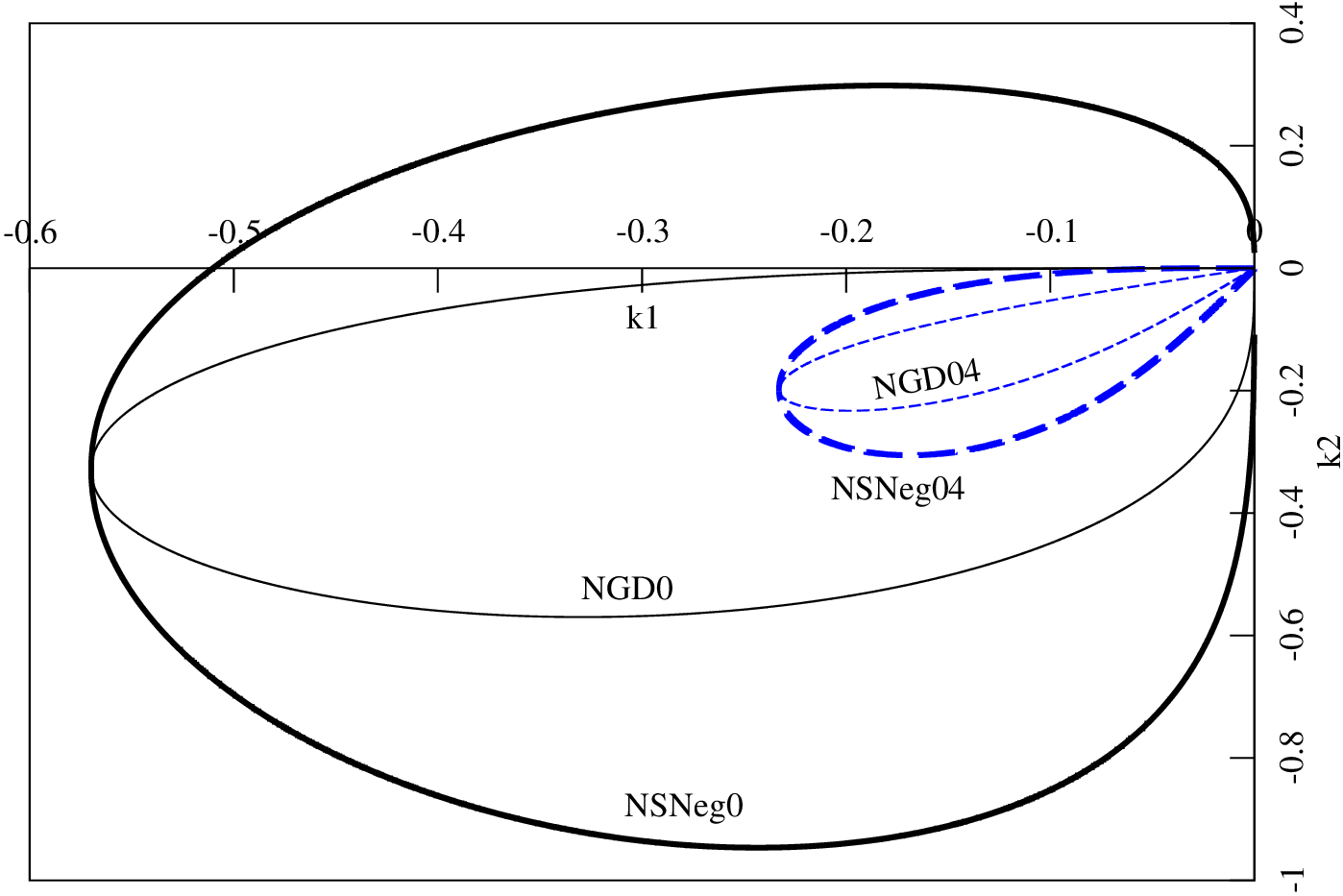}
\vskip -3mm
\caption{The geometries of $\partial\MaxSupportNbkpNegVoid(\sigma)$ with $\sigma=0$, $0.4$;
 the latter is intended for comparison.
 Also sketched are the curves of $\partial\MExtremeDomain$, $\sigma=$ $0$, $0.4$, for the purpose of comparison.}
\label{NSupportMax}
\end{figure}
The above support estimate helps us to fix $\LAsy(\bk')$ numerically if $\LAsy(\bk')$ is taken as the control variable under \eqref{N_ControlVariable}.

\item
With the above estimate of $\MaxSupportNbkpNegVoid(\sigma)$, (\ref{N_ControlVariable}$)_2$ is met automatically.
 We notice that this estimate is obtained by focusing on the variations of $M(\bk;k'_2)$ and $\LAsy(\bk')$ along the axis of $k'_2$ 
 under fixed $k_1 \in (\konemin, 0)$. To be comprehensive in the support estimate, we may also take into account the variation of $M(\bk;k'_2)$ along the $k_1$ direction,
 considering that $M(\bk;k'_2)$ is relatively small in the region where $|k_1|$ is small. For example,
   the predominantly negative values of $\LAsy(\bk')$ should be achieved in an appropriate range
   of $k_1$  and $k'_2$ in a neighborhood of $\MValley(k_1,\sigma)$; it then follows that (\ref{N_ControlVariable}$)_3$
 may be met even by a lower bound of $k_1$ beyond $\konemin$ with $\LAsy(\bk')$ being positive in the associated region of expansion,
 possibly resulting in a more robust numerical computation and larger turbulent energy.
 This expansion beyond $\konemin$ is compatible with \eqref{SupportEstimateFromsigma=0} and \eqref{M_Extreme_Consequence_sigma_UpperBound}.
 We may also enlarge the estimate of 
\eqref{PossibleSupportOfNkp} by taking, say $\big[2 \MValley',\,\max_{k_1}(2 \MPeak', |\MPeak'|)\big]$, the specific choice to be fixed through numerical simulation experiments.
 
\end{enumerate}

\subsubsection{Supports of $\dAtocAsy(\bk,\bl)$ and $\gsocAsy(\bk)$ Estimated}
\ \ \ \
For the convenience of numerical simulations, it is helpful to have estimates about the supports of $\dAtocAsy(\bk,\bl)$ and $\gsocAsy(\bk)$.  

The relationship between $\LAsy(\bk')$ and $\dAtocAsy(\bk',\bl)$ of \eqref{DefiningNkp} and the above analysis indicate the advantage to have
\begin{align}
\big\{\bk':\ \dAtocAsy(\bk',\bl)\not=0\ \text{for some}\ \bl\big\}\subseteq \MaxSupportNbkp
\end{align}
in order to satisfy the constraint of \eqref{NonNegativityofSoc11_Reduced}.
   Moreover, the symmetries of \eqref{HST_DivergenceFreeInPhysicalSpace_wiwjwk_fs_Transf_Asymp} require that $\gAtocAsy(\bk',\bl)=\gAtocAsy(\bl,\bk')$ for the arguments $\bk'$ and $\bl$. Therefore, the support of $\gAtocAsy(\bk',\bl)$, $\MaxSupportGbkpbl$, is taken as a subset of $\MaxSupportNbkp\times\MaxSupportNbkp$,
\begin{align}
\MaxSupportGbkpbl=\big\{(\bk',\bl):\ \gAtocAsy(\bk',\bl)\not=0\big\}
\subseteq\MaxSupportNbkp\times\MaxSupportNbkp
\end{align}
 Considering that we will approximate $\gAtocAsy(\bk',\bl)$ through certain numerical interpolation scheme, we adopt
\begin{align}
\MaxSupportGbkpbl=\MaxSupportNbkp\times\MaxSupportNbkp
\end{align}
in order to have the robustness in the optimization procedure.

\begin{figure}[t]
\centering
\psfrag{k1}{$k_1$}
\psfrag{k2}{$k_2$}
\psfrag{beta0}{$\partial\MaxSupportbetaVoid(0)$}
\psfrag{NSNeg0}{$\partial\MaxSupportNbkpNegVoid(0)$}
\psfrag{beta04}{$\partial\MaxSupportbetaVoid(0.4)$}
\psfrag{NSNeg04}{$\partial\MaxSupportNbkpNegVoid(0.4)$}
\includegraphics[width = 5.5in]{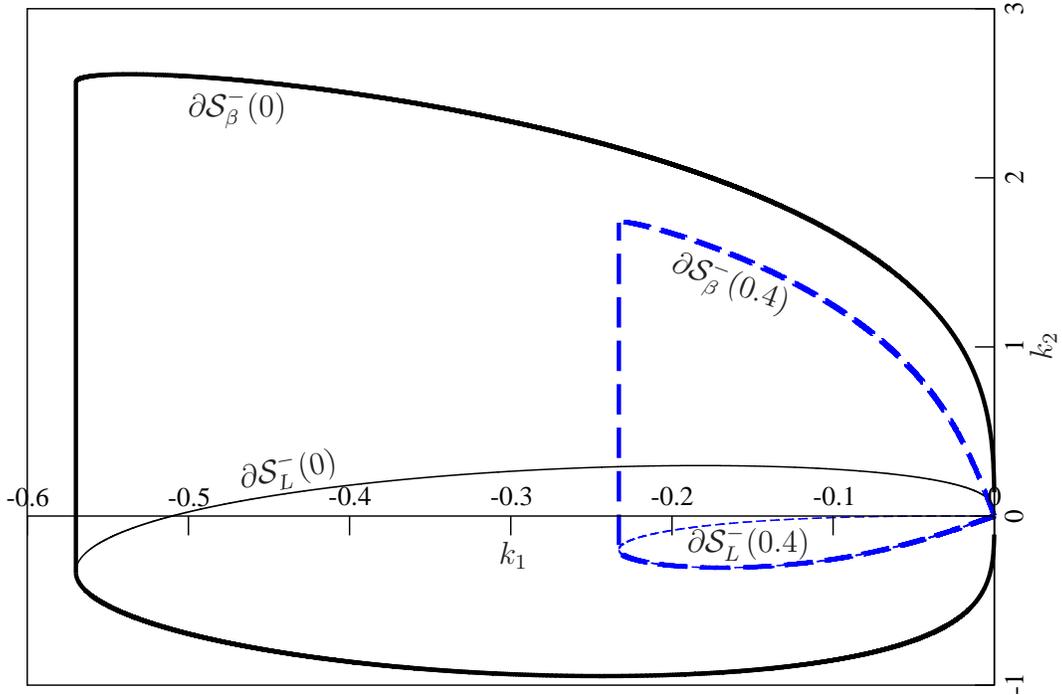}
\caption{The boundaries of the estimated $\MaxSupportbeta$ under $\sigma=0$, $0.4$, respectively;
 the latter is presented for comparison. 
The estimates are compatible with those of
 \eqref{SupportEstimateFromsigma=0} and \eqref{M_Extreme_Consequence_sigma_UpperBound},
 with possible enlargements along the negative direction of $k_1$.
 $\partial\MaxSupportNbkpNegVoid(\sigma)$, $\sigma=$ $0$, $0.4$, are also sketched for the purpose of comparison.
}
\label{betaSupportMax}
\end{figure}

The relationship between $\LAsy(\bk')$ and $\dsocAsy(\bk)$ of \eqref{HST_CLMInPhysicalSpace_w1w1_fs_Transf_Asymp_Sol_Reduced} provides us a ground to estimate the support for $\gsocAsy(\bk)$. Under $k_1\in(\konemin,\, 0)$ and $k_2\geq \MPeak'(k_1,\sigma)$, \eqref{HST_CLMInPhysicalSpace_w1w1_fs_Transf_Asymp_Sol_Reduced} gives
\begin{align}
\gsocAsy(\bk)
\leq &\,
   \frac{|k_1|^4}{|\bk|^4}\,\exp\s\bigg[-\frac{2\,k_2}{|k_1|}\,\bigg(\sigma+(k_1)^2+\frac{1}{3}\,(k_2)^2\bigg)\bigg]\,\gsocAsy(k_1,0)
\notag\\[4pt] &
+\frac{2}{|k_1|\,|\bk|^4}
   \exp\s\bigg[-\frac{2\, (k_2-\MPeak')}{|k_1|}\bigg(\sigma+(k_1)^2+\frac{1}{4}\,(k_2)^2 \bigg) \bigg]
\notag\\[4pt] &\hskip 10mm \times
      \int^{\MPeak'}_{0}\s dk'_2
          \exp\s\bigg[-\frac{2\,(k_2-k'_2)}{3\,|k_1|}\,\Big(k'_2+\frac{1}{2}\,k_2\Big)^2\bigg] 
               \l|\bk'\r|^2\l|\LAsy(\bk')\r| 
\label{gsoc_UpBoundFork2_a}
\end{align}
holding for $\MPeak'(k_1,\sigma)>0$, and
\begin{align}
\gsocAsy(\bk)
\leq &\,
 \frac{2}{|k_1|\,|\bk|^4}
   \exp\s\bigg[-\frac{2\,(k_2-\MPeak')}{|k_1|}\,\bigg(\sigma+(k_1)^2+\frac{1}{4}\,(k_2)^2 \bigg)\bigg]
\notag\\[4pt] &\hskip 10mm \times
  \int^{\MPeak'}_{\MValley'}\s dk'_2 
        \exp\s\bigg[-\frac{2\,(k_2-k'_2)}{3\,|k_1|}\,\Big(k'_2+\frac{1}{2}\,k_2\Big)^2\bigg] 
              \l|\bk'\r|^2 \l|\LAsy(\bk')\r|
\label{gsoc_UpBoundFork2_b}
\end{align}
which holds for $\MPeak'(k_1,\sigma)\leq 0$. We assume below that $\LAsy(\bk')$ has an adequate behavior in the limit of $k_1\rightarrow 0^-$, as those of \eqref{CVLgamma_Limits_Reduced}, \eqref{CVL_Limitsk2zero_Reduced}
  and \eqref{CVL_Limitsk2k1_Reduced},
 so as to ensure the moderate orders of magnitude for the quantities involved besides that of \eqref{TheExpQuantity}.
   Considering that
\begin{align}
|k_1|\leq |\konemin|\leq |\koneminzero|< 0.56988,\quad
\l|\MPeak'(k_1,\sigma)\r|< |\MPeak(k_1,\sigma)|<0.56988
\end{align}
we take $k_2$ in \eqref{gsoc_UpBoundFork2_a} and \eqref{gsoc_UpBoundFork2_b}, say $k_2=$ $3$, $4$, respectively, to get
\begin{align}
\exp\s\bigg[-\frac{2\, (k_2-\MPeak')}{|k_1|}\bigg(\sigma+(k_1)^2+\frac{1}{4}\,(k_2)^2 \bigg) \bigg]
<
 \exp(-21.95),\
 \exp(-52.06),\ \ \MPeak'>0
\label{TheExpQuantity}
\end{align}
and
\begin{align}
\exp\s\bigg[-\frac{2\, (k_2-\MPeak')}{|k_1|}\bigg(\sigma+(k_1)^2+\frac{1}{4}\,(k_2)^2 \bigg) \bigg]
<
 \exp(-27.11),\
 \exp(-60.71),\ \ \MPeak'\leq 0
\end{align}
These inequalities provide us a numerical ground for the estimate of upper bound of $k_2$ for the support of $\gsocAsy(\bk)$.
 For instance, we may take the value of $k_2=3$ as the upper bound. Furthermore, instead of this uniform upper bound of $k_2$, we may find a tighter estimated upper bound $\MaxSupportbetaUB(k_1,\sigma)$ if we take it as a function of $\sigma$ and $k_1$, such as
\begin{align}
\frac{2\, (\MaxSupportbetaUB-\MPeak')}{|k_1|}\bigg(\sigma+(k_1)^2+\frac{1}{4}\,(\MaxSupportbetaUB)^2 \bigg)=20
\label{gsoc_UpBoundFork2}
\end{align}
Therefore, we have a support estimate of $\gsocAsy(\bk)$ under $k_1<0$, 
\begin{align}
\MaxSupportbeta=\cup_{k_1\in[\konemin,\, 0]} \Big\{\{k_1\}\s\times\s\big[\MValley',\,\MaxSupportbetaUB\big]\Big\}
\label{gsoc_EffectiveSupport}
\end{align}
which is sketched in Fig. \ref{betaSupportMax} with $\sigma=0$ and $0.4$, respectively.  
  This estimate will help us to implement the constraint of \eqref{NonNegativityofSoc11_Reduced}.
    We may also estimate $\MaxSupportbeta$ as follows. We take $\MaxSupportDbkpblbm$ $=$ $\MaxSupportNbkp\s\times\s\MaxSupportNbkp\s\times\s\MaxSupportNbkp$ as the support for $\gfocAsy(\bk,\bl,\bm)$.
     With the technique above, we can then apply this support to 
     \eqref{HST_CLMInPhysicalSpace_w1w1w1_fs_Transf_Asymp_Sol} to get an estimate of $\MaxSupportGbkpbl$
      which may be greater than $\MaxSupportNbkp\times\MaxSupportNbkp$. 
     Finally, we determine $\MaxSupportbeta$, as done above.

The decreases of the domains $\MExtremeDomain$ in Fig. \ref{NGD}, $\MaxSupportNbkpNegVoid(\sigma)$ in Fig. \ref{NSupportMax} and $\MaxSupportbeta$ in Fig. \ref{betaSupportMax} under the increase of $\sigma \in [0,\sigma_{\max}]$ apparently reflect the meaning of $\sigma$. For instance, the case of lower $\sigma$ has greater supports for $\gsocAsy(\bk)$ and $\gAtocAsy(\bk,\bl)$ which contain greater subdomains of higher wave numbers, which in turn tend to dissipate more of the turbulent energy according to the term of $2\,|\bk|^2\,\tW_{\underlinej \underlinej}(\bk)$ in \eqref{HST_CLMInPhysicalSpace_ww_fs} and result in the slower growth rate $2\sigma$ of the turbulent energy. This feature may also imply the complicity of $\sigma<0$ and other non-asymptotic decaying cases of the
 homogeneous shear turbulence. 

We notice that the above estimates for the supports of $\LAsy(\bk)$,
   $\dAtocAsy(\bk,\bl)$ and $\dsocAsy(\bk)$ are expected to hold adequately for the fourth-order model too, 
since we have obtained them without resorting to any approximations to $\dAtocAsy(\bk,\bl)$.

\subsubsection{Feasible Solution and Determination of $\sigma_{\max}$}\label{Subsubsec:sigmamax}
\ \ \ \
As part of the solution, we need to determine the value of the upper bound for the exponential rate of growth $\sigma_{\max}$ within the reduced model.
 It may be inferred from the establishment of \eqref{SufficientCondition_TrivialSolution} 
that the satisfaction of \eqref{NonNegativityofSoc11_Reduced} underlies the existence of such a bound.
 To explore the issue in detail, we consider the case of \eqref{NonNegativityofSoc11_Reduced} 
under \eqref{SufficientCondition_TrivialSolution_Argument00}. 

 We discuss first the simpler equivalent constraints of \eqref{NonNegativityofSoc11_Reduced}
 under the equality condition of \eqref{SufficientCondition_TrivialSolution_Argument00},
\begin{align}
&
\int^{k_2}_{-\infty}\s d k'_2 \, \big(|k^0_1|^2+(k'_2)^2\big)\, 
   \exp\s\bigg[\frac{2\, k'_2}{|k^0_1|}\, \bigg(\sigma+|k^0_1|^2+\frac{1}{3}\,\big(k'_2\big)^2\bigg) \bigg]\, \LAsy(k^0_1,k'_2) \geq 0,
\notag\\[4pt] &
\int_{\mathbb{R}}\s dk'_2\,\LAsy(k^0_1,k'_2)= 0,\ \
\LAsy(k^0_1,k'_2)\not=0,\  \ 
|\LAsy(k^0_1,k'_2)|\leq 1,\ \
k^0_1\in(\konemin,0),\ \
\sigma\in[0,0.5)
\label{NonNegativityofSoc11_Reduced_ForSigmaMax}
\end{align}
Within the reduced model in which $\LAsy(\bk)$ is the control variable,
 we can construct mathematically a $\LAsy(k^0_1,k'_2)$ whose positive and negative values, respectively, distribute only in the peak and valley regions of $M((k^0_1,k_2);k'_2)$. For the sake of simple illustration, we take a discontinuous distribution of
\begin{align}
&
\LAsy(k^0_1,k'_2)
=
\l\{
\begin{array}{ll}
 L_P(k^0_1,\sigma), & k'_2 \in \big(\MPeak(k^0_1,\sigma)-\delta_P(k^0_1,\sigma), \ \MPeak(k^0_1,\sigma)+\delta_P(k^0_1,\sigma)\big);\\[4pt]
 -L_V(k^0_1,\sigma), & k'_2 \in \big(\MValley(k^0_1,\sigma)-\delta_P(k^0_1,\sigma), \ \MValley(k^0_1,\sigma)+\delta_P(k^0_1,\sigma)\big);\\[4pt]
 0, & \text{others},
\end{array}
\r.
\notag\\[6pt] & \hskip 20mm
0<L_V(k^0_1,\sigma)=L_P(k^0_1,\sigma)\leq 1,\quad 
0<\delta_P(k^0_1,\sigma)<<|\MValley(k^0_1,\sigma)|
\label{LAsyAsCV_LAsyApproxDistribution}
\end{align}
It is then trivial to verify that (\ref{NonNegativityofSoc11_Reduced_ForSigmaMax}$)_1$ is satisfied automatically, due to $M\big((k^0_1,k_2);\MPeak(k^0_1,\sigma)\big)> M\big((k^0_1,k_2);\MValley(k^0_1,\sigma)\big)$, etc.
 This specific example has demonstrated that, for the reduced model with $\LAsy(\bk)$ as the continuous control variable,
 we can construct mathematically a non-trivial feasible solution with
\begin{align}
\int_{\mathbb{R}}\s dk'_2\,\LAsy(\bk')= 0,\ \ \forall k_1\leq 0;\ \ 
|\LAsy(\bk')|\leq 1;\ \
\forall\sigma\in[0,0.5)
\end{align}
Consequently, we conclude that the issue of $\sigma_{max}$ cannot be resolved within the context of the reduced model with $\LAsy(\bk')$ as the control variable; We need to seek a solution possibly with $\dAtocAsy(\bk',\bl)$ as the control variable.  

In the reduced model with $\dAtocAsy(\bk',\bl)$ as the control variable,
 $\LAsy(\bk')$ is a derived quantity defined by \eqref{DefiningNkp} and the intrinsic equality of \eqref{Integrand_ZeroBalanceHalfPlane} holds automatically.
 In contrast to the case above, there are more possible ways for $\LAsy(\bk')$ to behave: 
 i) The predominant positive and negative values of $\LAsy(\bk')$ may not lie, respectively, in the peak and valley regions of $M(\bk;k'_2)$ for some $k_1<0$.
 ii) It may take positive and negative values alternately, with the number of peaks and valleys more than that of \eqref{LAsyAsCV_LAsyApproxDistribution}.
 iii) It may allow the occurrence of
\begin{align}
\int_{\mathbb{R}}\s dk'_2\,\LAsy(\bk')< 0,\ \ 
\text{for some } k_1\in(\konemin,0),\ \ 
\forall\sigma\in[0,0.5)
\end{align}
These possible mathematical behaviors introduce possible ways to violate (\ref{NonNegativityofSoc11_Reduced_ForSigmaMax}$)_1$. 
 For example, let us consider
\begin{align}
&
\int^{k_2}_{-\infty}\s d k'_2 \, \big(|k^0_1|^2+(k'_2)^2\big)\, 
   \exp\s\bigg[\frac{2\, k'_2}{|k^0_1|}\, \bigg(\sigma+|k^0_1|^2+\frac{1}{3}\,\big(k'_2\big)^2\bigg) \bigg]\, \LAsy(k^0_1,k'_2) \geq 0,
\notag\\[4pt] &
\int_{\mathbb{R}}\s dk'_2\,\LAsy(k^0_1,k'_2)< 0,\ \
\text{for some } k^0_1\in(\konemin,0)
\label{NonNegativityofSoc11_Reduced_ForSigmaMax_B}
\end{align}
We may understand its consequence by adopting \eqref{LAsyAsCV_LAsyApproxDistribution} with $0<L_P(k^0_1,\sigma)<L_V(k^0_1,\sigma)$ and obtaining from (\ref{NonNegativityofSoc11_Reduced_ForSigmaMax_B}$)_1$ 
\begin{align}
 \frac{M((k^0_1,k_2);\MValley(k^0_1,\sigma))}{M((k^0_1,k_2);\MPeak(k^0_1,\sigma))}
\leq \frac{L_P(k^0_1,\sigma)}{L_V(k^0_1,\sigma)}
=\frac{\int_{\mathbb{R}}\s dk'_2\,\max(\LAsy(k^0_1,k'_2),0)}{\int_{\mathbb{R}}\s dk'_2\,\max(-\LAsy(k^0_1,k'_2),0)}
\label{SigmaMax_Criterion}
\end{align}
This inequality may be violated for all the $k_2\geq\MValley(k^0_1,\sigma)+\delta_P(k^0_1,\sigma)$ in a scenario as follows: $\sigma$ is close to $0.5$ so that it cannot produce the sufficient height contrast between the peak and the valley, and the ratio on the left-hand side of \eqref{SigmaMax_Criterion} is lower than $1$.
 
That $\sigma_{\max}>0$ is allowed in the reduced model implies
 that a transient solution may approach an asymptotic state of $\sigma>0$ in the reduced model,
 which is impossible in the fourth-order model due to the constraint of \eqref{wiwj_Deviation_Asymp}.
 Therefore, if a transient solution does not decay,
 the reduced model may predict that it may have a exponential growth rate of $\sigma>0$.
 That is, the reduced model
 may not be suitable for the simulation of transient solutions which do not decay.

\subsubsection{$\LAsy(\bk)$ as Control Variable and Linear Programming}
\ \ \ \
As mentioned above, we may treat $\LAsy(\bk)$ as the control variable to determine $\dsocAsy(\bk)$ through optimization. 
    A direct search of an optimal solution of $\LAsy(\bk)$ in a space of functions poses a challenge.
 A simpler strategy is to adopt a specific form for $\LAsy(\bk)$ constructed with the help of certain function bases and symmetries; the unknown parameters contained in the specific form will be determined through the
 objective maximization under the constraint of inequality.
 Similar to the Galerkin method in the calculus of variations, such a treatment transforms the optimal control problem into an optimization problem in a finite-dimensional vector space whose dimension is equal to the number of unknown parameters involved.

 There are a few possible ways to deal with $\LAsy(\bk)$. The first is to adopt $\LAsy(\bk)$ simply as the control variable.
The second is to incorporate the limits of \eqref{CVLgamma_Limits_Reduced}, \eqref{CVL_Limitsk2zero_Reduced} and \eqref{CVL_Limitsk2k1_Reduced} and the support estimate of $\MaxSupportNbkpNeg$ by taking the special transformation of 
\begin{align*}
 \LAsy(\bk)=(k_1)^3\,\dot{L}^{(a)}(\bk),\quad |\dot{L}^{(a)}(\bk)|\leq C
\end{align*}
where $\dot{L}^{(a)}(\bk)$ is the control variable and the result of Sub-subsection~\ref{Subsubsec:sigmamax} holds. One can also adopt different transformations to meet the limit constraints.
 Considering that, to avoid the apparent singularity at $k_1=0$ in numerical simulations, the point of $\bk=\mathbf{0}$ on the support boundary may be relocated to $\bk=(k^-_1,0)$ with $k^-_1<0$ and $|k^-_1|$ small (along with its neighborhood points on the boundary), we present the first possibility here for discussion,
   and the others can be worked out in a similar fashion. They are to be tested numerically for the sake of comparison.

 Considering that expression \eqref{HST_CLMInPhysicalSpace_w1w1_fs_Transf_Asymp_Sol_Reduced} for $\dsocAsy(\bk)$ is restricted to $k_1<0$,
 a triangle mesh over $\MaxSupportNbkpNeg$ will be constructed with $\NodeNumberTotalSNNeg$ nodes and  
   $\TriangleNumberTotalSNNeg$ linear triangle elements whose collections are denoted, respectively, by
\begin{align}
\MaxSupportNbkpNegNodes=\big\{\NodeSNNeg_j:\ j=1,2,\cdots,\NodeNumberTotalSNNeg\big\},\qquad
\MaxSupportNbkpNegTriangles=\big\{\TriangleSNNeg_j:\ j=1,2,\cdots,\TriangleNumberTotalSNNeg\big\}
\label{NodesTriangles_LNeg}
\end{align}
 There are a point matrix $\PointMatrixSNNeg$ and a connectivity matrix $\ConnectivityMatrixSNNeg$ associated with the mesh. The point matrix is of $2\times\NodeNumberTotalSNNeg$ which stores in its $j$-th column the coordinates of Node $\NodeSNNeg_j$, $(k_1, k_2)$;
  The connectivity matrix is of $3\times\TriangleNumberTotalSNNeg$ whose $j$-th column contains the numbers of the three nodes in Triangle $\TriangleSNNeg_j$, the three nodes ordered in a counterclockwise sense.

The values of $\LAsy(\bk)$ at the nodes of $\MaxSupportNbkpNegNodes$ are denoted as
\begin{align}
\MaxSupportLbkLvalues
=
\Big\{\LAsy\s\l(\NodeSNNeg_i\r): \ i\in\{1,2,\cdots, \NodeNumberTotalSNNeg\} \Big\}
\label{toc_L_ValuedAtNodes}
\end{align} 
 The distribution of $\LAsy(\bk)$ in $\MaxSupportNbkpNeg$ can be approximated through the linear interpolation of
\begin{align}
\LAsy(\bk^{-})
=
\CharacteristicFunctionTriangleSNNegj(\bk^{-})\,
 \sum_{i\,=1}^{3}  \LAsy\big([\ConnectivityMatrixSNNeg]_{ij}\big)\,\ShapeFunction_i\big(\bk^{-};\TriangleSNNeg_j\big)
\label{L_ValuedAtkn}
\end{align}
Here, $\CharacteristicFunctionTriangleSNNegj(\bk^{-})$ is
   the characteristic function, and $\ShapeFunction_i\s\l(\bk^{-};\TriangleSNNeg_j\r)$, $i=1$, $2$, $3$, are the linear interpolation shape functions associated with Triangle $\TriangleSNNeg_j$.
  The distribution of $\LAsy(\bk)$ in $\MaxSupportNbkpPos$ can be found through 
 $\LAsy(-\bk)=\LAsy(\bk)$ of \eqref{N_ControlVariable}.

For the sake of computational convenience below, we recast \eqref{L_ValuedAtkn} in the form of
\begin{align}
\LAsy(\bk^{-})
\overset{\text{a.e.}}{=}
 \sum_{j}^{\TriangleNumberTotalSNNeg} \CharacteristicFunctionTriangleSNNegj(\bk^{-})\,
   \sum_{i\,=1}^{3}  \LAsy\big([\ConnectivityMatrixSNNeg]_{ij}\big)\,\ShapeFunction_i\big(\bk^{-};\TriangleSNNeg_j\big)
\label{L_ValuedAtkn_ae}
\end{align}
Here, $\overset{\text{a.e.}}{=}$ stands for `almost everywhere', since the equality may not hold when $\bk^{-}$ is in a common edge between two neighboring triangles or coincides with a node. This approximation will not have significant effects on the computations of $\gsocAsy(\bk)$, $\K^{\text{hom}(a)}(\sigma)$ and the intrinsic equality, with adequate mesh distributions to be explained below.

Substituting \eqref{L_ValuedAtkn_ae} into \eqref{HST_CLMInPhysicalSpace_w1w1_fs_Transf_Asymp_Sol_Reduced}, we obtain
\begin{align}
\gsocAsy(\bk)
=
-\frac{2}{k_1\,|\bk|^4}\,
\sum_{j\,=\,1}^{\TriangleNumberTotalSNNeg}\,
\sum_{i\,=1}^{3}  \LAsy\big([\ConnectivityMatrixSNNeg]_{ij}\big)\,
\int^{k_2}_{-\infty}\s dk'_2\,M(\bk;k'_2)\,\,
\ShapeFunction_i(\bk';\TriangleSNNeg_j)\,
\CharacteristicFunctionTriangleSNNegj(\bk')
\label{HST_CLMInPhysicalSpace_w1w1_fs_Transf_Asymp_Sol_Reduced_TriangleMesh_L01}
\end{align}
  Here, we should point out that the a.e. property of \eqref{L_ValuedAtkn_ae} might cause a potential problem
    in the integration with respect to $k'_2$,
  due to the possible double counting in the summation of \eqref{L_ValuedAtkn_ae} and \eqref{HST_CLMInPhysicalSpace_w1w1_fs_Transf_Asymp_Sol_Reduced_TriangleMesh_L01}
 for $\bk$ located in a common edge between two neighboring triangles;
  This double counting affects the validity of \eqref{HST_CLMInPhysicalSpace_w1w1_fs_Transf_Asymp_Sol_Reduced_TriangleMesh_L01} only if it contributes to the line integration. We can eliminate this problem by one of two ways: (i) to generate the triangle mesh in $\MaxSupportNbkpNeg$ such that no common edge is parallel to the axis of $k_2$;
  (ii) to choose $k_1$ in 
  \eqref{HST_CLMInPhysicalSpace_w1w1_fs_Transf_Asymp_Sol_Reduced_TriangleMesh_L01}
 such that it does not lie in any common edge parallel to the axis of $k_2$.
     The latter can be easily implemented since we need to impose the constraint of non-negativity only
   at a finite number of points inside $\MaxSupportbeta$ to be discussed.

Equation \eqref{HST_CLMInPhysicalSpace_w1w1_fs_Transf_Asymp_Sol_Reduced_TriangleMesh_L01} can be rewritten, through rearrangement and combination, as
\begin{align}
\gsocAsy(\bk)
=
\sum_{i\,=\,1}^{\NodeNumberTotalSNNeg}
       \LPConstraintCoefficient\s\l(\NodeSNNeg_i; \bk\r)\,\LAsy\s\l(\NodeSNNeg_i\r)
\label{HST_CLMInPhysicalSpace_w1w1_fs_Transf_Asymp_Sol_Reduced_TriangleMesh_L}
\end{align}
It is linear in $\LAsy\s\l(\NodeSNNeg_i\r)$ with the coefficients $\LPConstraintCoefficient\s\l(\NodeSNNeg_i; \bk\r)$s as continuous functions of $\bk$.
  Now, \eqref{ObjectiveFunction_01_Asymp_Reduced} becomes
\begin{align}
&
\K^{\text{hom}(a)}(\sigma)
\notag\\
=\,&
-4\,
\sum_{j\,=\,1}^{\TriangleNumberTotalSNNeg}\,
\sum_{i\,=1}^{3}  \LAsy\big([\ConnectivityMatrixSNNeg]_{ij}\big)
\int_{-\infty}^0\s dk_1\s \int_{\mathbb{R}}\s dk_2\,
\frac{1}{k_1\,|\bk|^2}\,
\int^{k_2}_{-\infty}\s dk'_2\,M(\bk;k'_2)\,
\ShapeFunction_i(\bk';\TriangleSNNeg_j)\,
\CharacteristicFunctionTriangleSNNegj(\bk')
\label{ObjectiveFunction_01_Asymp_Reduced_TriangleMesh_L}
\end{align}
The approximate nature of \eqref{L_ValuedAtkn_ae} should not affect the validity of
   \eqref{ObjectiveFunction_01_Asymp_Reduced_TriangleMesh_L} since the area measure of all the edges 
   is zero and the values of $\LAsy\big([\ConnectivityMatrixSNNeg]_{ij}\big)$  are supposedly finite. 
 The equation can be recast as
\begin{align}
\K^{\text{hom}(a)}(\sigma)
=\sum_{i\,=\,1}^{\NodeNumberTotalSNNeg}
       \LPObjectiveFunctionCoefficient\s\l(\NodeSNNeg_i\r)\,\LAsy\s\l(\NodeSNNeg_i\r)
\label{LP_ObjectiveFunction_LasCV}
\end{align}
This objective function is linear in $\LAsy\s\l(\NodeSNNeg_i\r)$. Following from \eqref{ObjectiveFunction_01_Asymp_Reduced}, 
   it is to be maximized under the constraints of \eqref{NonNegativityofSoc11_Reduced} and \eqref{N_ControlVariable}, whose consequences are as follows.

Firstly, combining \eqref{NonNegativityofSoc11_Reduced} and \eqref{HST_CLMInPhysicalSpace_w1w1_fs_Transf_Asymp_Sol_Reduced_TriangleMesh_L} gives
\begin{align}
\sum_{i\,=\,1}^{\NodeNumberTotalSNNeg}
       \LPConstraintCoefficient\s\l(\NodeSNNeg_i; \bk\r)\, \LAsy\s\l(\NodeSNNeg_i\r)
\geq 0
\label{LP_Constraint_Continuous_LasCV}
\end{align}
The coefficients are functions of $\bk$ and the above inequality needs to hold in the estimated support of $\gsocAsy(\bk)$, $\MaxSupportbeta$ given by \eqref{gsoc_EffectiveSupport} and sketched in Fig. \ref{betaSupportMax}. 
 Due to the equivalence between (\ref{NonNegativityofSoc11_Reduced_ForSigmaMax}$)_1$ and  \eqref{NonNegativityofSoc11_Reduced},
it is sufficient to enforce the inequality in $\MaxSupportNbkpNeg$ of \eqref{PossibleSupportOfNkp_Subdomains}.
 Considering the above-mentioned requirement of $k_1$ not-located at any common edge of the triangular meshes parallel to the axis of $k_2$,
  we can select adequately a finite set of collocation points inside the support $\MaxSupportNbkpNeg$,
\begin{align}
\Big\{
\big(k_1(\M_1,\M_2),\ k_2(\M_1,\M_2)\big):\ \M_1,\, \M_2
\Big\}
  \subset\MaxSupportNbkpNeg
\label{LP_Constraint_CollocationPoints_LasCV}
\end{align}
on which we apply \eqref{LP_Constraint_Continuous_LasCV} so as to approximate it with a finite number of linear constraints of
\begin{align}
-\sum_{i\,=\,1}^{\NodeNumberTotalSNNeg}
      \,\LPConstraintCoefficient\s\l(\NodeSNNeg_i; \bk(\M_1,\M_2)\r) \LAsy\s\l(\NodeSNNeg_i\r)
\leq 0,
\ \ 
\forall \M_1,\ \M_2
\label{LP_Constraint_AtCollocationPoints_LasCV}
\end{align}

Secondly, the intrinsic equality of (\ref{N_ControlVariable}$)_3$ requires that
\begin{align}
 \sum_{j}^{\TriangleNumberTotalSNNeg}\,
   \sum_{i\,=1}^{3}  \LAsy\big([\ConnectivityMatrixSNNeg]_{ij}\big)
\int_{-\infty}^0\s dk_1\, \int_{\mathbb{R}}\s dk_2\,
   \ShapeFunction_i\big(\bk;\TriangleSNNeg_j\big) \,
    \CharacteristicFunctionTriangleSNNegj(\bk)
=0
\end{align}
or
\begin{align}
\sum_{i\,=\,1}^{\NodeNumberTotalSNNeg}
      b\s\l(\NodeSNNeg_i\r)\, \LAsy\s\l(\NodeSNNeg_i\r)=0
\label{LP_Constraint_IntrinsicEquality_LasCV}
\end{align}

Thirdly, the support of $\LAsy(\bk)$ implies that
\begin{align}
\LAsy\s\l(\NodeSNNeg_i\r)=0,\ 
\text{if $\NodeSNNeg_i$ is a boundary node}
\label{LP_Constraint_BoundaryNodes_LasCV}
\end{align}
And finally, the bounds of \eqref{N_ControlVariable} can be represented in the equivalent form of
\begin{align}
&
\LAsy\s\l(\NodeSNNeg_i\r)\leq 1,\quad
-\LAsy\s\l(\NodeSNNeg_i\r)\leq 1,\ \
     \forall i \in\{1,2,\cdots, \NodeNumberTotalSNNeg\}
\label{LP_Constraint_LowerBounds_LasCV}
\end{align} 

We have a linear programming problem of the objective \eqref{LP_ObjectiveFunction_LasCV} to be maximized under the  sets of the linear constraints of \eqref{LP_Constraint_AtCollocationPoints_LasCV},  and \eqref{LP_Constraint_IntrinsicEquality_LasCV} through  \eqref{LP_Constraint_LowerBounds_LasCV}.

\subsubsection{$\dAtocAsy(\bk',\bl)$ as Control Variable and Linear Programming}\label{dtocAsyasCVLP}
\ \ \ \
The above treatment of $\LAsy(\bk)$ as the control variable has the advantage of computations only in the wave number space $\bk$. However, it does not provide any detailed information about the third order correlations $\tW^{(Ia)}_{ijk}(\bk,\bl)$. Also, it cannot resolve the issue of $\sigma_{\max}$.
  We now study the case that $\dAtocAsy(\bk,\bl)$ (or effectively its anti-symmetric part) is used as the control variable. A comprehensive distribution of $\dAtocAsy(\bk,\bl)$ should be determined with the fourth order model.

Motivated by the structure of \eqref{HST_CLMInPhysicalSpace_w1w1w1_fs_Transf_Asymp_Sol},
   the definition of \eqref{LAsy_Defn}
  and the limiting constraints of \eqref{CVLgamma_Limits_Reduced}, \eqref{CVL_Limitsk2zero_Reduced}
       and \eqref{CVL_Limitsk2k1_Reduced},   
    and  the symmetry of \eqref{HST_DivergenceFreeInPhysicalSpace_wiwjwk_fs_Transf_Asymp}, we present, amongst several choices, a partition form of
\begin{align}
\gtocAsy(\bk',\bl)=\,&
 \chi_{\MaxSupportNbkp}(\bk')\,\chi_{\MaxSupportNbkp}(\bl)\,\chi_{\MaxSupportNbkp}(\bk'+\bl)
  \big[k_1\,l_1\,(k_1+l_1)\big]^2 \,
\notag\\[4pt]&\hskip 10mm\times\s
    \big[ \GAsy(\bk',\bl)+\GAsy(\bl,-\bk'-\bl)+\GAsy(-\bk'-\bl,\bk') \big] 
\label{toc_GeneralSymmetries_d}
\end{align}
Here, $\chi_{\MaxSupportNbkp}$ is the characteristic function
 and $\GAsy(\bk',\bl)$ supposedly has the symmetry of
\begin{align}
\GAsy(\bk',\bl)=\GAsy(\bl,\bk')=\GAsy(-\bk',-\bl),\quad
\GAsy(\bk',-\bk')=0
\label{toc_GeneralSymmetries_d_Symmetry}
\end{align}
The support of $\GAsy(\bk',\bl)$ is taken the same as that of $\dtocAsy(\bk',\bl)$,
\begin{align}
\MaxSupportGGbkpbl=\MaxSupportGbkpbl=\MaxSupportNbkp\times\MaxSupportNbkp
\label{toc_G_Support}
\end{align}
 The symmetries of \eqref{toc_GeneralSymmetries_d_Symmetry} are less stringent than those of
    \eqref{HST_DivergenceFreeInPhysicalSpace_wiwjwk_fs_Transf_Asymp}.
In fact, without a partition as such or similar, it is difficult in numerical simulation to satisfy
    $\dtocAsy(\bk,\bl)=\dtocAsy(-\bk-\bl,\bl)$ of \eqref{HST_DivergenceFreeInPhysicalSpace_wiwjwk_fs_Transf_Asymp}. 
 The inclusion of the characteristic function in \eqref{toc_GeneralSymmetries_d} is to guarantee that the resultant $\gtocAsy(\bk',\bl)$ has the same support as the estimated $\MaxSupportGbkpbl$.

For numerical simulation of $\GAsy(\bk',\bl)$, we adopt a quasi-triangle mesh,
    i.e., a tensor-product of two triangle meshes over $\MaxSupportGGbkpbl$ in the fashion detailed below:
  First, we resort to the triangle mesh generated in \eqref{NodesTriangles_LNeg} over $\MaxSupportNbkpNeg$.
  Next, due to $\MaxSupportNbkpPos=-\MaxSupportNbkpNeg$,
           we have a corresponding triangle mesh over $\MaxSupportNbkpPos$ with
\begin{align}
&
\MaxSupportNbkpPosNodes=
\big\{\NodeSNPos_j:\ \NodeSNPos_j=\NodeSNNeg_j,\, j=1,2,\cdots,\NodeNumberTotalSNNeg\big\},
\notag\\[4pt] &
\MaxSupportNbkpPosTriangles=\big\{\TriangleSNPos_j:\ \TriangleSNPos_j=\TriangleSNNeg_j,\, j=1,2,\cdots,\TriangleNumberTotalSNNeg\big\}
\label{NodesTriangles_Pos}
\end{align}
 The corresponding point matrix $\PointMatrixSNPos$ and connectivity matrix $\ConnectivityMatrixSNPos$ are given by
\begin{align}
\PointMatrixSNPos\big|_{\text{column}\, j}=-\PointMatrixSNNeg\big|_{\text{column}\, j},\qquad
\ConnectivityMatrixSNPos\big|_{\text{column}\, j}=\ConnectivityMatrixSNNeg\big|_{\text{column}\, j}
\end{align}
which reflects that
\begin{align}
\bk(\NodeSNPos_j)=-\bk(\NodeSNNeg_j)
\label{Nodes_InversionSymmetry}
\end{align}
It then follows from above that $\MaxSupportNbkp$ is meshed by
\begin{align}
&
\MaxSupportNbkpNodes=\MaxSupportNbkpNegNodes\cup\MaxSupportNbkpPosNodes,\qquad
\MaxSupportNbkpTriangles=\MaxSupportNbkpNegTriangles\cup\MaxSupportNbkpPosTriangles,
\notag\\[4pt] &
\PointMatrixSN=\big\{\PointMatrixSNNeg,\, \PointMatrixSNPos\big\},\qquad
\ConnectivityMatrixSN=\big\{\ConnectivityMatrixSNNeg,\, \ConnectivityMatrixSNPos\big\}
\end{align}

We now adopt the tensor-product of the triangle meshes
\begin{align}
\MaxSupportGbkpblNodes=\MaxSupportNbkpNodes\times\MaxSupportNbkpNodes,\qquad
\MaxSupportGbkpblTrangles=\MaxSupportNbkpTriangles\times\MaxSupportNbkpTriangles
\label{BiTriangleMesh}
\end{align} 
over $\MaxSupportGGbkpbl$. This treatment is motivated mainly by its simple mesh generation, its easy implementations of the symmetry properties of \eqref{toc_GeneralSymmetries_d_Symmetry} and the notion of turbulent energy cascade if necessary.
   The values of $\GAsy(\bk,\bl)$ at the nodes of $\MaxSupportGbkpblNodes$ are denoted as
\begin{align}
\MaxSupportGbkpblFvalues
=\,&
\Big\{\GAsy\s\l(\NodeSNNeg_i;\NodeSNNeg_j\r),\ 
      \GAsy\s\l(\NodeSNNeg_i;\NodeSNPos_j\r),\ 
      \GAsy\s\l(\NodeSNPos_i;\NodeSNNeg_j\r),\
      \GAsy\s\l(\NodeSNPos_i;\NodeSNPos_j\r):
\notag\\ &
\hskip 90mm \ i, j \in\{1,2,\cdots, \NodeNumberTotalSNNeg\} \Big\}
\label{toc_G_ValuedAtNodes}
\end{align} 
We can take
\begin{align}
\MaxSupportGbkpblFvaluesNeg
=\Big\{\GAsy\s\l(\NodeSNNeg_i;\NodeSNNeg_j\r),\ 
      \GAsy\s\l(\NodeSNNeg_i;\NodeSNPos_j\r):\ i, j \in\{1,2,\cdots, \NodeNumberTotalSNNeg\} \Big\}
\label{FAtNodes_Primary}
\end{align} 
as the primary basis set, considering that $\GAsy\s\l(\NodeSNPos_i;\NodeSNNeg_j\r)$ and $\GAsy\s\l(\NodeSNPos_i;\NodeSNPos_j\r)$ can be found through
\begin{align}
&
\GAsy\s\l(\NodeSNPos_i;\NodeSNNeg_j\r)
=\GAsy\s\l(-\bk(\NodeSNPos_i);-\bk(\NodeSNNeg_j)\r)
=\GAsy\s\l(\NodeSNNeg_i;\NodeSNPos_j\r),
\notag\\ &
 \GAsy\s\l(\NodeSNPos_i;\NodeSNPos_j\r)
=\GAsy\s\l(\NodeSNNeg_i; \NodeSNNeg_j\r)
\label{G_ValuedAtNodes_InversionSymmetry}
\end{align}
due to \eqref{toc_GeneralSymmetries_d_Symmetry} and \eqref{Nodes_InversionSymmetry}.

Next, since \eqref{HST_CLMInPhysicalSpace_w1w1_fs_Transf_Asymp_Sol_Reduced} is of integral form and 
  the tensor-product of triangle meshes is adopted in \eqref{BiTriangleMesh}, we resort to a quasi-bilinear interpolation to find the distribution of $\GAsy(\bk,\bl)$ in $\MaxSupportGGbkpbl$,
\begin{align}
\GAsy(\bk^{-},\bl^{-})
=
 \sum_{i,\,j\,=1}^{3}  G\big([\ConnectivityMatrixSNNeg]_{ik};[\ConnectivityMatrixSNNeg]_{jl}\big)\,\ShapeFunction_i\big(\bk^{-};\TriangleSNNeg_k\big)\,\,\ShapeFunction_j\big(\bl^{-};\TriangleSNNeg_l\big),\ \ 
\bk^{-}\in\TriangleSNNeg_k,\ \bl^{-}\in\TriangleSNNeg_l 
\label{G_ValuedAtknln}
\end{align}
\begin{align}
\GAsy(\bk^{-},\bl^{+})
=
 \sum_{i,\,j\,=1}^{3}  G\big([\ConnectivityMatrixSNNeg]_{ik};[\ConnectivityMatrixSNPos]_{jl}\big)\,\ShapeFunction_i\big(\bk^{-};\TriangleSNNeg_k\big)\,\,\ShapeFunction_j\big(\bl^{+};\TriangleSNPos_l\big),\ \ 
\bk^{-}\in\TriangleSNNeg_k,\ \bl^{+}\in\TriangleSNPos_l 
\label{G_ValuedAtknlp}
\end{align}
and
\begin{align}
\GAsy(\bk^{+},\bl)
=\GAsy(-\bk^{+},-\bl),\ \ 
\bk^{+}\in\TriangleSNPos_k
\label{G_InversionSymmetry}
\end{align}
Here, \eqref{G_InversionSymmetry} comes from \eqref{toc_GeneralSymmetries_d_Symmetry}. We now need to discuss how the full content of  \eqref{toc_GeneralSymmetries_d_Symmetry} can be satisfied.
\begin{enumerate}
\item
The application of $\GAsy(\bk,\bl)=\GAsy(\bl,\bk)$ to the elements of \eqref{FAtNodes_Primary}, along with \eqref{G_ValuedAtNodes_InversionSymmetry}, yields 
\begin{align}
\GAsy\s\l(\NodeSNNeg_i;\NodeSNNeg_j\r)=\GAsy\s\l(\NodeSNNeg_j;\NodeSNNeg_i\r),\quad
\GAsy\s\l(\NodeSNNeg_i;\NodeSNPos_j\r)=\GAsy\s\l(\NodeSNNeg_j;\NodeSNPos_i\r)
\label{G_ValuedAtNodes_Symmetry}
\end{align} 
which will be imposed explicitly.
 Together with \eqref{G_ValuedAtknln} through \eqref{G_InversionSymmetry} and the shape function property, these constraints are also sufficient to guarantee 
\begin{align}
 \GAsy(\bk^{-},\bl^{-})=\GAsy(\bl^{-},\bk^{-}),\quad
 \GAsy(\bk^{+},\bl^{-})=\GAsy(\bl^{-},\bk^{+}),\quad
 \GAsy(\bk^{+},\bl^{+})=\GAsy(\bl^{+},\bk^{+})
\label{G_kl=lk_Symmetry}
\end{align}

\item
We now test for the symmetry $\GAsy(\bk,\bl)=\GAsy(-\bk,-\bl)$.
 Equation \eqref{G_InversionSymmetry} indicates that 
\begin{align*}
\GAsy(\bk^{+},\bl^{-})=\GAsy(-\bk^{+},-\bl^{-}),\quad
\GAsy(\bk^{+},\bl^{+})=\GAsy(-\bk^{+},-\bl^{+})
\end{align*}
 is automatically met through construction. For the rest two cases, we consider first $\GAsy(\bk^{-},\bl^{-})$. That 
  $\bk^{-}\in\TriangleSNNeg_k$ and  $\bl^{-}\in\TriangleSNNeg_l$ implies that
  $-\bk^{-}\in\TriangleSNPos_k$ and $-\bl^{-}\in\TriangleSNPos_l$ from the adopted mesh generation over $\MaxSupportNbkpPos$. Consequently,  the last equality above gives
\begin{align*}
\GAsy(-\bk^{-},-\bl^{-})=\GAsy(-(-\bk^{-}),-(-\bl^{-}))=\GAsy(\bk^{-},\bl^{-})
\end{align*}
as desired. In the case of $\GAsy(\bk^{-},\bl^{+})$, we apply \eqref{G_kl=lk_Symmetry} and \eqref{G_InversionSymmetry} to get
\begin{align*}
\GAsy(\bk^{-},\bl^{+})
=\GAsy(\bl^{+},\bk^{-})
=\GAsy(-\bl^{+},-\bk^{-})
=\GAsy(-\bk^{-},-\bl^{+})
\end{align*}
Therefore, the symmetry of $\GAsy(\bk,\bl)=-\GAsy(-\bk,-\bl)$ is satisfied automatically if \eqref{G_ValuedAtNodes_Symmetry} is enforced. 

\item
The last of \eqref{toc_GeneralSymmetries_d_Symmetry} requires that
\begin{align}
\GAsy\s\l(\NodeSNNeg_i;\NodeSNPos_i\r)=0
\label{G_ValuedAtNodes_bk&-bk}
\end{align}
which will be imposed explicitly.
\end{enumerate}

To help the computation of $\K^{\text{hom}(a)}(\sigma)$ and the implementation of $\gsocAsy(\bk)\geq 0$, we introduce a unified relation of
\begin{align}
&
\GAsy(\bl^{-},\bm)
\notag\\
\overset{\text{a.e.}}{=}\,&
 \sum_{l,\,m=1}^{\TriangleNumberTotalSNNeg} \sum_{i,\,j\,=1}^{3}
\Big[\GAsy\s\l(\l[\ConnectivityMatrixSNNeg\r]_{il};\l[\ConnectivityMatrixSNNeg\r]_{jm}\r)\,\ShapeFunction_i\big(\bl^{-};\TriangleSNNeg_l\big)\,\ShapeFunction_j\big(\bm;\TriangleSNNeg_m\big)\,\CharacteristicFunctionTriangleSNNegl(\bl^{-})\,\CharacteristicFunctionTriangleSNNegm(\bm)
\notag\\[4pt] &\hskip 20mm
     +\GAsy\s\l(\l[\ConnectivityMatrixSNNeg\r]_{il};\l[\ConnectivityMatrixSNPos\r]_{jm}\r)\,\ShapeFunction_i\big(\bl^{-};\TriangleSNNeg_l\big)\,\ShapeFunction_j\big(\bm;\TriangleSNPos_m\big)\,\CharacteristicFunctionTriangleSNNegl(\bl^{-})\,\CharacteristicFunctionTriangleSNPosm(\bm)
\Big]
\label{G_ValuedAtlnmpm_ae}
\end{align}
A remark like that of \eqref{L_ValuedAtkn_ae} can be made here.

Substituting \eqref{toc_GeneralSymmetries_d} and \eqref{G_ValuedAtlnmpm_ae} into
\eqref{HST_CLMInPhysicalSpace_w1w1_fs_Transf_Asymp_Sol_Reduced_Recast} 
 and using \eqref{NodesTriangles_Pos} and \eqref{G_ValuedAtNodes_Symmetry}, we obtain
\begin{align}
 \gsocAsy(\bk)
=\frac{2}{|k_1|\,|\bk|^4}\, \sum_{l,\, m\,=\,1}^{\TriangleNumberTotalSNNeg} \,\sum_{i,\,j\,=\,1}^{3}
\Big[&
     \hat{a}\s\l(\l[\ConnectivityMatrixSNNeg\r]_{il};\l[\ConnectivityMatrixSNNeg\r]_{jm};\bk\r)\,\GAsy\s\l(\l[\ConnectivityMatrixSNNeg\r]_{il};\l[\ConnectivityMatrixSNNeg\r]_{jm}\r)
\notag\\[4pt] &
     +\hat{a}\s\l(\l[\ConnectivityMatrixSNNeg\r]_{il};\l[\ConnectivityMatrixSNPos\r]_{jm};\bk\r)\,\GAsy\s\l(\l[\ConnectivityMatrixSNNeg\r]_{il};\l[\ConnectivityMatrixSNPos\r]_{jm}\r)
\Big]
\label{HST_CLMInPhysicalSpace_w1w1_fs_Transf_Asymp_Sol_Reduced_TriangleMesh}
\end{align}
where
\begin{align}
&
\hat{a}\s\l(\l[\ConnectivityMatrixSNNeg\r]_{il};\l[\ConnectivityMatrixSNNeg\r]_{jm};\bk\r)
\notag\\
=\,&
  \int^{k_2}_{\MValley'}\s d k'_2 \,M(\bk;k'_2)\,
\notag\\ &\hskip 10mm\times\s
\bigg[
 \int_{\TriangleSNNeg_l}\s d\bl\, \big[|\bl|^2-|\bk'+\bl|^2\big]\,(k_1\,l_2-k'_2\,l_1)\,
\chi_{\MaxSupportNbkpNeg}(\bk'+\bl)
\notag\\ &\hskip 35mm\times\s
      [k_1 l_1\,(k_1+l_1)]^2\,
\ShapeFunction_j\big(\bk';\TriangleSNNeg_m\big)\,
\ShapeFunction_i\big(\bl;\TriangleSNNeg_l\big)\,
\CharacteristicFunctionTriangleSNNegm(\bk')\,
\notag\\[4pt] &\hskip 18mm
-\int_{\TriangleSNNeg_l} d\bl\, |\bl|^2\,(k_1\,l_2-k'_2\,l_1)\,
    \chi_{\MaxSupportNbkpNeg}(\bk')\,
\notag\\ &\hskip 35mm\times\s
      [k_1 l_1\,(k_1-l_1)]^2\,
\ShapeFunction_i\big(\bl^{-};\TriangleSNNeg_l\big)\,
\ShapeFunction_j\big(\bk'-\bl;\TriangleSNNeg_m\big)\,
\CharacteristicFunctionTriangleSNNegm(\bk'-\bl)
\bigg]
\label{gsoc_SolutionCoefficient_Neg}
\end{align}
and
\begin{align}
&
\hat{a}\s\l(\l[\ConnectivityMatrixSNNeg\r]_{il};\l[\ConnectivityMatrixSNPos\r]_{jm};\bk\r)
\notag\\
=\,&
  \int^{k_2}_{\MValley'}\s d k'_2 \,M(\bk;k'_2)\,
\notag\\ &\hskip 10mm\times\s
\bigg[
 \int_{\TriangleSNNeg_l}\s d\bl\, \big(|\bk'-\bl|^2-|\bl|^2\big)\,(k_1\,l_2-k'_2\,l_1)\,
\chi_{\MaxSupportNbkp}(\bk'-\bl)  
\notag\\ &\hskip 35mm\times\s
      [k_1 l_1\,(k_1-l_1)]^2\,
\ShapeFunction_j\big(\bk';\TriangleSNNeg_m\big)\,
\ShapeFunction_i\big(\bl;\TriangleSNNeg_l\big)\,
\CharacteristicFunctionTriangleSNNegm(\bk')\,
\notag\\[4pt] &\hskip 18mm
+\int_{\TriangleSNNeg_l} d\bl\, |\bl|^2\,(k_1\,l_2-k'_2\,l_1)\,
    \chi_{\MaxSupportNbkpNeg}(\bk')\,
\notag\\ &\hskip 35mm\times\s
      [k_1 l_1\,(k_1+l_1)]^2\,
\ShapeFunction_i\big(\bl^{-};\TriangleSNNeg_l\big)\,
\ShapeFunction_j\big(\bk'+\bl;\TriangleSNNeg_m\big)\,
\CharacteristicFunctionTriangleSNNegm(\bk'+\bl)
\notag\\[4pt] &\hskip 18mm
-\int_{\TriangleSNNeg_l} d\bl\, |\bl|^2\,(k_1\,l_2-k'_2\,l_1)\,
    \chi_{\MaxSupportNbkpNeg}(\bk')\,
\notag\\ &\hskip 35mm\times\s
      [k_1 l_1\,(k_1-l_1)]^2\,
\ShapeFunction_i\big(\bl^{-};\TriangleSNNeg_l\big)\,
\ShapeFunction_j\big(\bl-\bk';\TriangleSNNeg_m\big)\,
\CharacteristicFunctionTriangleSNNegm(\bl-\bk')
\bigg]
\label{gsoc_SolutionCoefficient_Pos}
\end{align}

Next, substitution of \eqref{HST_CLMInPhysicalSpace_w1w1_fs_Transf_Asymp_Sol_Reduced_TriangleMesh} into \eqref{ObjectiveFunction_01_Asymp_Reduced} gives
\begin{align}
\K^{\text{hom}(a)}(\sigma)
=
4\,\sum_{l=1}^{\TriangleNumberTotalSNNeg}\sum_{m=1}^{\TriangleNumberTotalSNNeg}\sum_{i,\,j\,=1}^{3}
  \Big[&\,
      \hat{c}\s\l(\l[\ConnectivityMatrixSNNeg\r]_{il};\l[\ConnectivityMatrixSNNeg\r]_{jm}\r) \GAsy\s\l(\l[\ConnectivityMatrixSNNeg\r]_{il};\l[\ConnectivityMatrixSNNeg\r]_{jm}\r)
\notag\\[4pt] &
     +\hat{c}\s\l(\l[\ConnectivityMatrixSNNeg\r]_{il};\l[\ConnectivityMatrixSNPos\r]_{jm}\r)
            \GAsy\s\l(\l[\ConnectivityMatrixSNNeg\r]_{il};\l[\ConnectivityMatrixSNPos\r]_{jm}\r)
  \Big]
\label{LP_ObjectiveFunction_TriangleMesh_a}
\end{align}
where
\begin{align}
\hat{c}\s\l(\l[\ConnectivityMatrixSNNeg\r]_{il};\l[\ConnectivityMatrixSNNeg\r]_{jm}\r)
=   
 \int_{\MaxSupportbeta} d\bk\,\frac{1}{|k_1|\,|\bk|^2}\,
          \hat{a}\s\l(\l[\ConnectivityMatrixSNNeg\r]_{il};\l[\ConnectivityMatrixSNNeg\r]_{jm};\bk\r)\,
\end{align}
and
\begin{align}
\hat{c}\s\l(\l[\ConnectivityMatrixSNNeg\r]_{il};\l[\ConnectivityMatrixSNPos\r]_{jm}\r)
=   
 \int_{\MaxSupportbeta} d\bk\,\frac{1}{|k_1|\,|\bk|^2}\,
          \hat{a}\s\l(\l[\ConnectivityMatrixSNNeg\r]_{il};\l[\ConnectivityMatrixSNPos\r]_{jm};\bk\r)
\end{align}

We now have a linear programming problem to solve for $\GAsy\s\l(\NodeSNNeg_i;\NodeSNNeg_j\r)$ and $\GAsy\s\l(\NodeSNNeg_i;\NodeSNPos_j\r)$ as described below.

The objective function of \eqref{LP_ObjectiveFunction_TriangleMesh_a} can be recast, through rearrangement and combination, in the form of
\begin{align}
\K^{\text{hom}(a)}(\sigma)
=\sum_{i,\,j\,=\,1}^{\NodeNumberTotalSNNeg}
\Big[
       \LPObjectiveFunctionCoefficient\s\l(\NodeSNNeg_i ;\NodeSNNeg_j\r)
\GAsy\s\l(\NodeSNNeg_i ;\NodeSNNeg_j\r)
      +\LPObjectiveFunctionCoefficient\s\l(\NodeSNNeg_i ;\NodeSNPos_j\r)
\GAsy\s\l(\NodeSNNeg_i ;\NodeSNPos_j\r)
\Big]
\label{LP_ObjectiveFunction}
\end{align}
which is linear in $\GAsy\s\l(\NodeSNNeg_i ;\NodeSNNeg_j\r)$ and $\GAsy\s\l(\NodeSNNeg_i ;\NodeSNPos_j\r)$. This function is to be maximized following from \eqref{ObjectiveFunction_01_Asymp_Reduced}.

Equation \eqref{HST_CLMInPhysicalSpace_w1w1_fs_Transf_Asymp_Sol_Reduced_TriangleMesh} can also be recast in terms of $\GAsy\s\l(\NodeSNNeg_i ;\NodeSNNeg_j\r)$ and $\GAsy\s\l(\NodeSNNeg_i ;\NodeSNPos_j\r)$,
\begin{align}
  \gsocAsy(\bk)
=
\sum_{i,\,j\,=\,1}^{\NodeNumberTotalSNNeg}
\Big[
       \LPConstraintCoefficient\s\l(\NodeSNNeg_i ;\NodeSNNeg_j; \bk\r)\,\GAsy\s\l(\NodeSNNeg_i ;\NodeSNNeg_j\r)
      +\LPConstraintCoefficient\s\l(\NodeSNNeg_i ;\NodeSNPos_j; \bk\r)\,\GAsy\s\l(\NodeSNNeg_i ;\NodeSNPos_j\r)
\Big]
\end{align}
 and then, Equation \eqref{NonNegativityofSoc11_Reduced} results in the linear constraint of 
\begin{align}
\sum_{i,\,j\,=\,1}^{\NodeNumberTotalSNNeg}
\Big[
       \LPConstraintCoefficient\s\l(\NodeSNNeg_i ;\NodeSNNeg_j; \bk\r)\,\GAsy\s\l(\NodeSNNeg_i ;\NodeSNNeg_j\r)
      +\LPConstraintCoefficient\s\l(\NodeSNNeg_i ;\NodeSNPos_j; \bk\r)\,\GAsy\s\l(\NodeSNNeg_i ;\NodeSNPos_j\r)
\Big]
\geq 0
\label{LP_Constraint_Continuous}
\end{align}
 The coefficients are functions of $\bk$. We can apply the constraint to the collocation points of
    \eqref{LP_Constraint_CollocationPoints_LasCV} to obtain a finite number of linear constraints of
\begin{align}
-\sum_{i,\,j\,=\,1}^{\NodeNumberTotalSNNeg}
\Big[
      &\,\LPConstraintCoefficient\s\l(\NodeSNNeg_i ;\NodeSNNeg_j; \bk(\M_1,\M_2)\r) \GAsy\s\l(\NodeSNNeg_i ;\NodeSNNeg_j\r)
\notag\\ &
      +\LPConstraintCoefficient\s\l(\NodeSNNeg_i ;\NodeSNPos_j; \bk(\M_1,\M_2)\r) \GAsy\s\l(\NodeSNNeg_i ;\NodeSNPos_j\r)
\Big]
\leq 0,
\ \ 
\forall \M_1,\ \M_2
\label{LP_Constraint_AtCollocationPoints}
\end{align}

The support of $\GAsy(\bk',\bl)$ implies that
\begin{align}
\GAsy\s\l(\NodeSNNeg_i ;\NodeSNNeg_j\r)=\GAsy\s\l(\NodeSNNeg_i ;\NodeSNPos_j\r)=0,\ \text{if $\NodeSNNeg_i$ or $\NodeSNNeg_j$  
    is a boundary node}
\label{LP_Constraint_BoundaryNodes}
\end{align}
In addition, \eqref{G_ValuedAtNodes_Symmetry} and \eqref{G_ValuedAtNodes_bk&-bk} require that
\begin{align}
&
\GAsy\s\l(\NodeSNNeg_i;\NodeSNNeg_j\r)-\GAsy\s\l(\NodeSNNeg_j;\NodeSNNeg_i\r)=0,\quad
\GAsy\s\l(\NodeSNNeg_i;\NodeSNPos_j\r)-\GAsy\s\l(\NodeSNNeg_j;\NodeSNPos_i\r)=0,
\notag\\[4pt] &
\GAsy\s\l(\NodeSNNeg_i;\NodeSNPos_i\r)=0,\qquad
     \forall i, j \in\{1,2,\cdots, \NodeNumberTotalSNNeg\},\ \ i\not=j
\label{LP_Constraint_FromSymmetry}
\end{align} 
Constraint \eqref{tocBounds} is cast in the equivalent form of
\begin{align}
&
\GAsy\s\l(\NodeSNNeg_i;\NodeSNNeg_j\r)\leq 1,\quad
-\GAsy\s\l(\NodeSNNeg_i;\NodeSNNeg_j\r)\leq 1,\quad
\GAsy\s\l(\NodeSNNeg_i;\NodeSNPos_j\r)\leq 1,
\notag\\[4pt] &
-\GAsy\s\l(\NodeSNNeg_i;\NodeSNPos_j\r)\leq 1,\ \
     \forall i, j \in\{1,2,\cdots, \NodeNumberTotalSNNeg\}
\label{LP_Constraint_LowerBounds}
\end{align}

\end{document}